\documentclass[preprint2]{aastex61}

\newcommand{\WISE}{{\em WISE}\ }

\newcommand{\mko}{$_{\mathrm MKO}$}

\newcommand{\Ms}{CH$_{\mathrm 4}$s}
\newcommand{\Ml}{CH$_{\mathrm 4}$l}
\newcommand{\Msl}{\Ms--\Ml}

\shortauthors{Tinney {\it et~al.\/}}
\shorttitle{Y and T dwarfs identified by Methane Imaging}
\submitjournal{The Astrophysical Journal Supplement}
\received{29 December 2017}
\revised{21 February 2018}
\accepted{27 March 2018}
\begin{document}

\title{New Y and T dwarfs from WISE identified by Methane Imaging\footnote{This paper includes data gathered with the 3.9m Anglo-Australian Telescope located at Siding Spring Observatory, Coonabarabran, Australia.}\footnote{This paper includes data gathered with the 6.5 meter Magellan Telescopes located at Las Campanas Observatory, Chile.}}

\author[0000-0002-7595-0970]{C. G. Tinney}
\affiliation{Exoplanetary Science at UNSW, School of Physics, UNSW Sydney, NSW 2052, Australia.}
\affiliation{Australian Centre for Astrobiology, UNSW Sydney, NSW 2052, Australia}

\author{J. Davy Kirkpatrick}
\affiliation{IPAC, Mail Code 100-22, Caltech, Pasadena, CA 91125}

\author[0000-0001-6251-0573]{Jacqueline K. Faherty}
\affiliation{Department of Astrophysics, American Museum of Natural History, New York, NY 10023}

\author[0000-0001-7875-6391]{Gregory N. Mace}
\affiliation{McDonald Observatory and Department of Astronomy, University of Texas at Austin, Austin, TX 78712-1205, USA}

\author[0000-0001-7780-3352]{Mike Cushing}
\affiliation{Department of Physics and Astronomy, The University of Toledo, OH 43606, USA }

\author{Christopher R. Gelino}
\affiliation{IPAC, Mail Code 100-22, Caltech, Pasadena, CA 91125}

\author[0000-0002-6523-9536]{Adam J. Burgasser}
\affiliation{Center for Astrophysics and Space Science, University of California San Diego, La Jolla, CA 92093}

\author[0000-0003-3145-8682]{Scott S. Sheppard}
\affiliation{Department of Terrestrial Magnetism, Carnegie Institution for Science, Washington, DC 20015, USA}

\author[0000-0001-5058-1593]{Edward L. Wright}
\affiliation{Department of Physics and Astronomy, UCLA, Los Angeles, CA 90095-1547, USA}

\correspondingauthor{C. G. Tinney}
\email{c.tinney@unsw.edu.au}

\begin{abstract}
We identify new Y- and T-type brown dwarfs from the {\em WISE All Sky} data release using 
images  obtained in  filters that divide the traditional near-infrared H and J bands into 
two halves  -- specifically \Ms\ \& \Ml\ in the H and J2 \& J3 in the J. This 
proves to be very effective at identifying cool brown dwarfs via 
the detection of their methane absorption, as well as
providing preliminary classification using methane colours and \WISE-to-near-infrared colours. 
New and updated calibrations between T/Y spectral types
and \Msl\, J3--W2, and \Ms--W2 colours are derived, producing classification estimates good to a few spectral sub-types. 
We present photometry for a large sample of T and Y dwarfs in these filters, together with 
spectroscopy for 23 new ultra-cool dwarfs -- two Y dwarfs and twenty one T dwarfs.
We  identify a further 8 new cool brown dwarfs, which we have high confidence are T dwarfs based on 
their methane photometry. 
We find that, for objects observed on a 4m-class telescope at J band magnitudes of $\sim$20 or brighter,
\Msl\ is the more powerful colour for detecting objects and then estimating spectral types.
Due to the lower sky background in the J-band, the J3 and J2 bands are more useful for 
identifying fainter cool dwarfs at J$\gtrsim$22. The J3--J2
colour is poor at estimating spectral types. But fortunately, once J3--J2 confirms that
an object {\em is} a cool dwarf, the J3--W2 colour is very effective at estimating approximate spectral types.
\end{abstract}

\section{Introduction}
\label{intro}

Data from the NASA Wide-field Infrared Survey Explorer \citep[WISE;][]{wright2010,Mainzer11} have delivered  unprecedented advances 
in our understanding of the properties and space densities of the coldest compact astrophysical sources identified outside our 
Solar System -- the T- and Y-type brown dwarfs\citep[e.g.,][]{cushing2011,kirkpatrick2011,kirkpatrick2012,Kirkpatrick13a}. These very cold brown dwarfs have scientific impacts that span multiple astronomical arenas. In the field of star formation, they can deliver a historical record of the star formation process at very low masses and at epochs  billions of years prior to the star forming regions we observe today. In the field of planetary atmospheric theory, they represent low-temperature atmospheres that can be readily observed without the contaminating glare of a host star, and without the photochemical complications introduced by host star irradiation. In the field of exoplanet searches, they provide nearby, low-luminosity search targets potentially hosting planetary systems of their own, which would have implications for the debate on what differentiates a low-mass brown dwarf from a high-mass planet.\\

\WISE  readily identifies these very cold brown dwarfs as a result of their strong  thermal infrared methane absorption. 
The shortest wavelength \WISE band (hereafter W1) has a central wavelength of 3.4\,$\mu$m, which sits in the 
middle of the strong fundamental methane absorption band near 3.3\,$\mu$m. The second shortest \WISE band (hereafter W2), 
has a central wavelength of 4.6\,$\mu$m, where the photosphere is reasonably transparent and so detects flux from deeper, 
hotter layers in the brown dwarf. As a result, cold brown dwarfs can be identified via their very red W1--W2 colour. 
WISE's relatively uniform all-sky coverage, coupled with its ability to identify cool objects even when quite close to the Galactic plane, 
makes it the ideal data source from which to generate a complete thermal-infrared magnitude-limited sample of T and Y dwarfs in the Solar Neighbourhood. 
The generation of just such a T/Y-dwarf census is a key goal of the {\em WISE Science Team}  brown dwarf collaboration.
Large numbers of Y- and T-type brown dwarfs have been identified and spectroscopically confirmed to date by the {\em WISE Science Team} brown dwarf collaboration
using both colour selection \citep[see e.g.,][]{Mainzer11,cushing2011,kirkpatrick2011,kirkpatrick2012,tinney2012a,Mace13}, and more recently supplemented
by proper-motion selection from the {\em AllWISE} and {\em NEOWISE} samples \citep[see e.g.,][]{Kirkpatrick2014,Schneider2016,Kirkpatrick2016}. Additional cool brown dwarfs have been found by multiple teams independently working with public \WISE data \citep[e.g.,][]{Pinfield2014,Luhman2014}, while others have been identified from searches for cool companions to nearby stars  companions to nearby stars \citep[e.g.,][]{Luhman2011,Liu2012,Dupuy2015}.

The {\em WISE Science Team} brown dwarf collaboration has largely selected targets for follow-up on the basis of photometry consistent with
W1$-$W2 $>$ 2.9 over the full range of {\em WISE} W2 magnitudes, plus somewhat bluer 
objects (i.e. down to W1$-$W2 = 2.0) if  colours and magnitudes suggested a distance less than 20\,pc. 
In many cases these objects are non-detections at W1 (i.e. their W1$-$W2 colours are 3-$\sigma$ upper limits). This substantially
increases the observational phase space probed by this follow-up program, at the cost of increasing the number of ``false positives'' 
identified from the \WISE data, which must then be eliminated by subsequent observations.\\

Follow-up observations require the identification and spectral classification of a near-infrared counterpart to the \WISE
source. Unfortunately, the coldest objects (i.e. the Y dwarfs) can be {\em very} faint in the
near-infrared  (i.e. $J\sim H \gtrsim$ 22), making them challenging targets even for 8\,m-class ground-based telescopes. 
T dwarfs in \WISE  will be somewhat brighter in the near-infrared, and this gives  
medium-sized 4\,m-class telescopes a role in targeting them as part of the completion of the \WISE T/Y-dwarf census.\\

Methane imaging can play a useful role in this process. The discovery of very strong and broad methane absorption 
in the spectrum of the known T dwarf \citep[Gl\,229B][]{Nakajima1995,Oppenheimer1995} made the use of this spectral signature
for T-dwarf identification using specially designed imaging filters self-evidently obvious. This
was confirmed when \cite{MrSack1996} use a circularly variable filter (i.e. an adjustable narrow band filter) to
differentially re-detect Gl\,229B. Subsequent use of methane imaging (rather than spectroscopy) 
to ``winnow'' out T dwarfs from wide-field survey data was first carried out by \cite{Herbst1999}, while the first
use of methane filters in blind searches for new T dwarfs was carried out by \cite{Mainzer2003}. At around
the same time \cite{tinney2005} began using specially constructed methane filters in the IRIS2 instrument at 
the Anglo-Australian Telescope to identify T dwarfs from lists of candidates from the 2MASS survey.\\

The power of this technique for \WISE follow-up comes from the fact that the identification
of a near-infrared counterpart that shows methane absorption 
within a small distance on the sky from a \WISE {\em All-Sky} or {\em AlLWISE} position uniquely identifies this near-infrared
object as the counterpart to the \WISE source. Following which, the strength of the methane absorption can then provide a spectral type estimate, 
reducing the necessity for spectroscopic follow-up in all cases.\\

\section{Imaging -- Anglo-Australian Telescope (AAT)}
\label{AATimaging}

Imaging observations were carried out on the AAT on the nights listed (as local dates) in Table \ref{AATRuns} with  IRIS2 \citep{tinney2005}. 
IRIS2 is a near-infrared imager and spectrograph, with a single 1024$\times$1024 pixel detector giving an imaging field of view of
7.7$\arcmin$\ on a side at a pixel scale of 0.4486$\arcsec$/pixel. 

Data were obtained in seeing conditions ranging from 1.05$\arcsec$ to over 2.0$\arcsec$. Targets for observation were selected in 2011 from then extant catalogues using the pipeline software developed to produce the \WISE Preliminary Data Release (for details see the Explanatory Supplement\footnote{\url{http://wise2.ipac.caltech.edu/docs/release/prelim/expsup/}}), and in 2012 from the \WISE {\em All-Sky} release. As the optimal candidate selection procedures were being developed at the same time as improved versions of the \WISE pipeline, we report in Table \ref{ImageTable} the \WISE photometry for our observed sources as presented in the \WISE All-Sky data release made in March 2012 (rather than the photometry based on which they were originally selected). \\

In general the selected candidates satisfy the criteria that they were detected in W2 and had W1--W2$>$2.0. 
The \WISE processing pipeline records either detections at greater than 3-$\sigma$ significance, or 3-$\sigma$ upper limits.  Requiring a W1 detection (or 3-$\sigma$ upper limit) and
W1--W2$>$2.0, pushes the effective W2 detection floor much
brighter than that imposed by the pipeline's 3-$\sigma$ detection threshold. Our faintest T and Y dwarf candidates typically had W2 $\lesssim$ 15.5, which corresponds to a signal-to-noise ratio of $\sim$10.
Additional selection criteria have  been found to greatly assist in improving the rejection of non-brown dwarf contaminants. Broadly we require
the source was not identified as an artifact (i.e., known to be spurious), and that it wasn't flagged as blended (in which case it would have poorly determined photometry). For more details, including values
of the specific flags used, see \S2.1 of \cite{kirkpatrick2012}.\\

\begin{deluxetable}{llll}
\tabletypesize{\scriptsize}
\tablecaption{AAT \Ms,\Ml\ Methane Imaging.\label{AATRuns}} 
\tablewidth{0pt}
\tablehead{
\colhead{Run}          & \colhead{Prog.} & \colhead{Useful Nights}   & \colhead{Seeing} }
\startdata
2011 Jun 11-18			    & 11A/14	& \nodata					& \nodata	\\
2011 Sep 6-13	          & 11B/26	& Sep\,6,7,10,12,13		& 1.05-1.5\arcsec \\
2012 May 30-Jun 3		    & 12A/27	& May\,30,31				& 1.2-2.0\arcsec \\
2012 Jun 27-30			    & 12A/27	& Jun\,28,29,30			& 1.2-2.0\arcsec \\
2012 Sep	4-6		       & 12B/26   & Sep\,4,5,6				& 1.4-2.2\arcsec \\
2012 Dec 28-Jan 1        & 12B/26   & Dec\,28,29,30,31,Jan\,1 & 1.2-2.6\arcsec \\
\enddata
\end{deluxetable}

\subsection{J-band imaging}

Each candidate was initially observed in J-band with a planned exposure time of 54 minutes, broken up into thirty-six 90s pseudo-randomly dithered exposures. On-line data reduction using the ORACDR pipeline system \citep{cavanagh2008}\footnote{See also \tt\url{http://www.oracdr.org/}} would then produce a near-publication-quality processed sub-mosaic soon after the first 9 images of this dither pattern were completed.  This processed image was then analysed using purpose-built Perl scripts that automated the extraction of photometry for the image \citep[using SExtractor;][]{bertin96}, followed
by the  photometric and astrometric calibration of that data using the 2MASS Point Source Catalog \citep[PSC;][]{2MASS}. \\

This meant that within 5 minutes of the completion of the first 9-image sub-mosaic, we would know whether a J-band positional counterpart had been identified down to J$\approx$19.5. If a positionally matched candidate clearly emerged after this first sub-mosaic (when such an object emerged they usually did so at the $>$10-$\sigma$ level), then the dithering sequence would be truncated, and methane imaging observations  begun. In the absence of an ``early'' match the J-band imaging sequence was allowed to run to completion, providing imaging data to a depth of J$\approx$21.\\

\subsection{Methane imaging}

Once a plausible J-band counterpart had been identified, methane imaging observations were carried out with the \Ms\ and \Ml\ filters in IRIS2. The use of these filters for the study of cool brown dwarfs is described in \cite{tinney2005} --  we summarise just the salient points here. The IRIS2 methane filters (\Ms\ and \Ml) divide the H-band in half, sampling the wavelength ranges 1.520-1.620\,$\mu$m and 1.640-1.740\,$\mu$m (respectively). Flux in the \Ml\ filter is substantially depressed by methane absorption in T and Y dwarfs, and so the \Msl\ colour provides a powerful means of determining T and Y dwarf classifications. Methane filter observations are obtained by interleaving \Ms\ and \Ml\ observations (in an ABBA sense), while also dithering the telescope on sky. These data are also processed ``on the fly'' by ORACDR and deliver pairs of reduced images after every 7-9 pairs of images, which are then processed and differentially calibrated onto the \Msl\ photometric system of \cite{tinney2005}. As with the J-band imaging, purpose-built scripts are run as soon as the first pair of \Ms\ and \Ml\ mosaics are produced. Once again, if a clear methane signature was detected the observing sequence was truncated, and observations moved on to the next \WISE candidate.\\

\subsection{Photometry}

Following the initial processing done at the telescope, the ORACDR pipeline was used to reprocess these data for final analysis. A simple two-step process is followed: in the first step {\em all} images on each object are processed together to produce a flat-field and a first-pass mosaic, allowing the flattened images that went into this first-pass mosaic to be analysed to determine their photometric zero-point and image quality; in the second-pass, frames with poor image quality or poor photometric throughput are culled and the remaining images reprocessed to produce a final mosaic.
Photometry was then obtained from the final mosaics on each field, by differentially calibrating onto the J$_{\mathrm MKO}$ or 
\Msl\ systems using 2MASS PSC photometry for stars lying within the IRIS2 field of view. \\

For J band data this differential calibration is achieved by identifying objects in the field of view with 2MASS $-0.2 < J-K < 1.5$ and the converting those J magnitudes from the 2MASS photometric system to the MKO photometric system, using the equations determined by J. Carpenter in the 2MASS All-sky Survey Explanatory Supplement\footnote{\url{http://www.ipac.caltech.edu/2mass/releases/allsky/doc/sec6\_4b.html}}. As the IRIS2 J,H,K,Ks filters are MKO ones \citep[i.e. manufactured to the prescriptions of ][]{tokunaga2002}, the result is J photometry on the MKO system.\\

For \Ms,\Ml\ data, objects in the field-of-view for which transformations from the 2MASS to MKO photometric systems are reliable \citep[i.e. have 2MASS colours in ranges where the Carpenter transformations are valid -- see][]{tinney2005} are converted to MKO H, and then used to define zero-points for the \Ms\ and \Ml\ systems, and so determine \Msl. The resulting zero-points for a given field are typically determined to between $\pm$0.02-0.1 mag, and the scatter about the zero-point determination is used to determine a standard error in the mean, which is then propagated to the finally quoted photometric uncertainties. \\

Table \ref{ImageTable} presents J and \Msl\ photometry from the AAT, as well as \WISE W1 and W2 photometry from 
the {\em All-Sky}  release -- the official \WISE designation for all sources combines the prefix 
``WISE'' with the position designation in column 1 of the table -- e.g. WISE\,J001505.87-461517.6. These are abbreviated thereafter with the 
letter ``W'' and the first four digits of the right ascension and declination 
of the designation -- e.g. W0015-4615. Also listed in the table are spectral types resulting from 
spectroscopic observations in this paper (\S\ref{spectra}) and other programs, as well as estimated spectral types based on a new \Msl\ calibration derived in \S\ref{newcal}.

\begin{longrotatetable}
\startlongtable
\begin{deluxetable}{rclcllllllll}
\tabletypesize{\scriptsize}
\tablecaption{\WISE, J$_{\mathrm MKO}$ and \Msl photometry for candidate cool \WISE brown dwarfs\label{ImageTable}}
\tablewidth{0cm}
\tablehead{
\colhead{WISE Desig.\tablenotemark{a}} 
                    &\colhead{W1} 
                                   &\colhead{W2}
                                                  &\colhead{W1$-$W2}
                                                              &\colhead{J$_{\mathrm MKO}$\tablenotemark{b}}
                                                                           &\colhead{\Ms\tablenotemark{c}} 
                                                                                        &\colhead{\Msl\tablenotemark{c}}
                                                                                                      &\colhead{CH$_4$ SpT\tablenotemark{d}}
                                                                                                                 &\colhead{J$-$W2}
                                                                                                                              &\colhead{\Ms$-$W2}
                                                                                                                                          &\colhead{Spec}
                                                                                                                                               &\colhead{Notes}}
\startdata
\multicolumn{12}{l}{\bf T/Y dwarfs : Position matches with unresolved sources showing CH$_4$ absorption.}\\[6pt]
J001505.87$-$461517.6 & 17.02 0.15 & 14.25 0.06 & 2.77 0.16 & 17.76 0.06 & 17.45 0.06 &$-$1.24 0.16 &  T7.0 0.5& 3.51 0.09 & 3.23 0.08 & T8    & FIRE, this paper\\     
J003231.09$-$494651.4 & 18.08 0.34 & 15.07 0.09 & 3.01 0.35 & 18.54 0.09 & 18.20 0.07 &$-$1.96 0.23 &  T8.6 0.5& 3.47 0.13 & 3.13 0.11 & T8.5  & FIRE, this paper\\     
J014807.25$-$720258.7 &$>$18.94    & 14.69 0.05 &$>$4.25    & \nodata    & 18.77 0.05 &$-$2.34 0.20 &  T9.2 0.5& \nodata   & 4.08 0.07 & T9.5  & Kirkpatrick et al. 2012\\                                          
J024124.73$-$365328.0 & 16.89 0.10 & 14.34 0.04 & 2.55 0.11 & 16.59 0.04 & 16.55 0.03 &$-$1.06 0.06 &  T6.5 0.5& 2.25 0.06 & 2.21 0.08 & T7    & FIRE, this paper\\                           
J030919.67$-$501614.3 & 16.51 0.08 & 13.61 0.03 & 2.90 0.09 & 17.17 0.03 & 17.01 0.03 &$-$1.31 0.06 &  T7.2 0.5& 3.56 0.10 & 3.40 0.05 & \nodata & \\                                                             
J032504.33$-$504400.3 &$>$18.73    & 15.70 0.10 &$>$3.03    & 18.94 0.09 & 18.39 0.09 &$-$2.20 0.36 &  T9.0 0.5& 3.24 0.13 & 2.69 0.13 & T9    & Schneider et al. 2015 \\                 
J035000.32$-$565830.2 &$>$18.90    & 14.73 0.06 &$>$4.17    & 22.47 0.49 &\nodata     & \nodata     &\nodata   & 7.74 0.51 & \nodata   & Y1    & Kirkpatrick et al . 2012\\                   
J040443.48$-$642029.9 &$>$18.86    & 15.73 0.09 &$>$2.60    & 19.55 0.22 & 19.43 0.04 &$-$1.27 0.10 &  T7.1 0.5& 3.82 0.23 & 3.70 0.10 & T9    & Schneider et al. 2015\\ 
J041022.71$+$150248.4 &$>$18.33    & 14.18 0.06 &$>$4.15    & \nodata    & 20.47 0.07 & \nodata     &\nodata   & \nodata   & 6.29 0.09 & Y0    & Kirkpatrick et al. 2012 \\            
J062842.71$-$805725.0 &$>$18.78    & 15.45 0.08 &$>$3.32    & 18.71 0.16 & 18.34 0.04 &$-$2.30 0.13 &  T9.1 0.5& 3.26 0.18 & 2.89 0.09 & \nodata & \\ 
J064528.38$-$030248.2 &$>$18.18    & 14.94 0.10 &$>$3.24    & 16.91 0.04 & 16.88 0.04 &$-$1.00 0.09 &  T6.3 0.5& 1.97 0.12 & 1.94 0.15 & T6    & SpeX, this paper\\
J071322.55$-$291751.9 &$>$18.35    & 14.48 0.06 &$>$3.87    & 19.64 0.15 & 19.33 0.04 &$-$2.72 0.15 &  T9.7 0.5& 5.16 0.16 & 4.85 0.08 & Y0    & Kirkpatrick et al. 2012 \\            
J071301.84$-$585445.1 &$>$19.04    & 15.44 0.07 &$>$3.60    & \nodata    & \nodata    & \nodata     & \nodata  & \nodata   & \nodata   & T9     & FIRE, this paper        \\
J072227.27$-$054029.9\tablenotemark{e} & 15.19 0.05 & 12.21 0.03 & 2.98 0.06 & 16.52 0.02 & 16.39 0.02 &$-$2.12 0.04 &  T8.9 0.5& 4.31 0.03 & 4.18 0.03 & T9\tablenotemark{f} & Kirkpatrick et al. 2012\\                                            
J091408.96$-$345941.5 &$>$17.83    & 15.03 0.09 &$>$2.80    & 18.36 0.11 & 17.79 0.02 &$-$1.70 0.07 &  T8.2 0.5& 3.33 0.14 & 2.76 0.10 & T8    & FIRE, this paper\\   
J094020.10$-$220820.5 &17.01 0.14  & 14.57 0.07 & 2.43 0.16 & 17.36 0.05 & 17.30 0.04 &$-$1.52 0.11 &  T7.8 0.5& 2.79 0.17 & 2.73 0.08 & T8    & SpeX, this paper \\
J105553.59$-$165216.3 &$>$18.37    & 15.04 0.10 &$>$3.33    & 20.79 0.02 & 20.35 0.06 &$<-1.69$     &$>$T8.2   & 5.75 0.11 & 5.31 0.14 & T9    & FIRE, this paper \\
J111239.24$-$385700.7 & 17.97 0.40 & 14.36 0.06 & 3.61 0.40 & 20.26 0.15 & 19.94 0.05 &$-$1.72 0.12 &  T8.2 0.5& 5.90 0.16 & 5.58 0.08 & T9    & FIRE, this paper \\
J114156.71$-$332635.8 & 17.20 0.17 & 14.53 0.06 & 2.67 0.18 & 19.76 0.14 & 19.69 0.04 &($-$0.71 0.09)\tablenotemark{g} &  (T5.4 0.5)\tablenotemark{g}
                 																																& 5.23 0.15 & 5.16 0.07 & Y0    & FIRE, this paper \\                                       
J143311.42$-$083736.4 &$>$18.74    & 15.23 0.10 &$>$3.51    & 19.07 0.11 & \nodata    & \nodata     &\nodata   & 3.84 0.15 & \nodata   & T8\tablenotemark{h}
                                                                                                                                         & FIRE, this paper \\                     
J144806.48$-$253420.3 &$>$18.28    & 15.03 0.09 &$>$3.25    & 18.85 0.11 & 18.71 0.13 &$-$1.02 0.29 &  T6.4 0.5& 3.82 0.14 & 3.68 0.16 & T8    & FIRE, this paper\\  
J150115.92$-$400418.4 & 16.48 0.11 & 14.21 0.05 & 2.07 0.12 & 16.53 0.04 & 16.05 0.03 &$-$0.86 0.06 &  T5.9 0.5& 2.32 0.06 & 1.84 0.05 & T6    & SpeX, this paper\\  J173551.72$-$820900.1 & 15.61 0.06 & 13.73 0.04 & 1.88 0.07 & 16.58 0.03 & 16.41 0.03 &$-$0.76 0.08 &  T5.6 0.5& 2.85 0.05 & 2.68 0.05 & T6 & FIRE, this paper\\                                   
J201748.72$-$342102.5 &$>$18.21    & 15.09 0.13 &$>$3.12    & 20.89 0.24 & 20.19 0.06 &$-$1.46 0.15 &  T7.7 0.5& 5.80 0.27 & 5.10 0.14 & \nodata & \\                                                                                
J205628.91$+$145953.2 &$>$18.25    & 13.93 0.05 &$>$4.33    & \nodata    & 18.99 0.04 &$-$3.17 0.25 &  Y0.2 0.5& \nodata   & 5.06 0.06 & Y0    & Kirkpatrick et al . 2011\\                                                        
J210200.15$-$442919.5 & 16.94 0.17 & 14.12 0.05 & 2.83 0.18 & 18.25 0.06 & 18.06 0.06 &$-$2.20 0.24 &  T9.0 0.5& 4.13 0.08 & 3.94 0.08 & T9    & NIRSPEC, this paper\\                                                          
J215949.48$-$480854.9 & 17.76 0.29 & 14.54 0.07 & 3.19 0.30 & 18.89 0.08 & 18.64 0.09 &$-$1.31 0.28 &  T7.2 0.5& 4.35 0.11 & 4.10 0.13 & T9    & FIRE, this paper\\                     
J221140.52$-$475826.5 & 17.67 0.21 & 14.58 0.06 & 3.09 0.22 & 17.38 0.04 & 17.38 0.04 &$-$1.51 0.14 &  T7.8 0.5& 2.80 0.07 & 2.80 0.07 & \nodata &\\                                                                                  
J221216.33$-$693121.6 & 18.05 0.40 & 14.90 0.09 & 3.15 0.41 & 19.72 0.13 & 19.29 0.04 &$-$3.15 0.23 &  Y0.2 0.5& 4.82 0.16 & 4.39 0.10 & T9.5  & FIRE, this paper\\                     
J222055.31$-$362817.4 &$>$18.65    & 14.66 0.06 &$>$3.99    & 20.38 0.17 & 20.20 0.07 &$<-$1.52     &$>$T7.8   & 5.72 0.18 & 5.58 0.09 & Y0    & Kirkpatrick et al . 2012\\                                                        
J223204.50$-$573010.5 &$>$17.90    & 15.15 0.11 &$>$2.75    & 18.86 0.09 & 18.74 0.10 &$-$1.49 0.30 &  T7.7 0.5& 3.71 0.14 & 3.59 0.14 & T9    & FIRE, this paper\\                      
J230228.68$-$713441.6 & 17.14 0.14 & 14.28 0.05 & 2.86 0.14 & 16.97 0.04 & 17.52 0.07 &$-$1.14 0.15 &  T4.7 0.5& 2.69 0.06 & 3.24 0.09 & \nodata & \\                                                                                  
J233226.49$-$432510.6 &$>$17.88    & 14.99 0.09 &$>$2.89    & \nodata    & 19.14 0.04 &$<-$2.59     &$>$T9.5   & \nodata   & 4.15 0.09 & T9:   & NIRSPEC, this paper \\                          
J235425.33$-$564928.6 & 16.93 0.13 & 14.84 0.08 & 2.09 0.15 & 17.09 0.04 & 16.98 0.04 &$-$0.55 0.07 &  T7.8 0.5& 2.25 0.09 & 2.14 0.09 & T6 & FIRE, this paper\\    
J235447.80$-$814044.9 & 18.23 0.33 & 15.03 0.07 &$>$3.20    & 17.64 0.05 & 17.68 0.06 &$-$1.04 0.12 &  T6.4 0.5& 2.61 0.09 & 2.65 0.10 & \nodata & \\[6pt]                                             
\multicolumn{12}{l}{\bf T/Y dwarfs : Position matches ($<$1\arcsec) with an extended source, sources showing no methane absorption and/or non-detections.}\\[6pt]
J012102.92$-$190656.9 & 16.47 0.09  & 14.71 0.07 & 1.76 0.11 & 19.67 0.16 & 18.87 0.13 &  +0.17 0.17 &  \nodata  & 4.96 0.17 & 4.16 0.15 &\nodata& \tablenotemark{i}\\ 
J071939.54$-$173514.8 & 16.72 0.07  & 13.71 0.03 & 3.01 0.08 & \nodata    & 20.1  0.1  &  +0.33 0.15 &  \nodata  & \nodata   & 6.39 0.17 &\nodata& \tablenotemark{j}\\
J074551.79$-$015122.1 & 18.14 0.41  & 14.97 0.09 & 3.17 0.42 & 19.90 0.16 & 19.53 0.15 &  +0.25 0.20 &  \nodata  & 4.93 0.17 & 4.56 0.17 &\nodata& \tablenotemark{i} \\
J083942.85$-$402938.9 & 17.37 0.24  & 13.66 0.04 & 3.71 0.25 & 20.6  0.2  & 18.82 0.11 &  +0.40 0.15 &  \nodata  & 6.94 0.22 & 5.16 0.12 &\nodata& \tablenotemark{k}\\
J130740.45$-$463035.1 & 15.00 0.03  & 12.93 0.03 & 2.07 0.04 & \nodata    & 16.59 0.04 &  +0.24 0.05 &  \nodata  & \nodata   & 3.66 0.05 &\nodata& \tablenotemark{l}\\ 
J150711.06$-$344026.0 & 15.64 0.06  & 14.03 0.05 & 1.62 0.08 & 17.73 0.04 & 17.18 0.05 &  +0.26 0.07 &  \nodata  & 3.70 0.07 & 3.15 0.09 &\nodata& \tablenotemark{m} \\ 
J164445.19$-$645628.9 &$>$18.50     & 15.22 0.10 &$>$3.28    & \nodata    & 18.13 0.09 &  +0.15 0.11 &  \nodata  & \nodata   & 2.91 0.13 &\nodata& \tablenotemark{i}\\                                              
J185709.40$-$315345.5 & 16.48 0.12  & 14.42 0.07 & 2.06 0.14 & 18.58 0.10 & 17.84 0.09 &  +0.05 0.12 &  \nodata  & 4.16 0.12 & 3.42 0.11 &\nodata& \tablenotemark{i}\\ 
J190230.27$-$371246.1 & 17.18 0.21  & 14.46 0.07 & 2.72 0.22 & 20.27 0.29 & \nodata    & \nodata     &  \nodata  & 5.81 0.30 & \nodata   &\nodata& \tablenotemark{n}\\                                                              
J193441.70$-$490837.6 & 17.00 0.12  & 14.50 0.06 & 2.50 0.13 & 20.36 0.31 & \nodata    & \nodata     &  \nodata  & 5.86 0.32 & \nodata   &\nodata& See Table \ref{MagTable}\\
J203020.25$-$692043.0 & 18.06 0.43  & 14.83 0.09 & 3.23 0.44 & 19.51 0.17 & 18.83 0.15 &  -0.07 0.23 &  \nodata  & 4.68 0.20 & 4.00 0.17 &\nodata& \tablenotemark{o}\\ 
J203119.30$-$690500.3 &$>$17.58     & 14.41 0.07 &$>$3.17    & 20.44 0.22 & 19.18 0.03 &  -0.41 0.06 &  T4.2 0.5 & 6.03 0.23 & 4.77 0.08 &\nodata& \tablenotemark{p}\\         
J224245.85$-$201511.0 &$>$18.32     & 15.15 0.13 &$>$3.17    & 20.03 0.16 & \nodata    & \nodata     &  \nodata  & 4.88 0.21 & \nodata   &\nodata& \tablenotemark{q}\\ 
\enddata                                                                                                                                             
\tablenotetext{a}{\WISE object designations and photometry are from the March 2012 release of the All-sky Source Catalog, unless an ``A'' prefix is used in which case they come from the November 2013 AllWISE Source Catalog release. 
The All-Sky WISE designation combines the prefix ''WISE'' with the position designation (i.e. WISE\,J001505.87-461517.6 for an All-sky Source and WISEA\,J030237.53$-$581740 for an AllWISE Source). These are abbreviated with the first four digits of the right ascension and declination throughout this paper -- e.g. W0015-4615. }
\tablenotetext{b}{IRIS2 J photometry was obtained through a J$_{\mathrm MKO}$ filter, differentially calibrated using J$_{\mathrm 2MASS}$ photometry converted to J$_{\mathrm MKO}$ in each field.}
\tablenotetext{c}{IRIS2 \Ms and \Msl photometry calibrated (as described in the text) on the system of \citet{tinney2005}.}                                                                                                                                                                                                                                              
\tablenotetext{d}{Estimated spectral types using \Msl\ and the calibration of \S\ref{newcal}.}                                                                                                                        
\tablenotetext{e}{The discovery name for this object is UGPS\,J072227.51-054031.2 \citep{lucas2010} and J$_{\mathrm MKO}$ is from that paper.}                                                                                                                          
\tablenotetext{f}{While noting that the spectral type for this object has been the subject of some debate, we adopt the T9 type of \citet{kirkpatrick2012}}                                                                                                                          
\tablenotetext{g}{The methane colour of W1141 suggests a spectral type of $\approx$T5.5, however the J--W2 colour is indicative of a type of T9 or later. As discussed
in the text and shown in Fig. \ref{W1141}, the \Msl\ (and especially the \Ml) photometry is contaminated by a background galaxy.The J3$-$W2 photometry (\S\ref{J3W2_SpT}) and the FIRE spectrum (\S\ref{spectra})
clearly show this to be a Y0. We do not include this object in the \Msl\ calibration of \S\ref{newcal}.}                                                                                                                          
\tablenotetext{h}{143311.42$-$083736.4: A spectrum and spectral type for this object has also been  published by Lodieu et al. (2012). Those values are in agreement with those presented here.}                                                                                                                          
\tablenotetext{i}{J012102.92-190656.9, J074551.79-015122.1, J164445.19-645628.9 and J185709.40-315345.5: a bright source has a position match with the \WISE position. The \Ms-\Ml\ colours indicates they are not T dwarfs.}                                                                                                                          
\tablenotetext{j}{J071939.54-173514.8: extended source at position with no methane absorption}
\tablenotetext{k}{J083942.85-402938.9: is clearly an extended source (2.1\arcsec$\times$3.0\arcsec), with no methane absorption. WISE source is likely this galaxy.}
\tablenotetext{l}{J130740.45-463035.1: a bright \Ms=16.60 source has a position match with the \WISE position, and shows no proper motion since the 2MASS epoch.. The \Ms-\Ml\ colour indicates this object is not a T dwarf.}                                                                                                                          
\tablenotetext{m}{J150711.06-344026.0: this bright W2 source was observed as a poor conditions backup target and it has a position match with the \WISE position, but the \Ms-\Ml\ colour indicates this object is not a T dwarf. Shows no proper motion since the 2MASS epoch).}                                                                                                                          
\tablenotetext{n}{J190230.27-371246.1: very extended (3.5\arcsec) source at position, making this galaxy the likely WISE source.}
\tablenotetext{o}{J203020.23-692043.1: has excellent position match object with J=19.51, however this is marginally resolved (2.2\arcsec\ fwhm) in 2.0\arcsec\ seeing, and shows no methane absorption. It is likely the \WISE\ flux arises from this galaxy.}                                                                                                                          
\tablenotetext{p}{J203119.30-690500.3 has excellent position match object with J=20.44 and methane absorption. However the object is resolved at J (2.2\arcsec\ fwhm in 2.0\arcsec) and H (1.84\arcsec\ fwhm in 1.39\arcsec\ seeing). Moreover, the methane colour suggests a spectral type of T4 which is inconsistent with the J-W2 and \Ms-W2 colours (which suggest a much later spectral type of $>$T9). This identification is therefore considered tentative. }                                                                                                                          
\tablenotetext{q}{J224245.85-201511.0 has J=20.03 object 1.2\arcsec\ from the \WISE\ position. However, this object has an almost identical J and J2 magnitudes which suggests methane absorption would be extremely weak.}
\end{deluxetable}                                                                                                                                     
\end{longrotatetable}

\section{Imaging -- Magellan} \label{MAGimaging}

Imaging observations were obtained using the FourStar imaging camera \citep{persson2013} on the Magellan Baade telescope between March
10, 2012 and January 5, 2017. FourStar is a near-infrared mosaic imager with four 2048$\times$2048 pixel detectors giving an imaging field of view of
11$\arcmin$\ on a side at a pixel scale of 0.159$\arcsec$/pixel. It is equipped with a set of intermediate-band filters, originally specified
for the measurement of photometric redshifts. These filters turn out to be almost ideally suited for observing very cool 
brown dwarfs \citep[see Fig.1 in][]{tinney2012a}.
In particular,  the J3 filter ($\lambda_{cen} \approx 1.29\,\mu$m, 90\% of peak throughput range 1.210-1.366\,$\mu$m) collects 
almost all of the ``methane free'' J-band flux from late-T and Y dwarfs, while
the J2 filter ($\lambda_{cen} \approx 1.14\,\mu$m, 90\% of peak throughput range 1.067-1.224\,$\mu$m) 
is strongly impacted by methane absorption between 1.1\,$\mu$m and 1.2\,$\mu$m. All the observations described here  were
performed with the \WISE target positioned in FourStar's Chip\,2.\\

\begin{deluxetable}{cc}
\tabletypesize{\scriptsize}
\tablewidth{0pt}
\tablecaption{Magellan J2/J3 Methane Imaging.\label{MagRuns}}
\tablehead{
\colhead{UT Date} & \colhead{Median Seeing ($\arcsec$)}                
}
\startdata
2012 Mar 10	& 0.64 \\
2012 May 10	& 0.56 \\
2012 Jul 6	& 0.70 \\
2012 Jul 7	& 0.53 \\
2013 Jan 15	& 0.40 \\
2013 Mar 22	& 0.56 \\
2013 Apr 22	& 0.57 \\
2013 Aug 15 & 0.50 \\
2016 Nov 18 & 0.80 \\
2017 Jan 05 & 0.54 \\
\enddata
\end{deluxetable}

Image quality over the course of this program varied between 0.4$\arcsec$ and 0.7$\arcsec$  (see Table \ref{MagRuns}).  
Our observing and analysis techniques follow those previously described by us \citep{tinney2012a,tinney2014a}, and involve observing each 
target with the FourStar J3 and J2 filters in a sequence of 60-120s pseudo-randomly dithered exposures. 
Targets are observed with net integration times ranging from 5 minutes to 1.0h.
As with the AAT data,  images are processed at the telescope
using a modified version of the ORACDR\footnote{\url{http://www.jach.hawaii.edu/JACpublic/UKIRT/software/oracdr}} data reduction pipeline, and
examined during observing to determine whether a counterpart had been detected. As for the AAT data, the final analysis involves
running ORACDR twice for each jitter set, with individual exposures of poor image quality removed from the list used in the second-pass.\\

Photometric processing and calibration followed the procedures outlined above for  AAT data, except for the final stage
of calibration onto the J2$/$J3 photometric system, which followed that outlined in \cite{tinney2012a}. That is, 2MASS sources in the field-of-view 
were calibrated onto the J\mko\ system, and we used this MKO photometry for stars in the range $0.4 < $ (J--K)\mko\ $ < 0.8$
to define the zero-point for the J2 and J3 magnitude system. Table \ref{MagTable} presents J2 and J3 photometry from Magellan, 
as well as \WISE W1 and W2 photometry (as contained in the March 2012 \WISE  {\em All-Sky} data release).\\

\begin{longrotatetable}
\startlongtable
\begin{deluxetable}{rllllllllll}
\tabletypesize{\scriptsize}
\tablecaption{\WISE and Magellan J3,J2 photometry of candidate cool brown dwarfs from \WISE\label{MagTable}}
\tablewidth{0cm}
\tablehead{\colhead{WISE Designation\tablenotemark{a}} 
                    &\colhead{W1} 
                                   &\colhead{W2}
                                                  &\colhead{W1$-$W2}
																	&\colhead{J3\tablenotemark{b}} 
                                                                             &\colhead{J3-J2}
																						&\colhead{J3$-$W2}
                                                                                                         &\colhead{Est. SpT\tablenotemark{c}}
                                                                                                                    &\colhead{SpT}
                                                                                                                              &\colhead{Notes}}
\startdata
\multicolumn{10}{c}{\bf T/Y dwarfs : Position matches with unresolved sources showing CH$_4$ absorption.}\\
AJ004206.84$-$584023.9&18.86 0.49  & 15.60 0.10 & 3.26 0.50 & 17.76 0.04     & $-$1.10 0.06 & 2.16 0.11 &  T6.6     & \nodata & \\
 J014807.25$-$720258.7&$>$18.94    & 14.69 0.05 &$>$4.25    & 18.83 0.02 (9)\tablenotemark{d}
                                                                             & $-$1.72 0.12 & 4.16 0.07 &  T8.9     & T9.5    & Kirkpatrick et al. 2013 \\
AJ030237.53$-$581740.3&$>$19.26    & 15.81 0.09 &$>$3.45    & 20.76 0.11     & $-$1.52 0.27 & 4.95 0.14 &  T9.5     & Y0      & FIRE, this paper\\
 J064723.23$-$623235.5&$>$19.09    & 15.32 0.08 &           & 22.51 0.09 (2) & $-$1.21 0.20 & 7.19 0.12 &  Y0.9     & Y1      & Kirkpatrick et al. 2014 \\
 J071322.55$-$291751.9&$>$18.35    & 14.48 0.06 &$>$3.87    & 19.42 0.03 (8) & $-$1.58 0.06 & 4.94 0.0  &  T9.5     & Y0      & Kirkpatrick et al. 2012 \\  
 J081117.81$-$805141.3& 17.29 0.12 & 14.38 0.04 & 2.91 0.13 & 19.31 0.01 (6) & $-$1.71 0.09 & 4.95 0.04 &  T9.5     & T9.5    & Mace et al. 2013 \\
 J091408.96$-$345941.5&$>$17.83    & 15.03 0.09 &$>$2.80    & 17.73 0.02     & $-$1.29 0.03 & 2.70 0.09 &  T7.4     & T8      & FIRE, this paper\\
AJ102313.22$-$315126.7&$>$18.94	  & 15.76 0.12 &$>$3.18    & 18.24 0.03     & $-$1.12 0.06 & 2.48 0.13 &  T7.1     & T8      & FIRE, this paper\\
 J111239.24$-$385700.7& 17.97 0.40 & 14.36 0.06 & 3.61 0.40 & 20.11 0.04 (4) & $-$1.36 0.11 & 5.77 0.07 &  Y0.1     & T9      & FIRE, this paper \\ 
 J114156.71$-$332635.8& 17.20 0.17 & 14.53 0.06 & 2.67 0.18 & 19.63 0.05 (6) & $-$1.30 0.10 & 5.22 0.08 &  T9.7     & Y0      & FIRE, this paper \\ 
 J122152.28$-$313600.8& 15.86 0.06 & 13.85 0.04 & 2.01 0.07 & 15.80 0.03     & $-$1.14 0.05 & 1.95 0.05 &  T6.3     & T5.5    & Mace et al. 2013 \\
 J150115.92$-$400418.4& 16.48 0.11 & 14.21 0.05 & 2.07 0.12 & 15.95 0.01     & $-$1.06 0.02 & 1.74 0.05 &  T5.8     & T6      & SpeX, this paper\\  
 J163940.83$-$684738.6& $>$17.89   & 13.65 0.05 &$>$4.24    & 20.62 0.08     & $-$1.65 0.12 & 6.98 0.09 &  Y0.8     & Y0      & Tinney et al. 2012\\
AJ172907.10$-$753017.0& 18.74 0.38 & 15.55 0.08 & 3.19 0.39 & 17.71 0.02     & $-$1.09 0.04 & 2.16 0.08 &  T6.6     & T7      & FIRE, this paper\\
 J173551.72$-$820900.1& 15.61 0.06 & 13.73 0.04 & 1.88 0.07 & 16.17 0.01     & $-$1.09 0.01 & 2.44 0.05 &  T6.9     & T6      & FIRE, this paper\\                                   
AJ204928.59$-$443143.9&$>$18.94    & 15.64 0.11 &$>$3.30    & 18.22 0.03     & $-$1.39 0.06 & 2.58 0.11 &  T7.3     & \nodata & \\
 J210200.15$-$442919.5& 16.94 0.17 & 14.12 0.05 & 2.83 0.18 & 18.08 0.01 (11)& $-$1.52 0.14 & 3.96 0.06 &  T8.7     & T9      & NIRSPEC, this paper\\                                          
 J233226.49$-$432510.6&$>$17.88    & 14.99 0.09 &$>$2.89    & 19.13 0.02 (6) & $-$1.64 0.18 & 4.19 0.10 &  T8.9     & T9:     & NIRSPEC, this paper\\[6pt]                                           
\multicolumn{10}{c}{\bf T/Y dwarfs : Position matches ($<$1\arcsec) with an extended source, sources showing no methane absorption and/or non-detections.}\\[6pt]
 J080811.36$-$682521.2& $>$19.27   & 16.18 0.11 &$>$3.08    & 20.97 0.07     & +0.12 0.09 &           &           &         &\tablenotemark{e}    \\
 J083440.09$-$643616.4&$>$18.79    & 14.75 0.05 &$>$4.04    & 20.29 0.03     & -0.21 0.05 &           &           &         &\tablenotemark{f}    \\
 J092958.73$-$712733.0& $>$19.29   & 16.40 0.14 &$>$2.89    & \nodata        &  \nodata   & \nodata   & \nodata   & \nodata &\tablenotemark{g}    \\
 J093730.46$-$735454.0& $>$18.97   & 16.15 0.12 &$>$2.81    & 19.47 0.05     & -0.29 0.07 &           &           &         &\tablenotemark{h}    \\
 J101242.66$-$482842.3& $>$18.66   & 15.73 0.12 &$>$2.93    & 20.86 0.06     & -0.26 0.10 &           &           &         &\tablenotemark{i}    \\
 J102719.82$-$443656.7&  18.12 0.47& 15.20 0.09 & 2.92 0.47 & 20.48 0.04     & -0.38 0.07 &           &           &         &\tablenotemark{j}    \\
 J103059.32$-$373140.3& $>$18.25   & 15.78 0.17 &$>$2.48    & 20.41 0.08     & -0.13 0.11 &           &           &         &\tablenotemark{k}    \\
 J120906.70$-$520034.5& $>$18.82   & 15.85 0.13 &$>$2.97    & 21.44 0.10     & -0.07 0.13 &           &           &         &\tablenotemark{l}    \\
 J122329.62$-$480051.3& $>$18.51   & 15.85 0.16 &$>$2.66    & 20.78 0.08     & -0.07 0.11 &           &           &         &\tablenotemark{m}    \\
 J123252.86$-$800525.4& $>$18.47   & 15.88 0.15 &$>$2.59    & \nodata        &  \nodata   & \nodata   & \nodata   & \nodata &\tablenotemark{n}    \\
 J124156.18$-$370345.5& 17.78 0.22 & 14.87 0.06 & 2.91 0.23 & 20.49 0.03     & -0.11 0.05 &           &           &         &\tablenotemark{o}    \\
 J130348.35$-$460959.4& $>$18.79   & 16.09 0.18 &$>$2.71    & 21.60 0.06     & +0.08 0.08 &           &           &         &\tablenotemark{p}    \\
 J143032.81$-$323321.7& 17.11 0.17 & 14.72 0.08 & 2.39 0.19 & 20.57 0.04     & -0.49 0.07 &           &           &         &\tablenotemark{q}    \\
 J153802.79$-$450936.5& $>$17.96   & 14.90 0.10 &$>$3.06    & See note       &            &           &           &         &\tablenotemark{r}    \\
 J171507.17$-$552116.8& $>$17.23   & 14.60 0.08 &$>$2.63    & 20.74 0.05     & -0.21 0.08 &           &           &         &\tablenotemark{s}    \\
 J193441.70$-$490837.6& 17.00 0.12 & 14.50 0.06 & 2.50 0.13 & 20.50 0.05     & -0.12 0.08 &           &           &         &\tablenotemark{t}    \\
 J211213.88$-$552855.5& $>$18.55   & 15.83 0.18 &$>$2.71    & 21.46 0.08     & -0.36 0.13 & \nodata   & \nodata   & \nodata &\tablenotemark{u}    \\
 J212505.20$-$323412.9& $>$18.48   & 15.06 0.11 &$>$3.42    & 20.21 0.03     & -0.11 0.05 &           &           &         &\tablenotemark{v}    \\
 J220011.60$-$544914.2& $>$17.37   & 14.86 0.11 &$>$2.51    & 19.78 0.03     & -0.11 0.04 &           &           &         &\tablenotemark{w}    \\
 J221342.90$-$294908.7& 17.87 0.43 & 14.59 0.09 & 3.28 0.44 & 20.73 0.01     & -0.14 0.04 &           &           &         &\tablenotemark{x}    \\[6pt]
\multicolumn{10}{c}{\bf Unconfirmed methane absorbers within 30\arcsec\ of \WISE position}\\[6pt]                                                  
 J011020.86$-$132313.0& $>$18.247  & 14.85 0.18&$>$3.39     & \nodata        &  \nodata   &  \nodata  & \nodata   & \nodata &\tablenotemark{aa}   \\
 J092320.42$-$731704.1& $>$19.124  & 16.68 0.19&$>$2.45     & \nodata        &  \nodata   &  \nodata  & \nodata   & \nodata &\tablenotemark{bb}   \\
AJ183437.34$-$584119.4& $>$18.714  & 15.67 0.11&$>$3.04     & \nodata        &  \nodata   &  \nodata  & \nodata   & \nodata &\tablenotemark{cc}   \\
\enddata                                                                                                                                             
\tablenotetext{a}{\WISE designations follow Table \ref{ImageTable}}
\tablenotetext{b}{FourStar J3 and J2 photometry processed as described in \citet{tinney2014a,tinney2012a}. J3 observations with numbers in parentheses are the average of the indicated number of independent observations, and the uncertainty is the standard error in the mean. For other observations the uncertainty is the photon-counting and calibration zero-point uncertainties combined in quadrature.}                                                                                                                          
\tablenotetext{c}{Estimated spectral type from J3$-$W2 colour and the calibration described in \S\ref{J3W2_SpT}.}                                                                                                                          
\tablenotetext{d}{The J3 photometry reported here corrects the incorrect value in Table 1 of \citet{tinney2012a}.}                                                                                                                        
\tablenotetext{e}{J080811.36$-$682521.2: Object 0.5\arcsec\ from \WISE position with no methane absorption. No methane absorbers within 30\arcsec\ to J3=22}
\tablenotetext{f}{J083440.09$-$643616.4: Galaxy (resolved in 0.48\arcsec\ seeing) 0.2\arcsec\ from \WISE. No methane absorbers within 30\arcsec\ to J3=22.}
\tablenotetext{g}{J092958.73$-$712733.0: No objects with methane absorption within 30\arcsec\  to J3=21}
\tablenotetext{h}{J093730.46$-$735454.0: Object 0.5\arcsec\ from \WISE position with no methane absorption. No methane absorbers within 30\arcsec\ to J3=21.5}
\tablenotetext{i}{J101242.66$-$482842.3: Object 0.6\arcsec\ from \WISE position with no methane absorption. No methane absorbers within 30\arcsec\ to J3=22}
\tablenotetext{j}{J102719.82$-$443656.7: Unresolved object (in 0.5\arcsec\ seeing) with no methane absorption 0.2\arcsec\ from \WISE, plus J3=21.0 galaxy 1.7\arcsec" from WISE position. This pair would be confused in WISE \& likely cause of W1--W2 colour. No methane absorbers within 30\arcsec\ to J3=22}
\tablenotetext{k}{J103059.32$-$373140.3: Object with no methane absorption 1.9\arcsec\ from \WISE position. No methane absorbers within 30\arcsec\ to J3=21}
\tablenotetext{l}{J120906.70$-$520034.5: Object with no methane absorption 0.5\arcsec\ from \WISE position. No methane absorbers within 30\arcsec\ to J3=21}
\tablenotetext{m}{J122329.62$-$480051.3: Object with no methane absorption 1.5\arcsec\ from \WISE position. No methane absorbers within 30\arcsec\ to J3=21.5}
\tablenotetext{n}{J123252.86$-$800525.4: No sources within 10" of WISE position to J=21. No methane absorbers within 30\arcsec.}
\tablenotetext{o}{J124156.18$-$370345.5: Galaxy (resolved in 0.48\arcsec\ seeing) 0.3\arcsec from \WISE position. No methane absorbers within 30\arcsec\ to J3=22.5}
\tablenotetext{p}{J130348.35$-$460959.4: Galaxy (1.0$\times$0.8\arcsec\ in 0.55\arcsec\ seeing) at WISE position. No methane absorbers within 30\arcsec\ to J3=22.5}
\tablenotetext{q}{J143032.81$-$323321.7: Galaxy (resolved in 0.47\arcsec\ seeing) 0.3" from \WISE position. No methane absorbers within 30\arcsec\ to J3=22}
\tablenotetext{r}{J153802.79$-$450936.5: Crowded field with multiple objects (J3=19.3, 20.4,17.1,21.4,21.5,21.0) within 5\arcsec\ of \WISE position, none with methane absorption. No methane absorbers within 30\arcsec\ to J3=21.5}
\tablenotetext{s}{J171507.17$-$552116.8: Unresolved source (0.6\arcsec\ seeing) with no methane 0.5\arcsec\ from \WISE. No methane absorbers within 30\arcsec\ to J3=22.5}
\tablenotetext{t}{J193441.70$-$490837.6: J3=20.5 galaxy (1.1\arcsec\ resolved in 0.9\arcsec\ seeing) 0.2\arcsec\ from \WISE. No methane absorbers within 30\arcsec\ to J3=21.5}
\tablenotetext{u}{J211213.88$-$552855.5: Object 2.4\arcsec\ from \WISE position with no methane absorption. No methane absorbers within 30\arcsec\ to J3=21.5}
\tablenotetext{v}{J212505.20$-$323412.9: Galaxy 0.7\arcsec from \WISE position. One methane absorber 21\arcsec\ from \WISE at 21:25:06.3 -32:33:56 with J3=21.3, J3$-$J2=$-$0.84$\pm$0.12, but galaxy is the more likely \WISE source.}
\tablenotetext{w}{J220011.60$-$544914.2: Galaxy (1.0\arcsec\ resolved in 0.6\arcsec\ seeing) 0.4\arcsec\ from \WISE position. No methane absorbers within 30\arcsec\ to J3=22}
\tablenotetext{x}{J221342.90$-$294908.7: Galaxy (0.7$\times$0.6\arcsec)  0.3\arcsec\ from WISE position with no methane. No methane absorbers within 30\arcsec\ to J3=22}
\tablenotetext{aa}{J011020.86$-$132313.0: Only candidate methane absorber is J3=21.0, J3$-$J2=$-$0.88$\pm$0.08, but 32.6\arcsec away from \WISE position at 01:10:18.7 $-$13:2322  requires $\sim11$\arcsec$/$yr motion to be the \WISE source. If a match would have J3$-$W2=6.17 suggesting a possibly Y0.}
\tablenotetext{bb}{J092320.42$-$731704.1: Only candidate methane absorber within 30\arcsec\ is J3=21.1, J3$-$J2=$-$0.96$\pm$0.18, 19.4\arcsec\  from \WISE position at 09:23:20.4 $-$73:16:44. If a match would would require a $\sim$3\arcsec$/$yr motion, and J3$-$W2=4.42 would suggest a T9-T9.5.}
\tablenotetext{cc}{J183437.34$-$584119.4: Only candidate methane absorber within 30\arcsec\ is J3=22.4, J3$-$J2=$-$1.2$\pm$0.2,  14.0\arcsec\  from \WISE position at 18:34:36.5 -58:41:06. If a match  would would require a 
$\sim$3\arcsec$/$yr motion, and J3$-$W2=6.8$\pm$0.2 which would suggest a Y0.}
\end{deluxetable}                                                                                                                                     
\end{longrotatetable}

\section{Imaging Summary}

The \WISE candidates observed naturally fall into a few classes: 
\begin{itemize}
\item[(a)] a bright (i.e. $J<19.5$) object is readily detected as a positional match with the \WISE source. In 100\% of cases, these objects are subsequently confirmed by methane imaging as T dwarfs; 
\item[(b)] a fainter (i.e. $19.5<J<21.5$) object is detected as a positional match that (in some cases) we have been able to verify  is a T- or Y-dwarf using either methane imaging, or spectroscopy obtained on other telescopes in parallel with this methane imaging program);
\item[(c)] either a fainter (i.e. $19.5<J<21.5$) object is detected as a positional match with photometry that is inconsistent with this object being a T/Y dwarf, {\em or} we observe the object to be resolved in good imaging at Magellan indicating the \WISE source is most likely a galaxy; or 
\item[(d)] no counterpart is detected down to J$\sim$21.5-22 -- these could either be contaminant sources (i.e. non-cool-brown-dwarfs whose \WISE photometry mimics those of brown dwarfs due to confusion, photometric scatter, etc) or cool brown dwarfs fainter than $J \sim 22$. 
\end{itemize}

Finding charts for objects newly identified as being a T- or Y-dwarfs are provided in Fig. 1.
Figure \ref{W1141} shows expanded images at the position of the new Y0 dwarf W1141-3326. These show how the 2012 \Msl\ photometry for this object was contaminated
by confusion with a background source. The images from 2014 (by which time W1141-2236 had moved $\approx$2\arcsec\ to the west) show W1141-2236 clearly separated from this background source. It is also clear that in the 2012 images, \Ms\ is more contaminated that \Ml\, making \Msl\ more positive than it should be.  We therefore exclude W1141-3336 from our updated \Msl-to-spectral-type calibration in \S\ref{newcal}. \\

\subsection{Brown dwarf proper motions}

Figure 1 highlights the complicating factor of proper motion
when seeking to identify cool brown dwarfs. The charts for W2049-4431, 
W2159-4808, W1112-3857 and W1141-3326 (for example) all show substantial motion between the epoch of the
\WISE detection (mean \WISE epochs are in the range 2010 Apr-Jun for these four targets) and the epoch of 
follow-up (from 2011 Sep to 2017 Jan). \\

The observed motions of our new WISE T and Y dwarfs are summarised in Table \ref{PMTable}. 
We obtained astrometry for our J, \Ms, or J3 images by cross-matching against
the 2MASS PSC and relying on \WISE and the PSC being
on the same co-ordinate system. The individual positional uncertainty of the \WISE objects (0.3-0.5\arcsec) 
is the dominant source of uncertainty, so this is a good assumption. Note that
we distinguish between the ``apparent'' motion of our targets from just two epochs
as observed by \WISE and our IRIS2 or FourStar follow-up imaging (here denoted $\mu^\prime$), 
and the true proper motion from an astrometric solution that includes parallax.
The large \WISE position uncertainties mean that the uncertainty in the right ascension and
declination components of $\mu^\prime$ are very similar, so only a single uncertainty
is quoted for both in the Table. This uncertainty includes an additional term added in quadrature
of 100\,mas to account for possible parallax motion not being accounted for in objects that
could be as close as 10\,pc away. Five of these brown dwarfs are the subject of our
on-going astrometric program with FourStar, so we provide preliminary proper motion
solutions (in this case including a parallax solution) for comparison. In each case the
solutions agree, although the FourStar astrometric solutions are between 10 and 20 times better
than the \WISE--IRIS2/FourStar estimates.\\

Motions of 0.5\arcsec\,yr$^{-1}$ are not uncommon, and the motions observed for the most rapidly moving 
brown dwarfs (W0032-4946 $\sim$1.5\arcsec\,yr$^{-1}$,
W2159-4808 $\sim$1.4\arcsec\,yr$^{-1}$, W1141-3326 $\sim$0.95\arcsec\,yr$^{-1}$, W1112-3857 $\sim$0.9\arcsec\,yr$^{-1}$,  
 W2049-4431 $\sim$0.7\arcsec\,yr$^{-1}$) highlight  how identification via ``blind'' spectroscopy becomes substantially
more difficult after just a few years. After 5 years the positional error box that must be searched can be as large as 20\arcsec\ on a side,
and can contain multiple objects that would have to be spectroscopically observed in turn to find the correct T/Y-dwarf.  
In this situation methane imaging becomes an even more powerful tool to pick out the correct counterpart.\\

\begin{deluxetable}{cRRRRRR}
\tabletypesize{\scriptsize}
\tablecaption{Apparent ($\mu^\prime$) and Proper ($\mu$) Motions for new cool brown dwarfs from \WISE\label{PMTable}}
\tablewidth{0cm}
\tablehead{
\colhead{Object\tablenotemark{a}} & \multicolumn{3}{c}{\WISE--IRIS2/FouStar} & \multicolumn{3}{c}{FourStar alone}\\
\colhead{} 
                        &\colhead{$\mu^\prime_\alpha$} 
                                   &\colhead{$\mu^\prime_\delta$}
                                                &\colhead{Unc.}
																		 &\colhead{$\mu_\alpha$} 
                                   									&\colhead{$\mu_\delta$}
                                                						&\colhead{Unc.} \\
\colhead{} 
                        &\multicolumn{3}{c}{(mas\,yr$^{-1}$)} 
																		 &\multicolumn{3}{c}{(mas\,yr$^{-1}$)} 
}
\startdata
W0015$-$4615 & +720 & -450 &\pm250 &  &  &    \\
W0032$-$4946 & -714 &-1300 &\pm250 &  &  &    \\
W0042$-$5840 & +122 & -140 &\pm120 &  &  &    \\
W0241$-$3653 & +350 & +100 &\pm230 &  &  &    \\
W0302$-$5817 &  -86 & -150 &\pm120 &  &  &    \\
W0309$-$5016 & +645 & +100 &\pm270 &  &  &    \\
W0404$-$6420 & -139 & +180 &\pm155 &  &  &    \\
W0628$-$8057 & +110 & -550 &\pm150 &  &  &    \\
W0645$-$0302 &  -43 & -430 &\pm160 &  &  &    \\
W0914$-$3459 & -160 &  -20 &\pm180 &  &  &    \\
W0940$-$2208 &  -70 &  +30 &\pm160 &  &  &    \\
W1023$-$3151 & -220 & +250 &\pm110 &  &  &    \\
W1055$-$1652 & -728 & +250 &\pm160 &  &  &    \\
W1112$-$3857 & +510 & +590 &\pm180 & +666 & +638 & \pm10  \\
W1141$-$3326 & -915 & +120 &\pm180 & -897 & -84  & \pm6   \\
W1221$-$3136 & +550 & +360 &\pm180 &  &  &    \\
W1433$-$0837 & -130 &    0 &\pm240 &  &  &    \\
W1448$-$2534 & +110 & -670 &\pm180 &  &  &    \\
W1501$-$4004 & +580 & -290 &\pm170 &  &  &    \\
W1729$-$7530 &  +30 & -230 &\pm120 &  &  &    \\
W1735$-$8209 & -350 & -540 &\pm230 &  &  &    \\
W2017$-$3421 & +385 &  +10 &\pm240 &  &  &    \\
W2049$-$4431 & +415 & -585 &\pm120 &  &  &    \\
W2102$-$4429 & -250 & -340 &\pm240 &  +41 & -354 &  \pm3   \\
W2159$-$4808 & +275 &-1380 &\pm240 &  &  &    \\
W2211$-$4758 & -170 & -160 &\pm240 &  &  &    \\
W2212$-$6931 & +600 & +200 &\pm240 &  &  &    \\
W2220$-$3628 & +320 &  +60 &\pm250 &  +283 & -94 &  \pm4   \\
W2232$-$5730 & +240 &  +80 &\pm170 &  &  &    \\
W2302$-$7134 & -220 &    0 &\pm160 &  &  &    \\
W2332$-$4325 & +200 & -315 &\pm250 &  +248 & -256 & \pm4   \\
W2354$-$5649 & +390 & -330 &\pm160 &  &  &    \\
W2354$-$8140 &  +30 &  -15 &\pm230 &  &  &    \\
\enddata                                                                                                                                             
\tablenotetext{a}{\WISE designations follow Table \ref{ImageTable}}
\end{deluxetable}

\begin{figure*}
\figurenum{1}
\includegraphics[width=7.0cm,trim=0.5cm 19.0cm 9.0cm 1.5cm,clip=true]{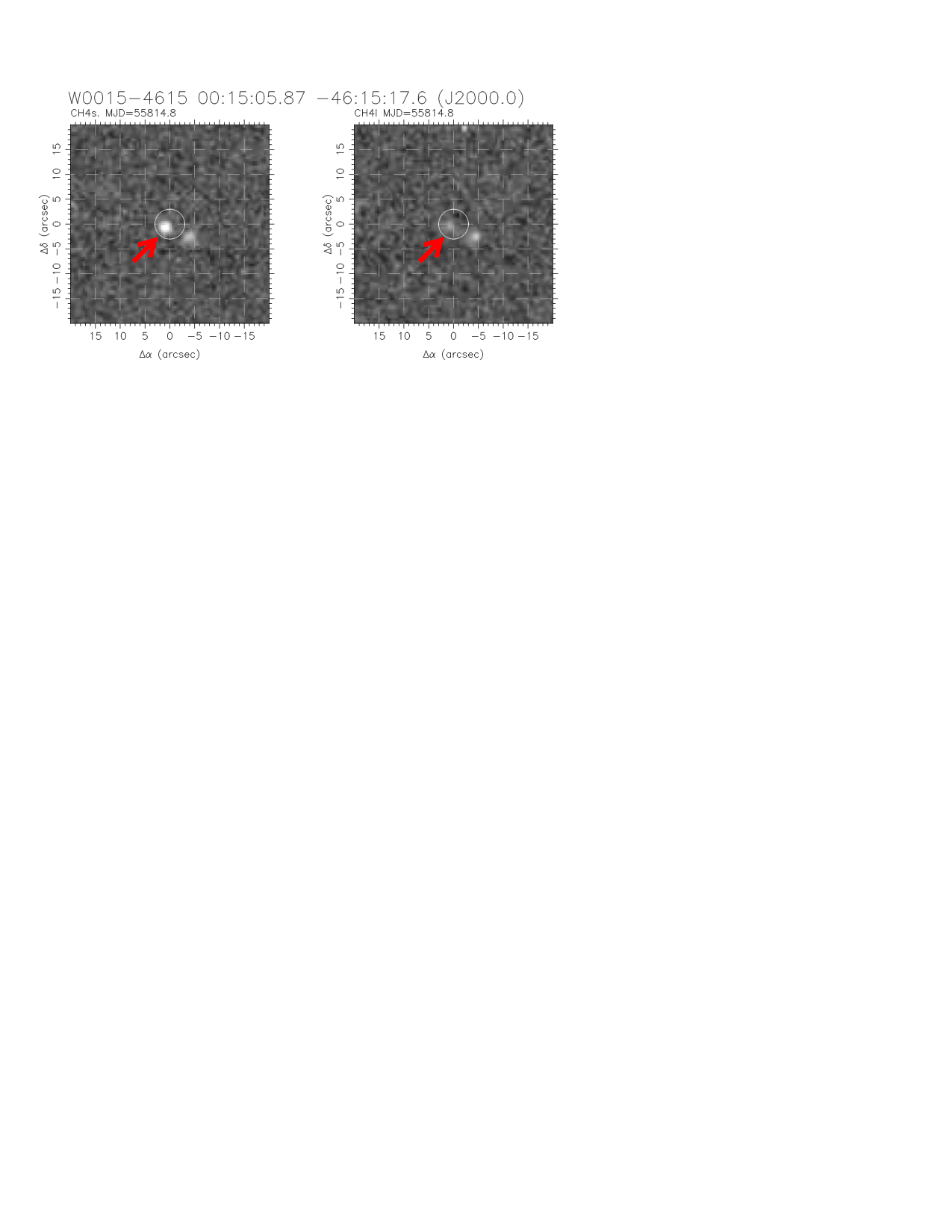}\qquad\includegraphics[width=7.0cm,trim=0.5cm 19.0cm 9.0cm 1.5cm,clip=true]{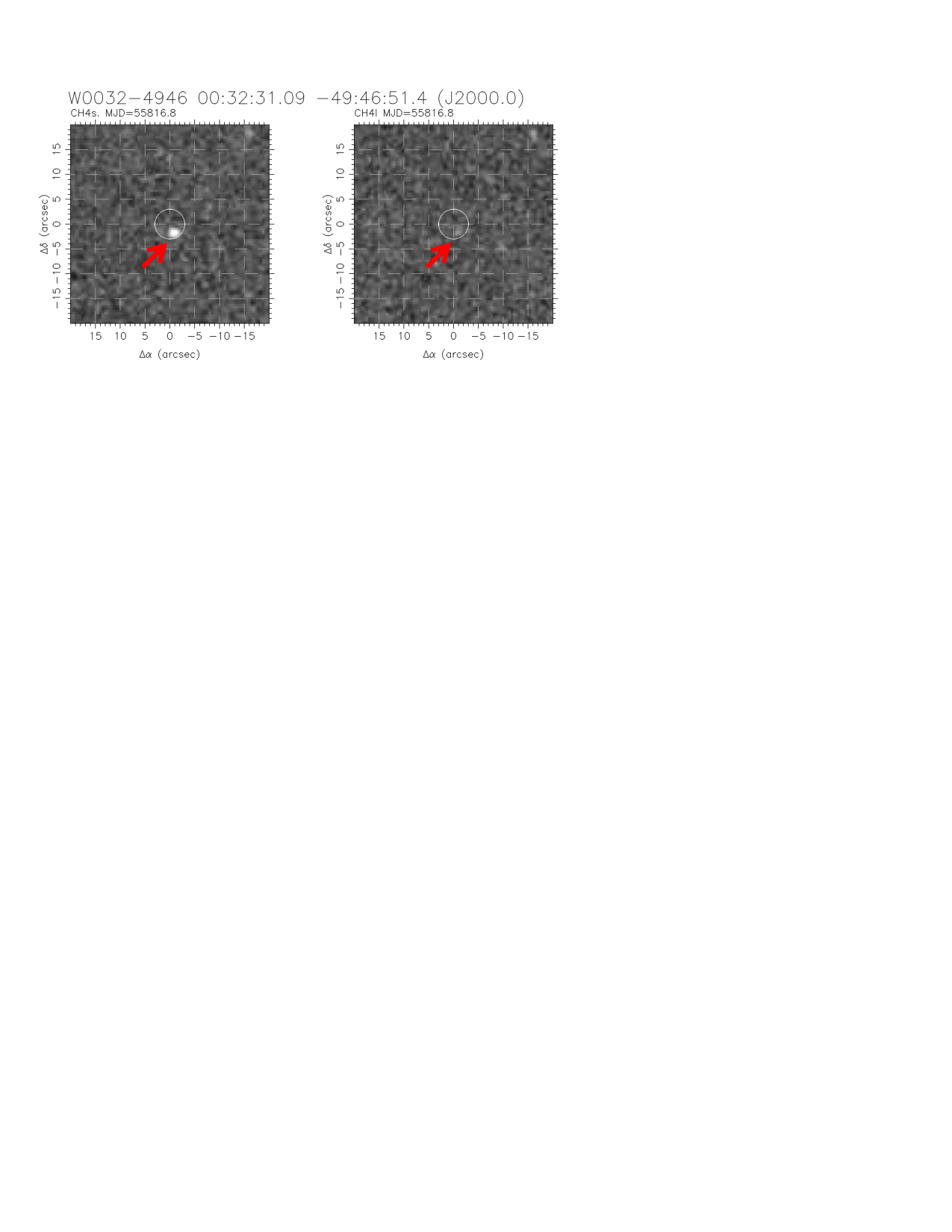}\\[-0.5cm]
\includegraphics[width=7.0cm,trim=0.5cm 19.0cm 9.0cm 1.5cm,clip=true]{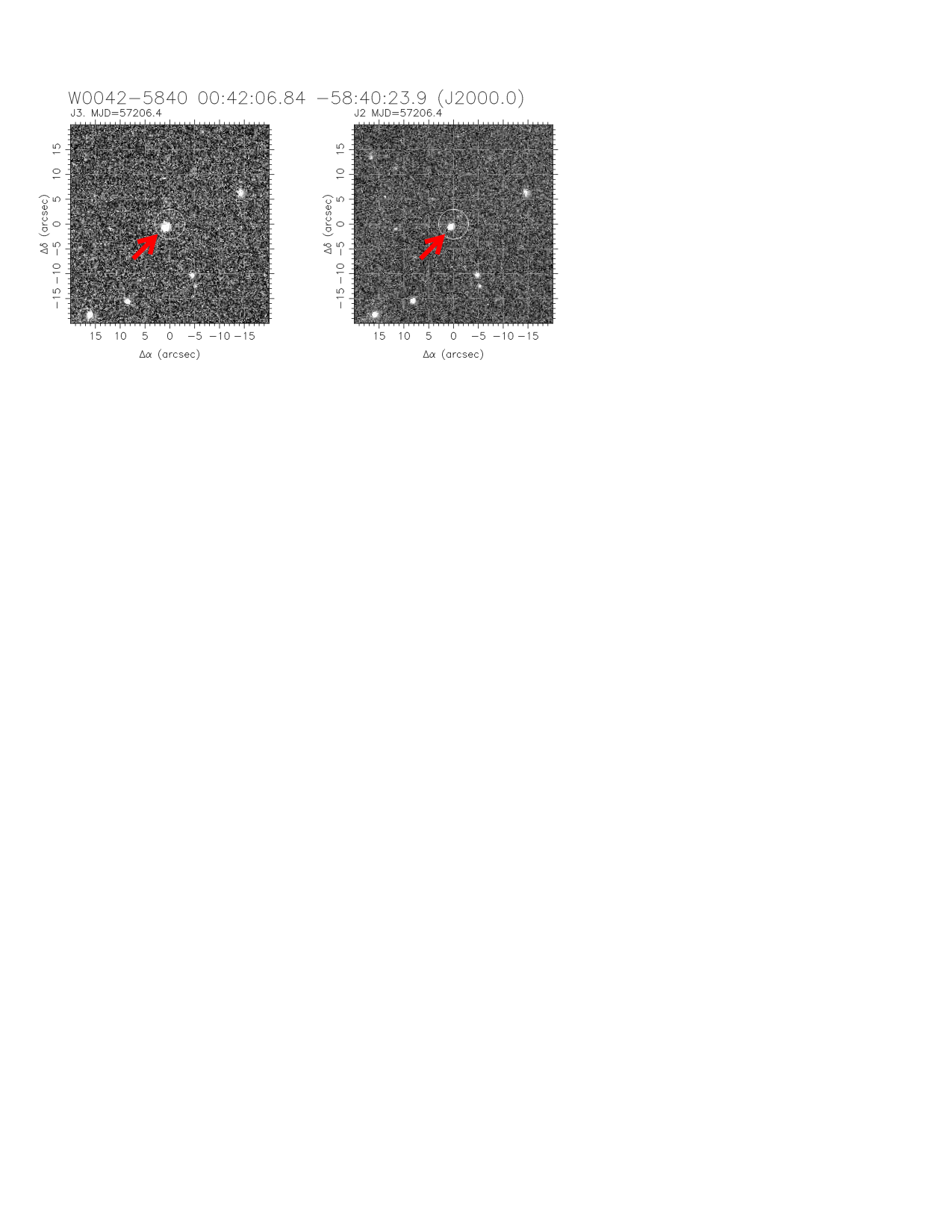}\qquad\includegraphics[width=7.0cm,trim=0.5cm 19.0cm 9.0cm 1.5cm,clip=true]{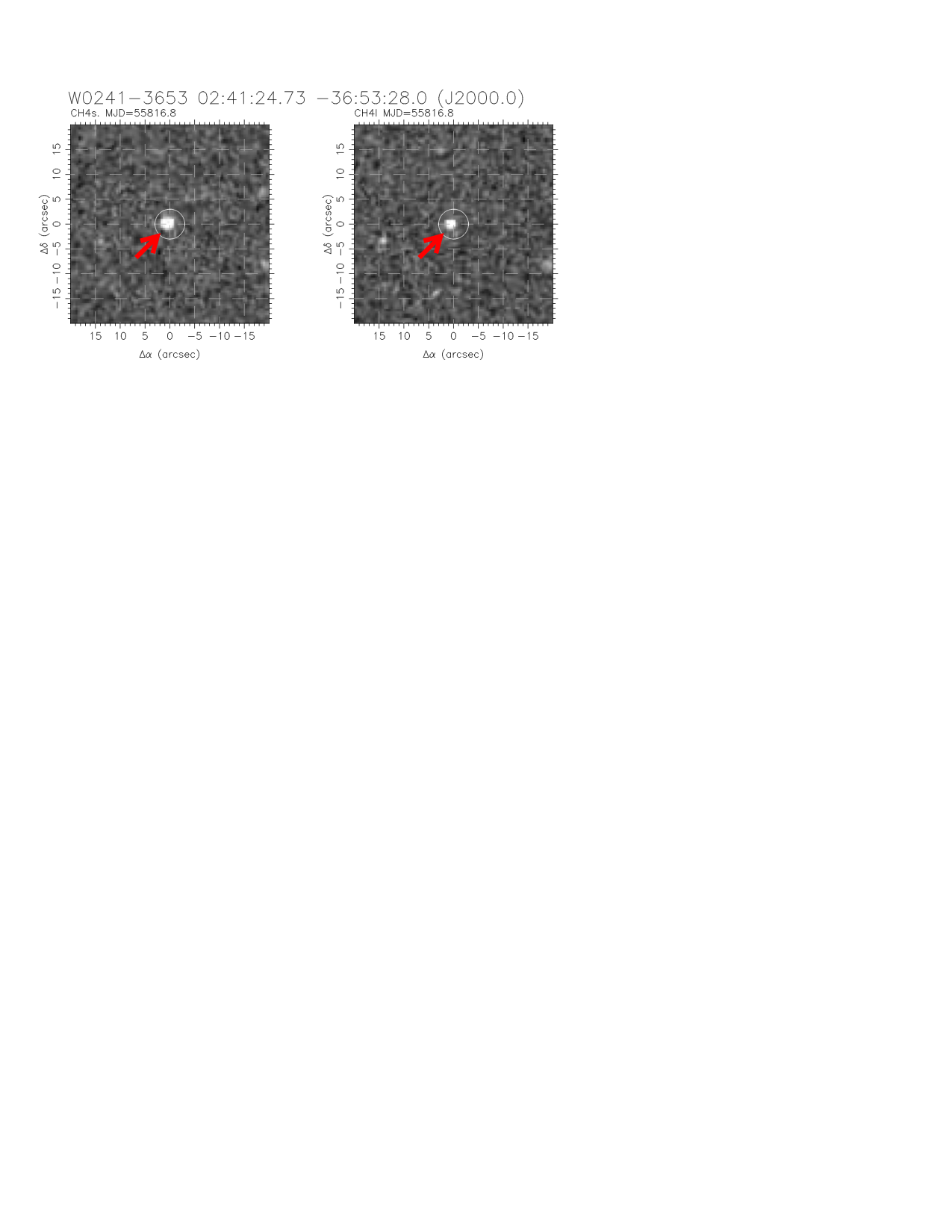}\\[-0.5cm]
\includegraphics[width=7.0cm,trim=0.5cm 19.0cm 9.0cm 1.5cm,clip=true]{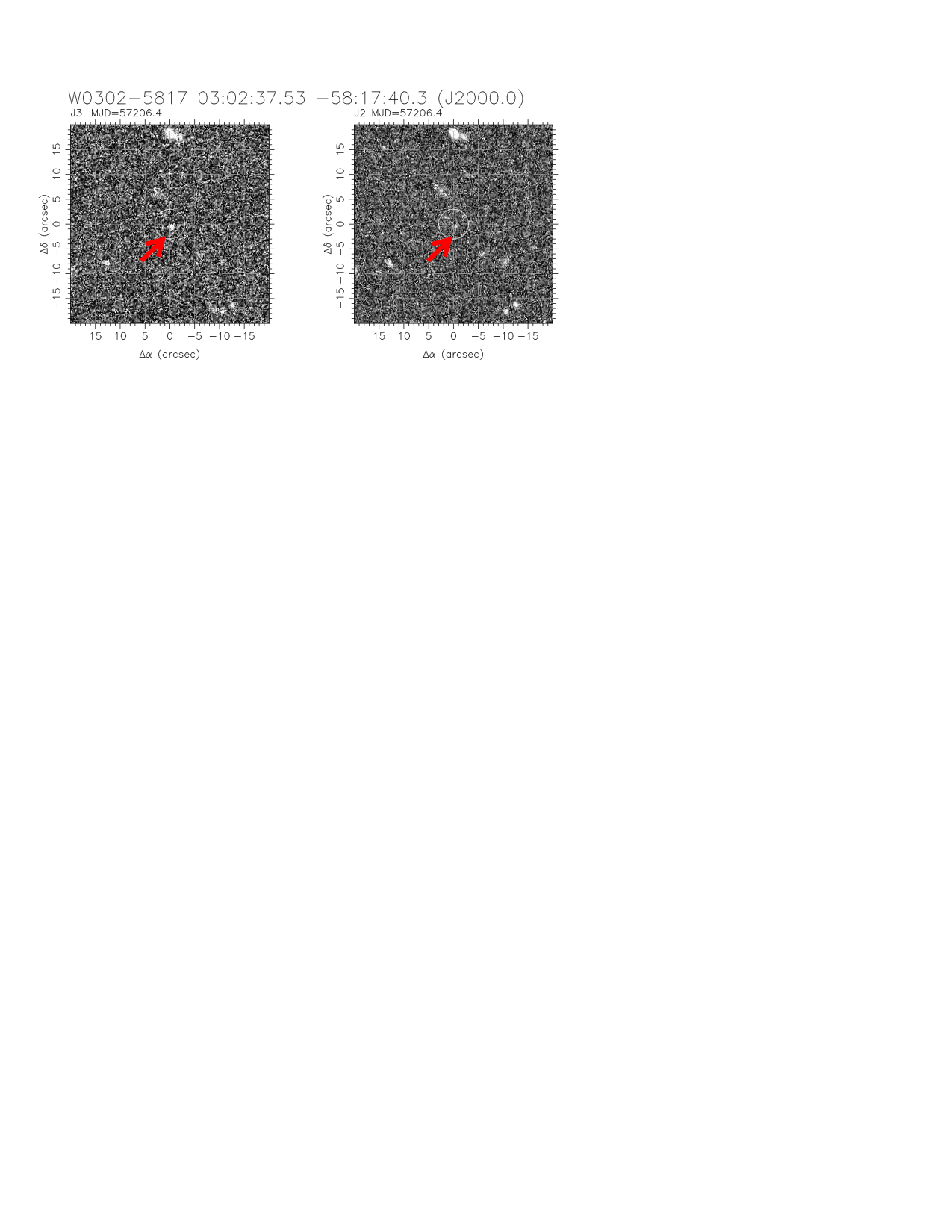}\qquad\includegraphics[width=7.0cm,trim=0.5cm 19.0cm 9.0cm 1.5cm,clip=true]{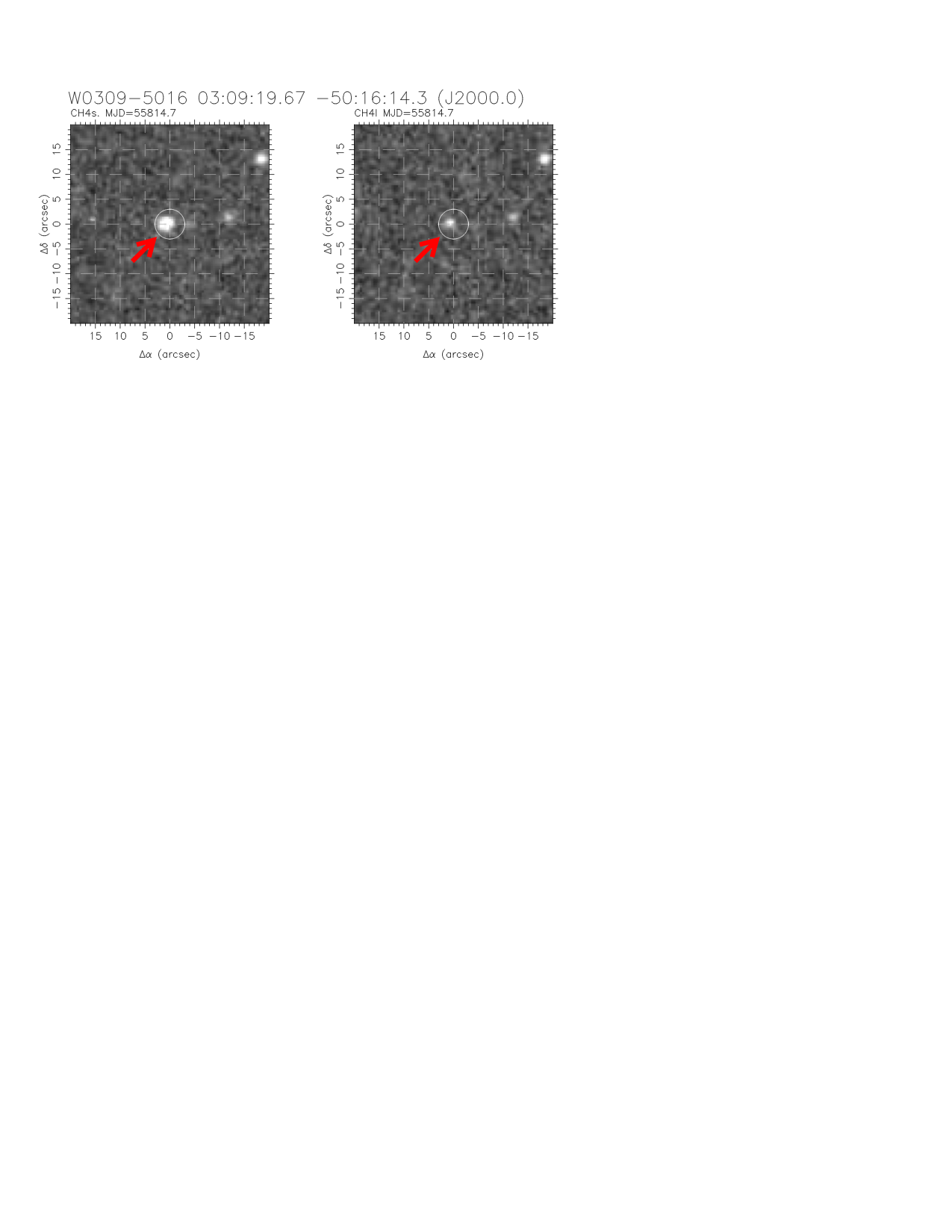}\\[-0.5cm]
\includegraphics[width=7.0cm,trim=0.5cm 19.0cm 9.0cm 1.5cm,clip=true]{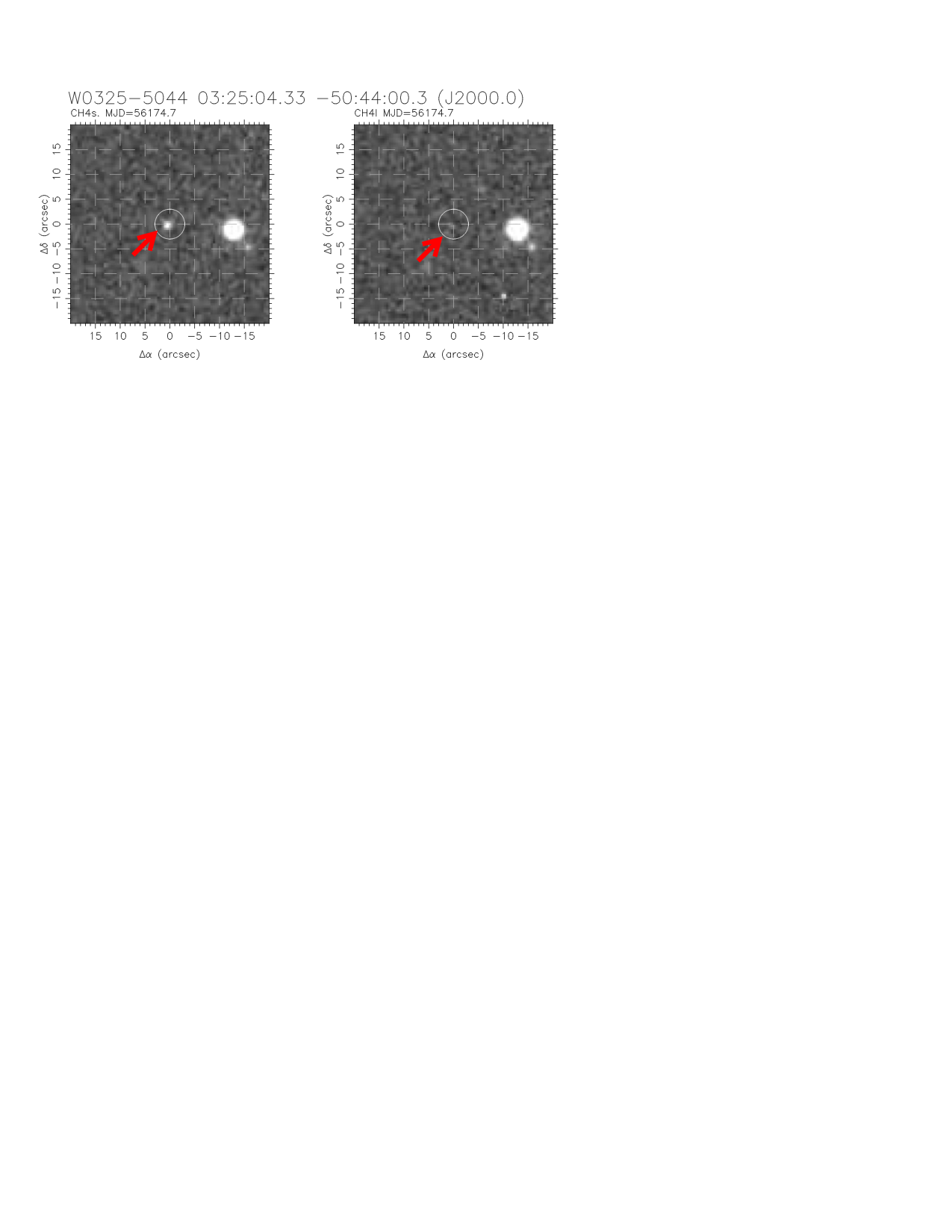}\qquad\includegraphics[width=7.0cm,trim=0.5cm 19.0cm 9.0cm 1.5cm,clip=true]{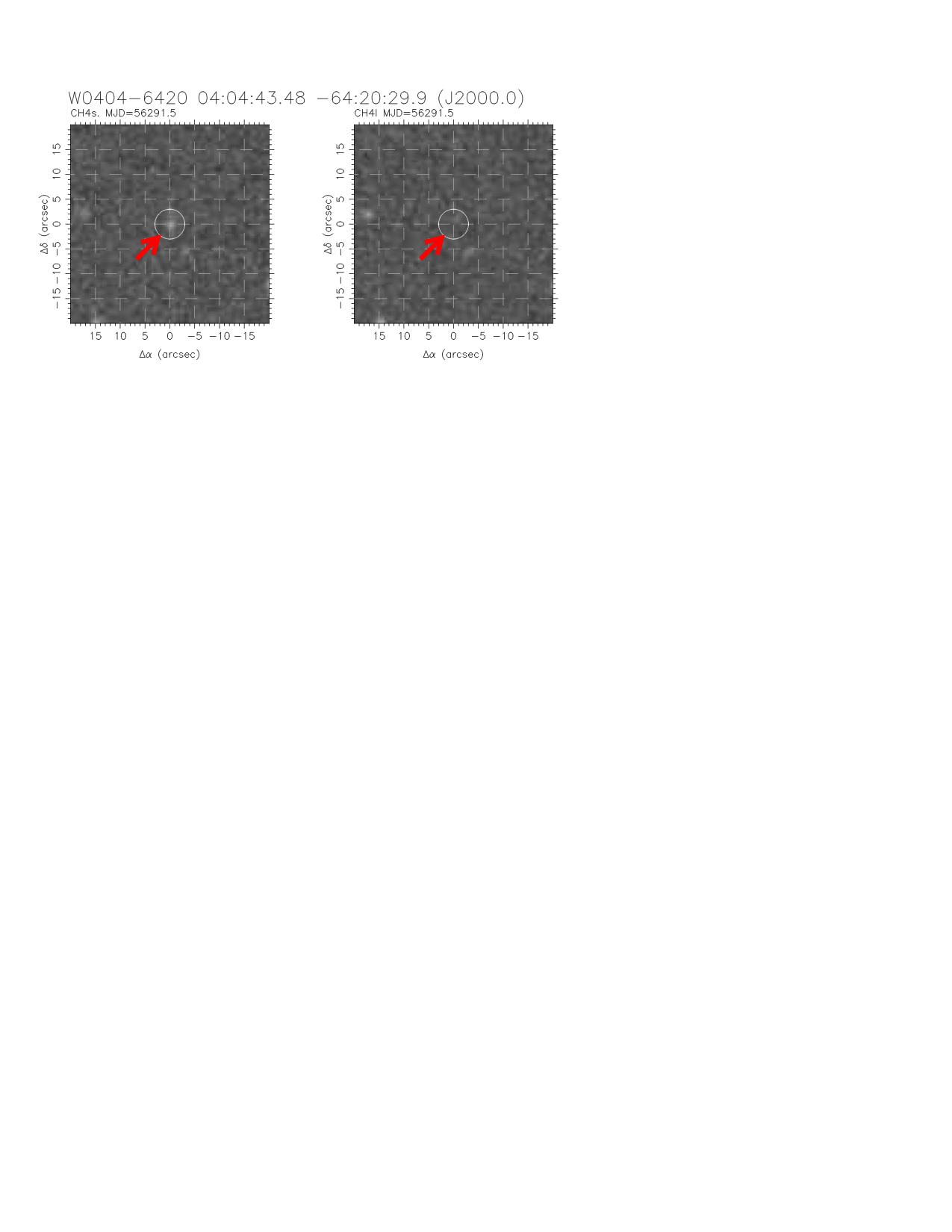}\\[-0.5cm]
\includegraphics[width=7.0cm,trim=0.5cm 19.0cm 9.0cm 1.5cm,clip=true]{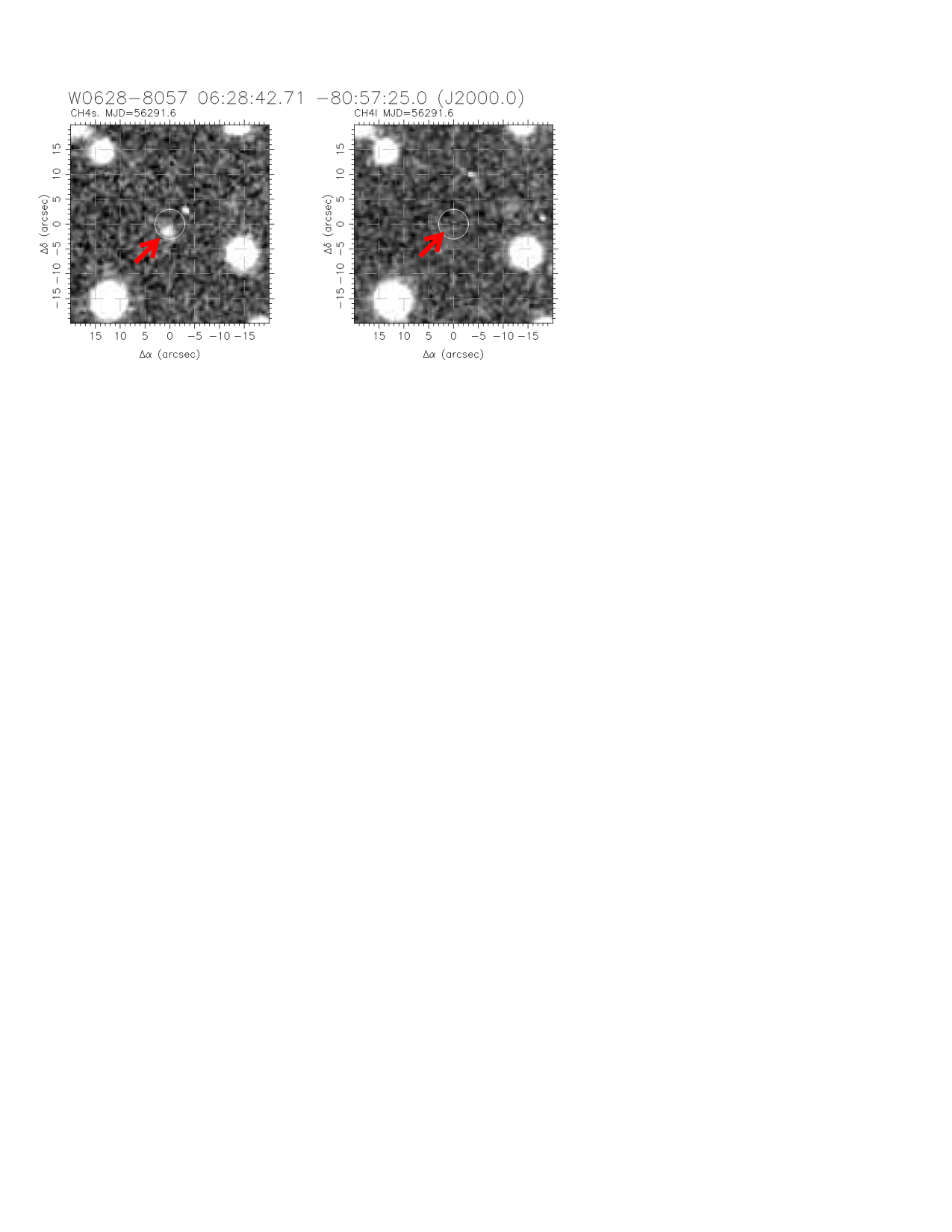}\qquad\includegraphics[width=7.0cm,trim=0.5cm 19.0cm 9.0cm 1.5cm,clip=true]{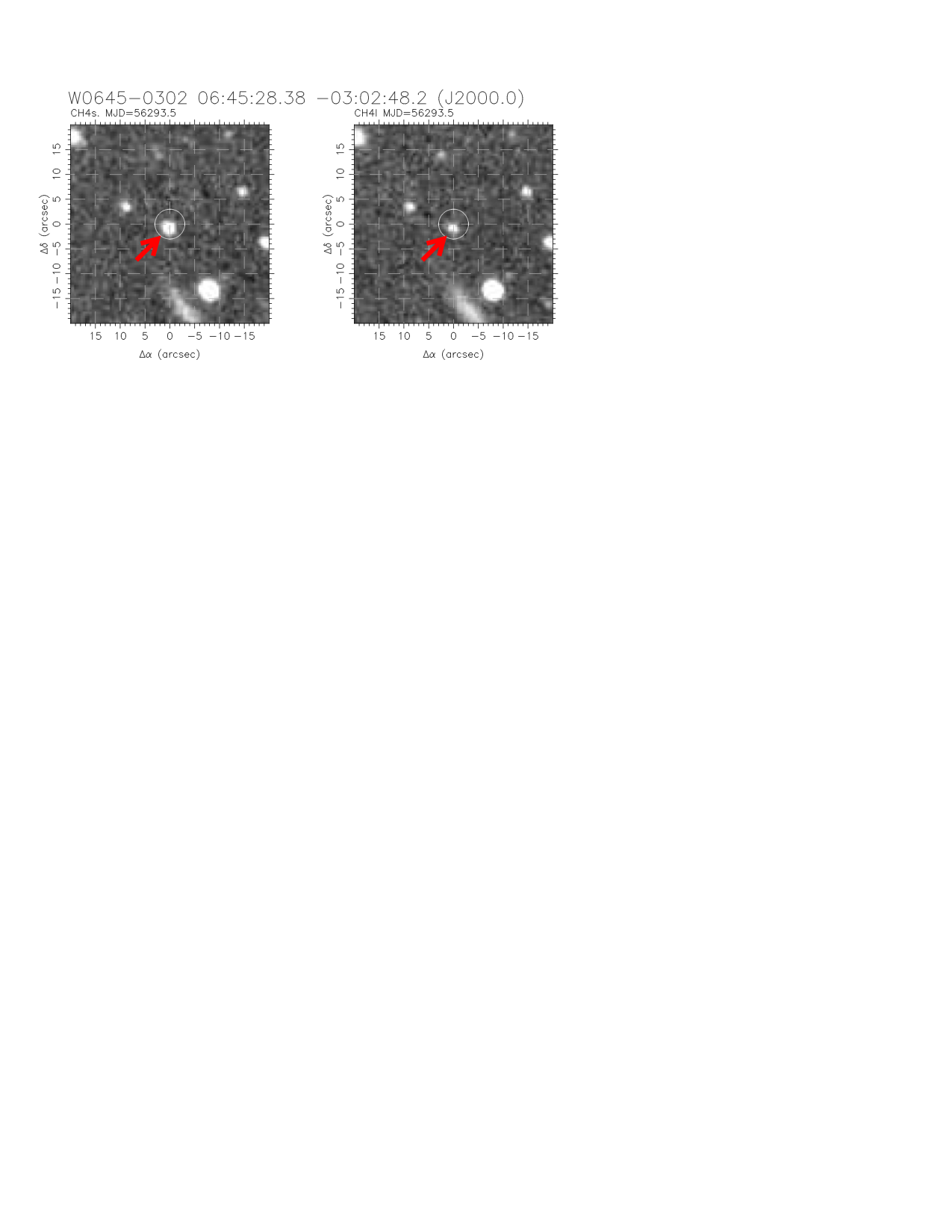}\\[-0.5cm]
\includegraphics[width=7.0cm,trim=0.5cm 19.0cm 9.0cm 1.5cm,clip=true]{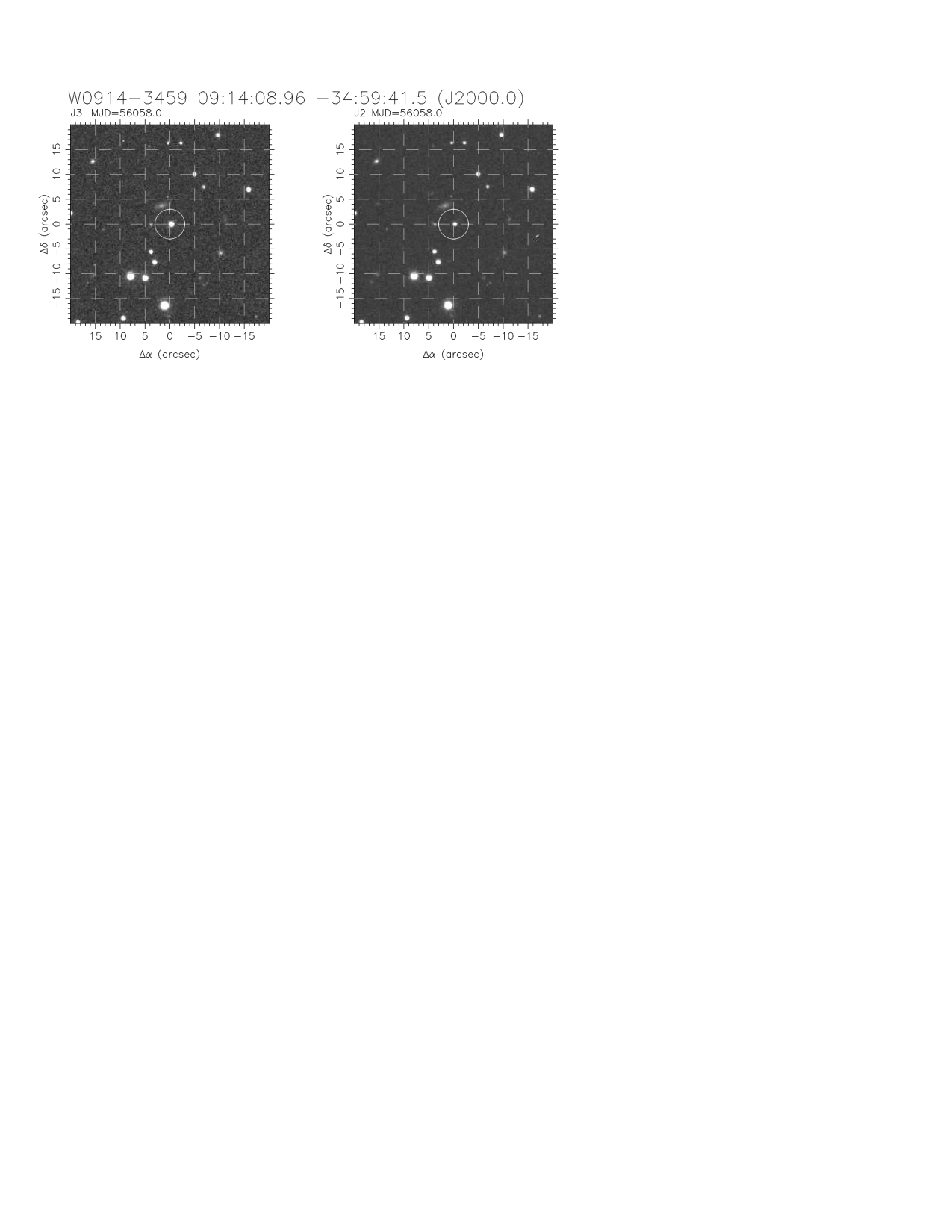}\qquad\includegraphics[width=7.0cm,trim=0.5cm 19.0cm 9.0cm 1.5cm,clip=true]{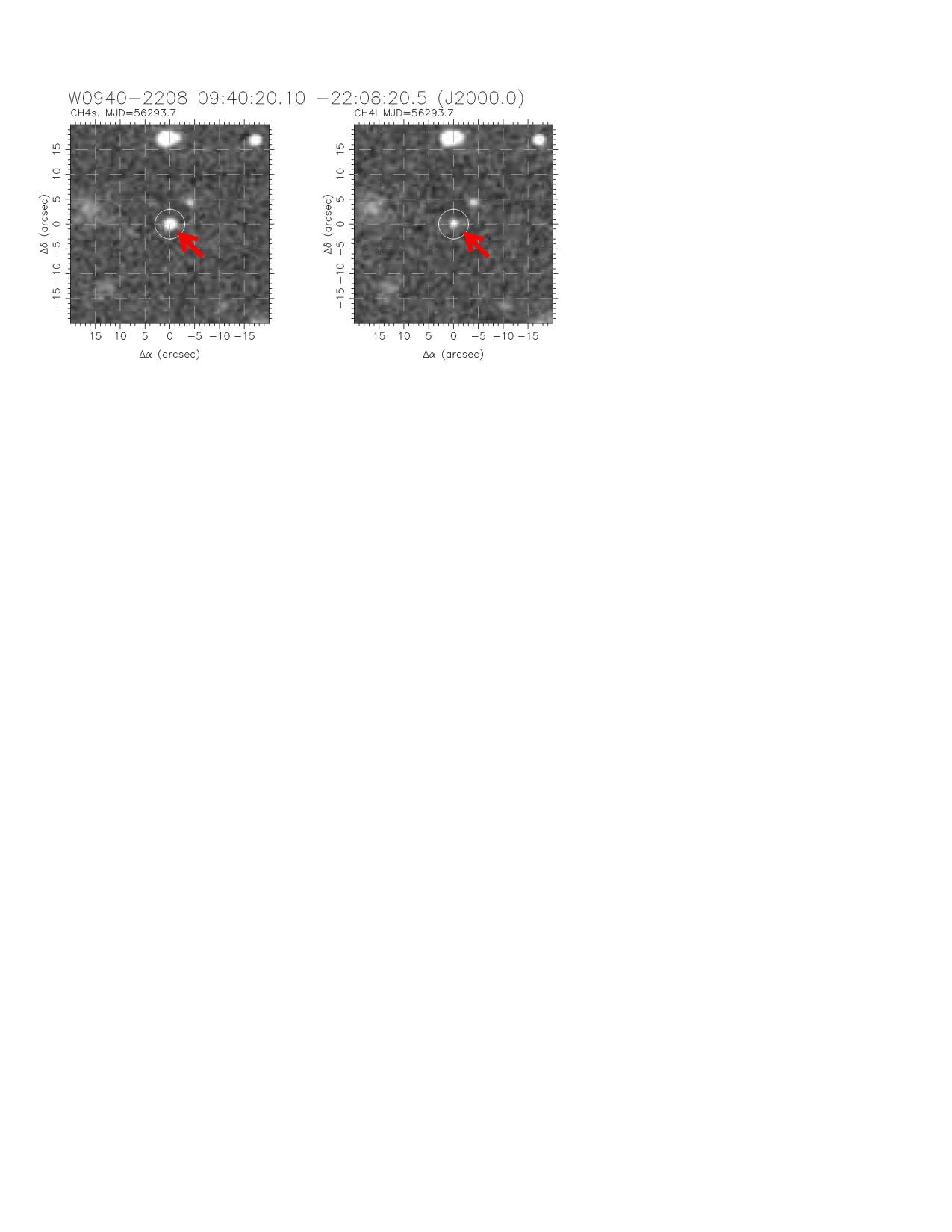}\\[-0.5cm]
   \caption{Pairs of finding charts in \Ms/\Ml\ or J3/J2 for the imaging observations tabulated in Tables \ref{ImageTable} and \ref{MagTable}. All charts are 40\arcsec\ on a side, N top the top, W to the right,
   centred on the \WISE source position and with a 3\arcsec\ circle drawn at the \WISE position. 
   Methane absorbing objects are brighter in the left panel of each pair of images. In ambiguous cases, an arrow is used to identify the cool brown dwarf.\label{FC1}}
\end{figure*}

\begin{figure*}
\figurenum{1 (cont)}
\includegraphics[width=7.0cm,trim=0.5cm 19.0cm 9.0cm 1.5cm,clip=true]{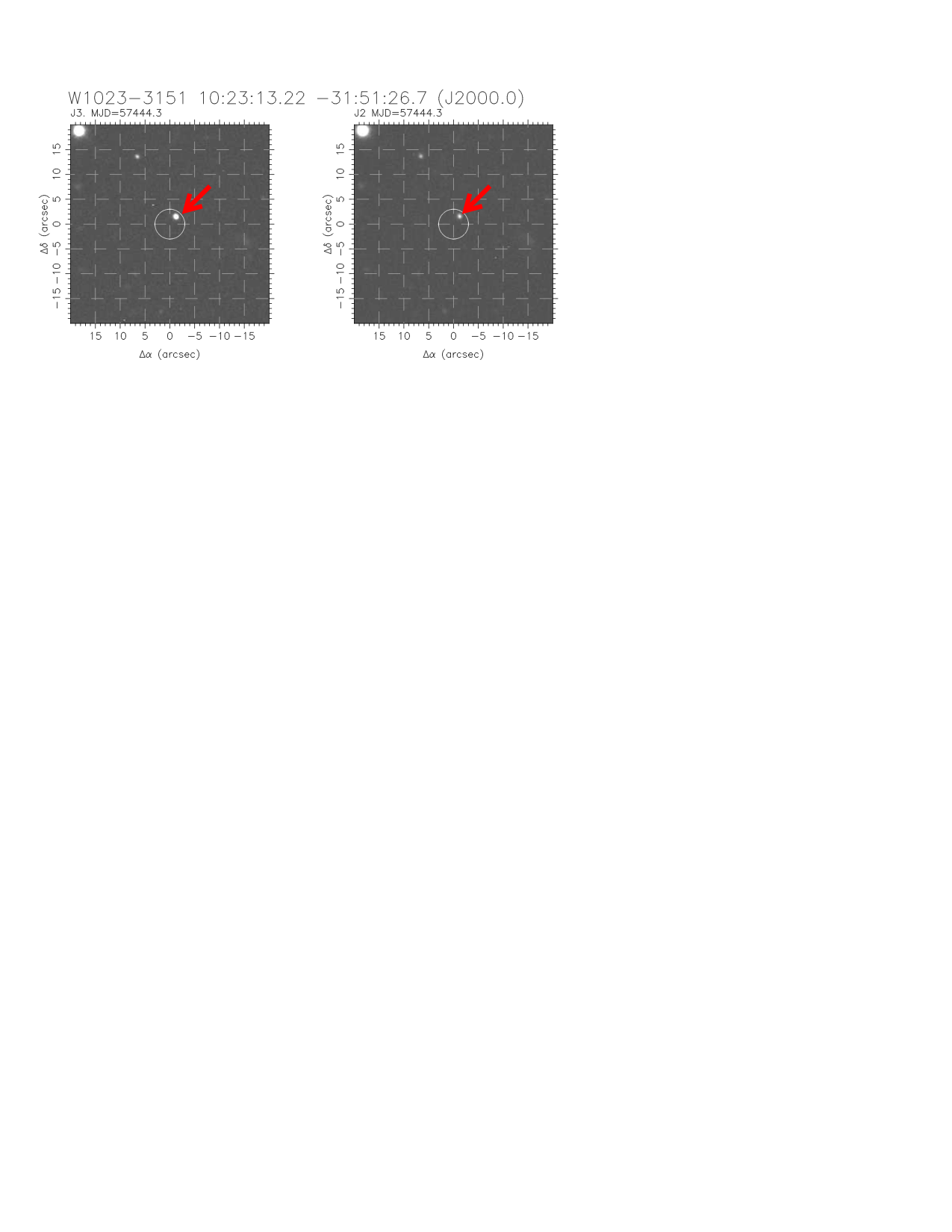}\qquad\includegraphics[width=7.0cm,trim=0.5cm 19.0cm 9.0cm 1.5cm,clip=true]{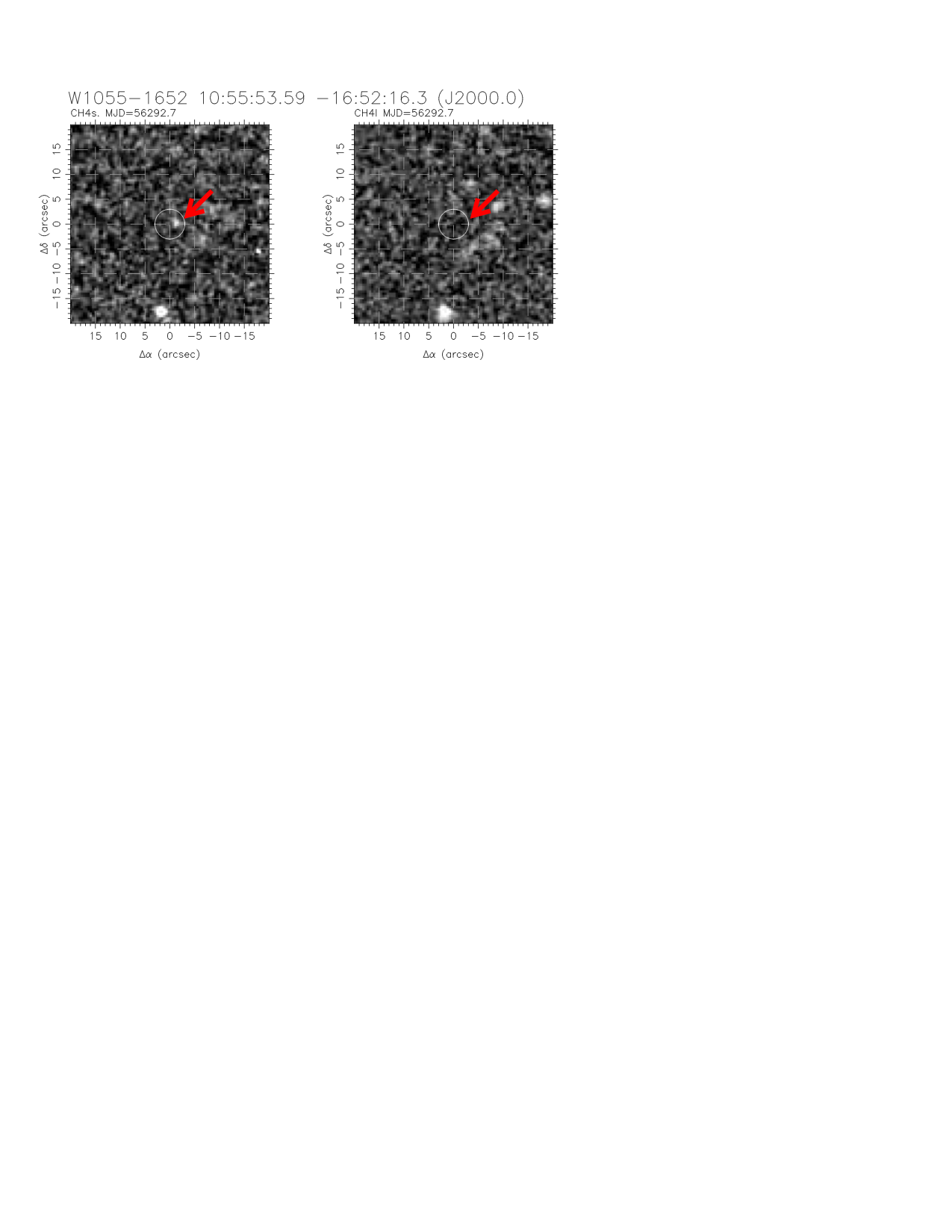}\\[-0.5cm]
\includegraphics[width=7.0cm,trim=0.5cm 19.0cm 9.0cm 1.5cm,clip=true]{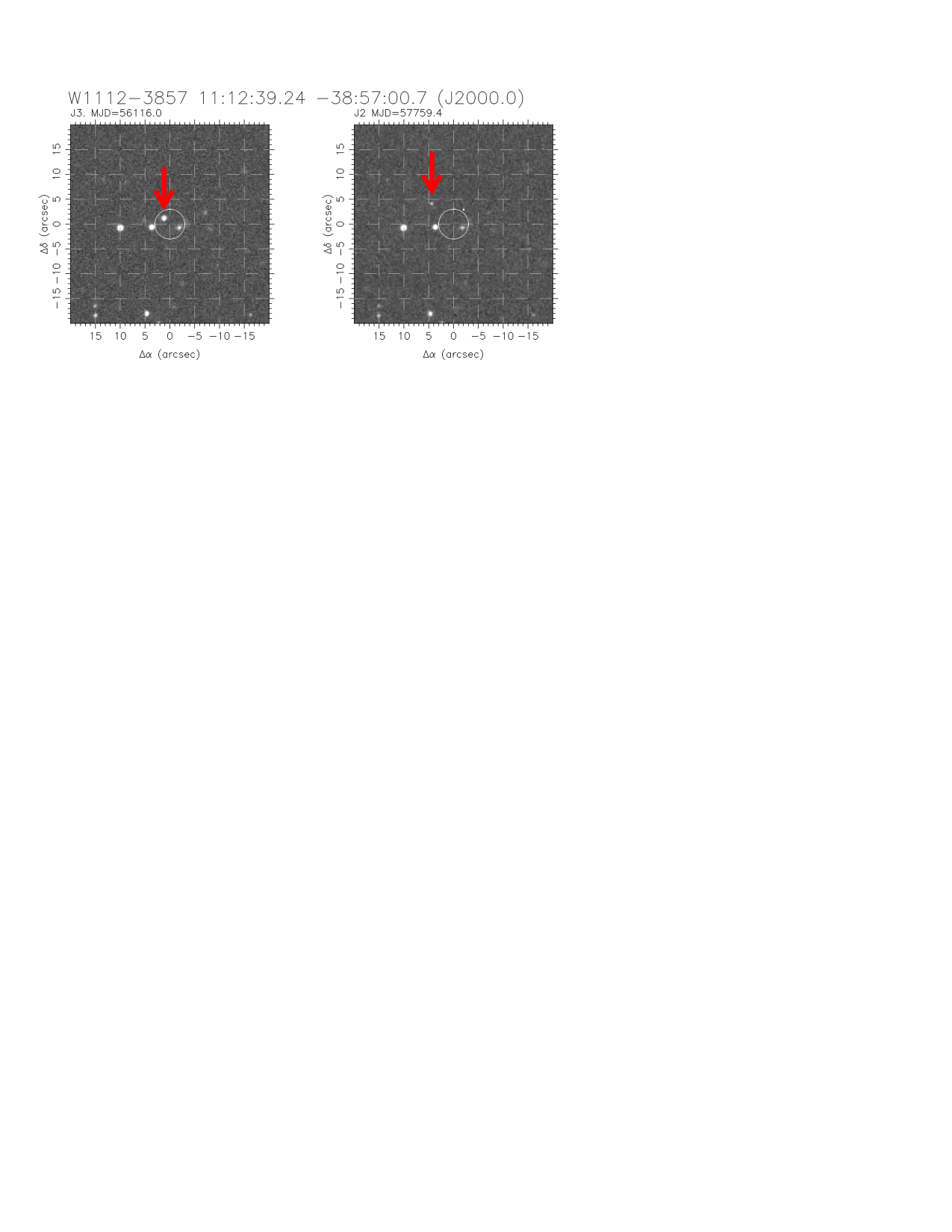}\qquad\includegraphics[width=7.0cm,trim=0.5cm 19.0cm 9.0cm 1.5cm,clip=true]{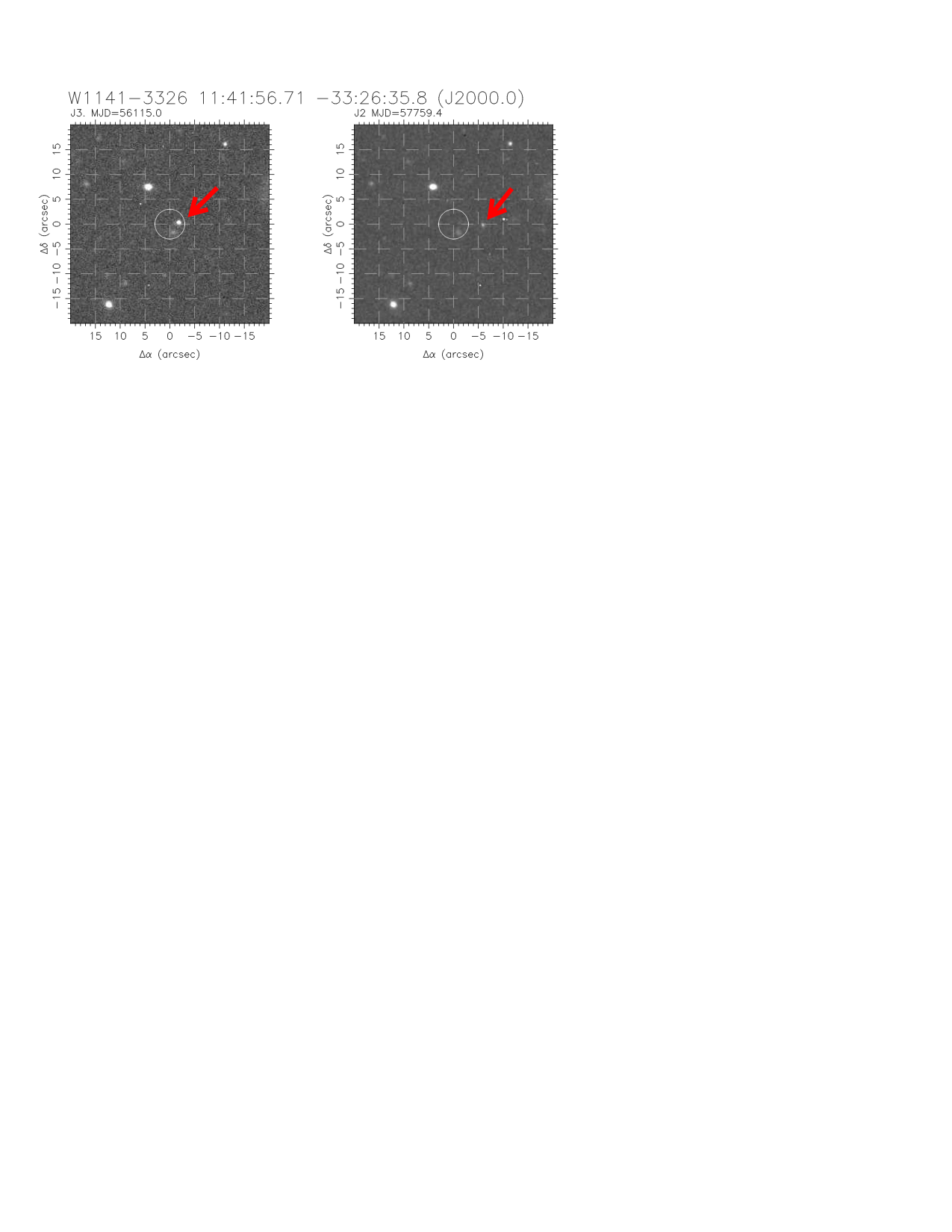}\\[-0.5cm]
\includegraphics[width=7.0cm,trim=0.5cm 19.0cm 9.0cm 1.5cm,clip=true]{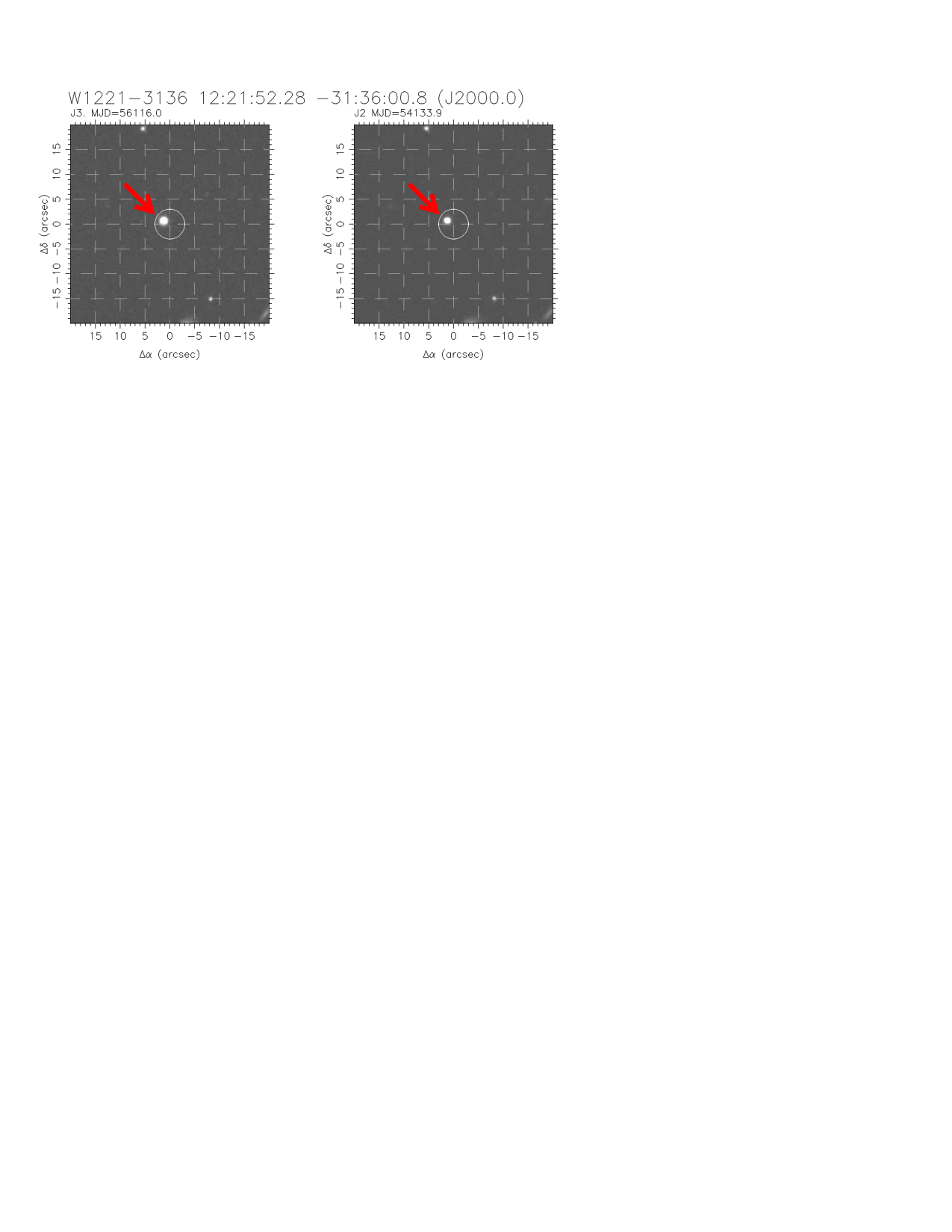}\qquad\includegraphics[width=7.0cm,trim=0.5cm 17.5cm 4.0cm 0.5cm,clip=true]{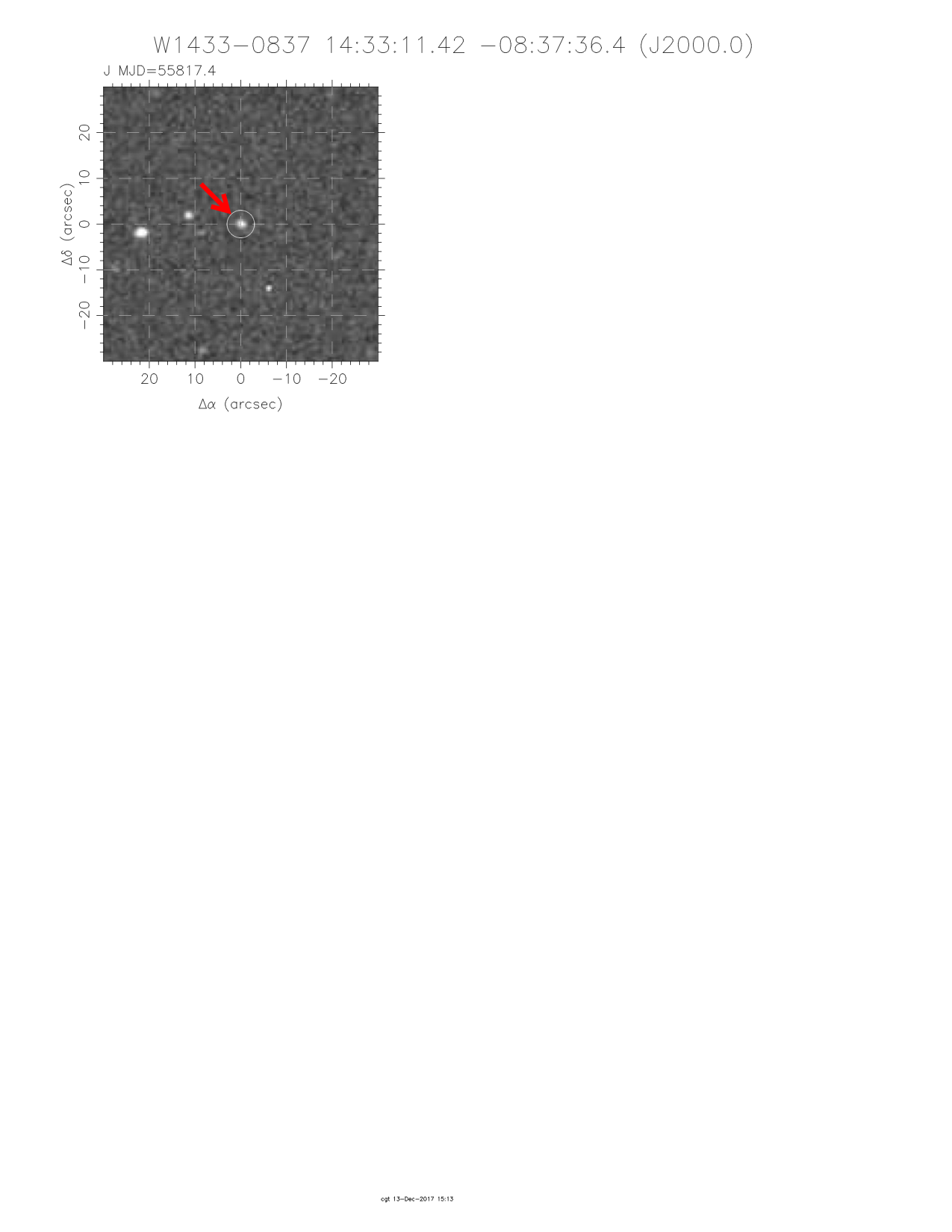}\\[-0.5cm]
\includegraphics[width=7.0cm,trim=0.5cm 19.0cm 9.0cm 1.5cm,clip=true]{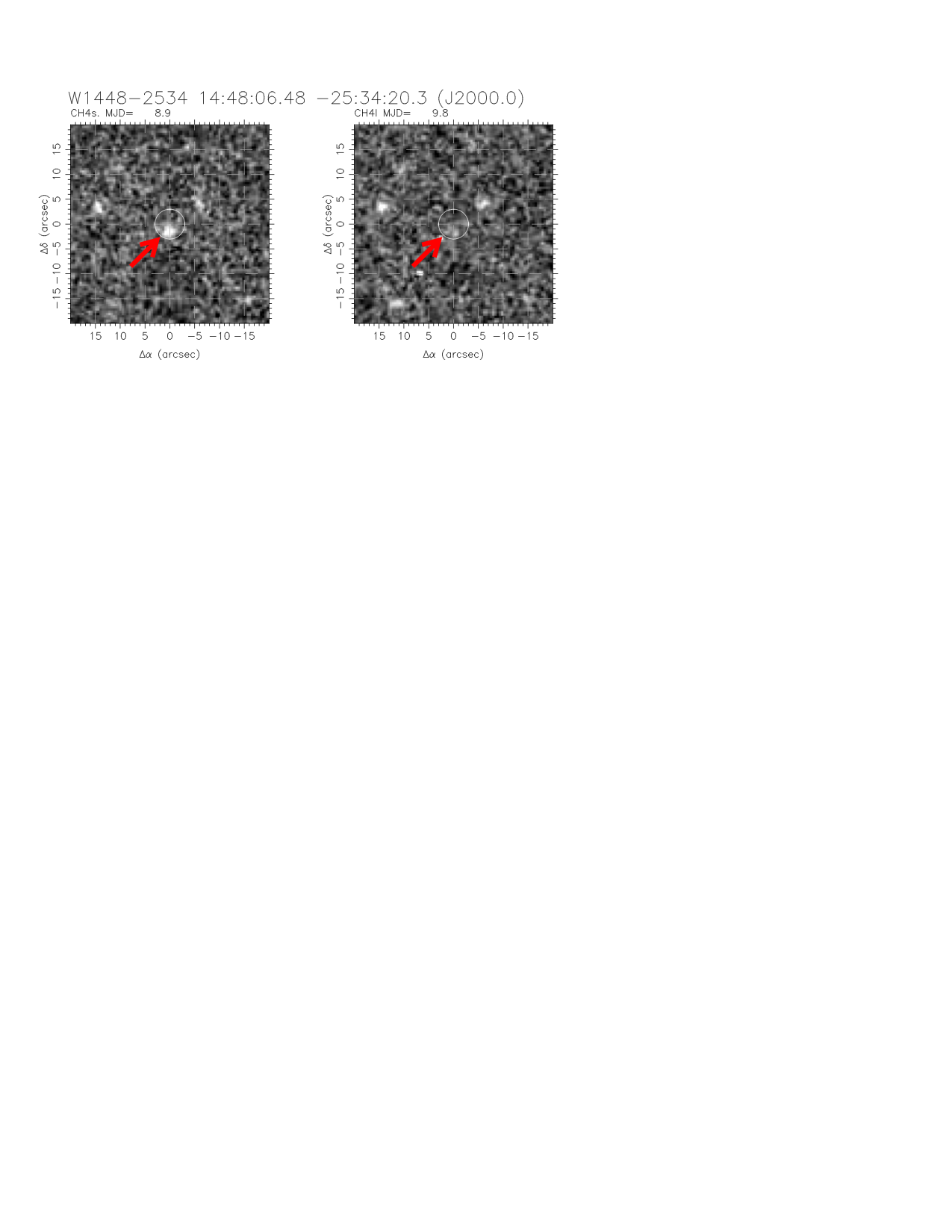}\qquad\includegraphics[width=7.0cm,trim=0.5cm 19.0cm 9.0cm 1.5cm,clip=true]{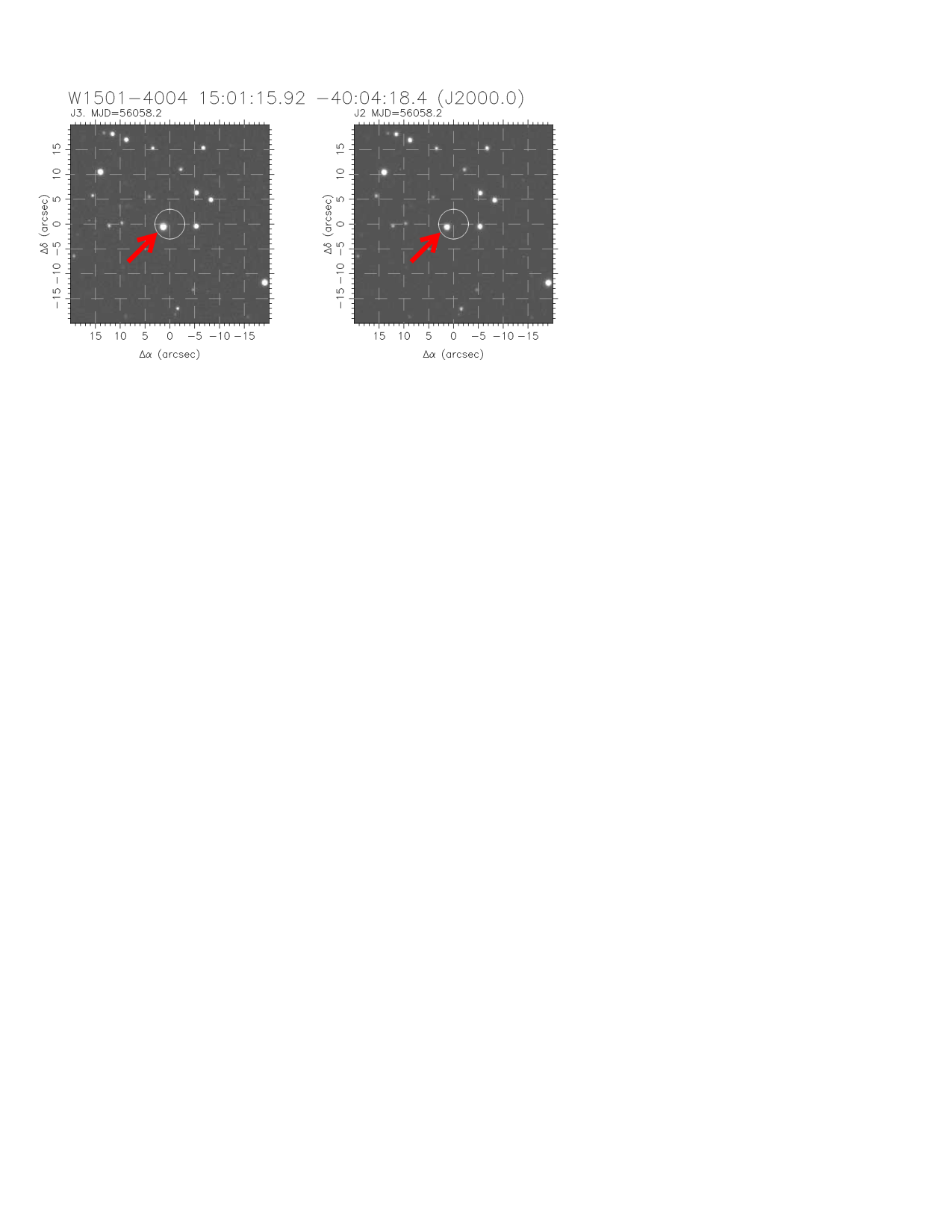}\\[-0.5cm]
\includegraphics[width=7.0cm,trim=0.5cm 19.0cm 9.0cm 1.5cm,clip=true]{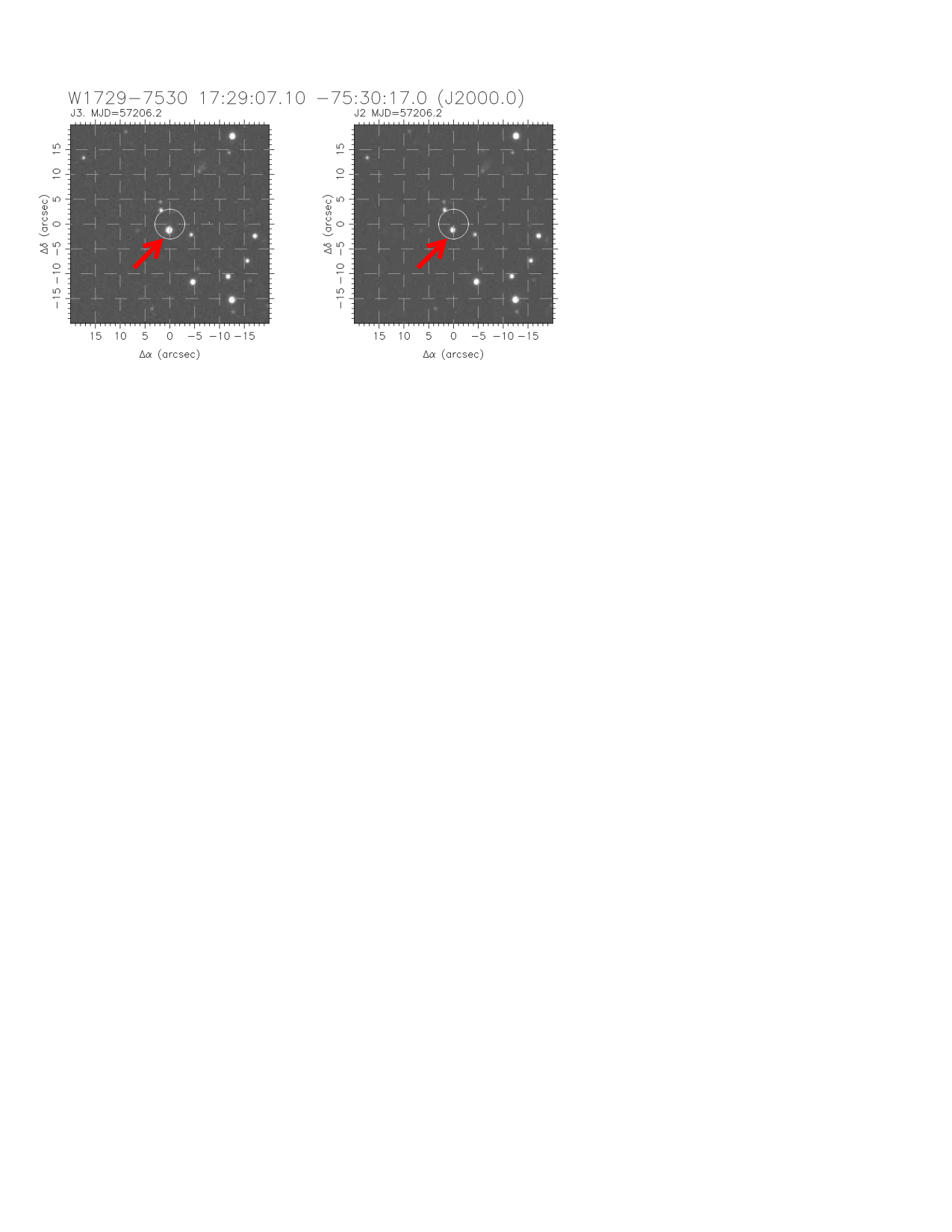}\qquad\includegraphics[width=7.0cm,trim=0.5cm 19.0cm 9.0cm 1.5cm,clip=true]{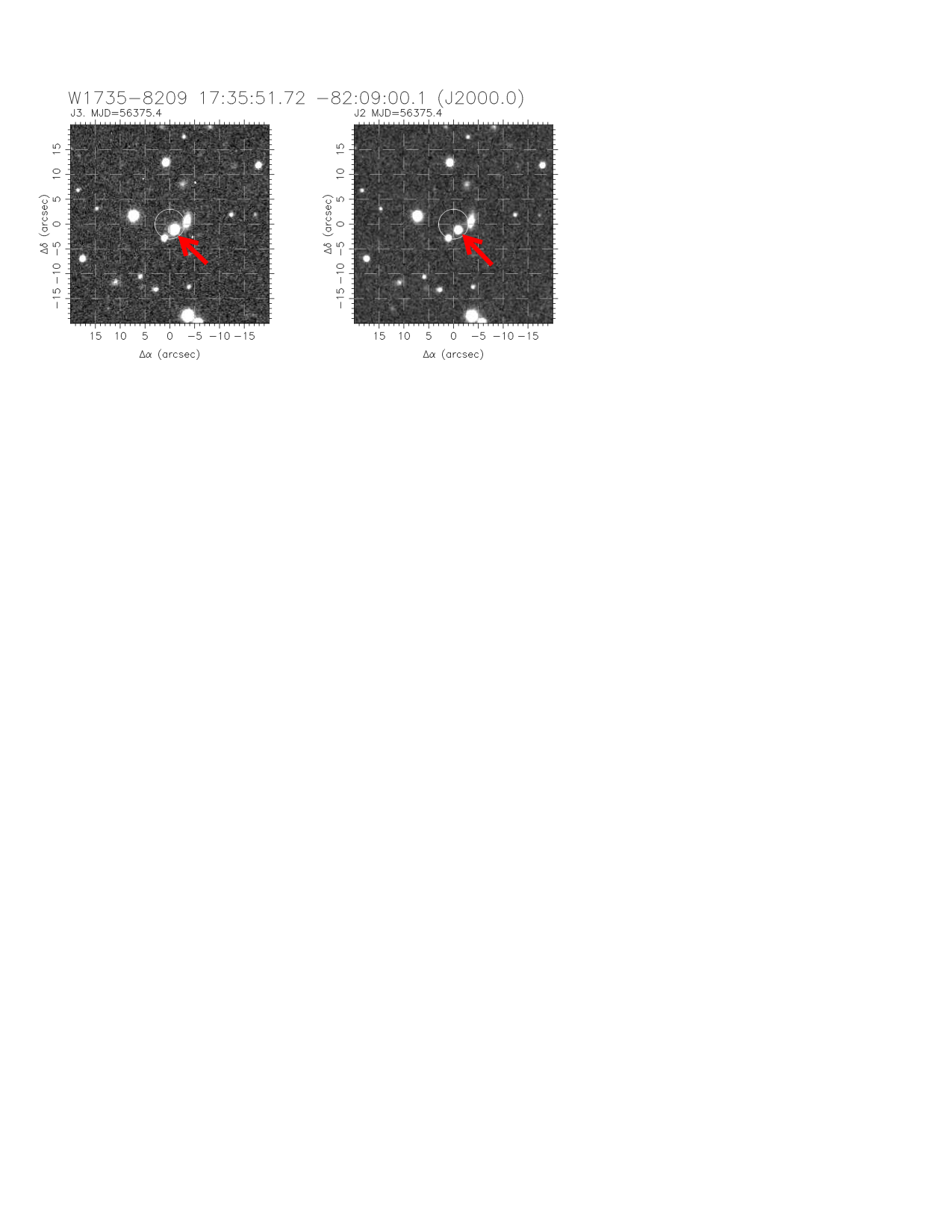}\\[-0.5cm]
\includegraphics[width=7.0cm,trim=0.5cm 19.0cm 9.0cm 1.5cm,clip=true]{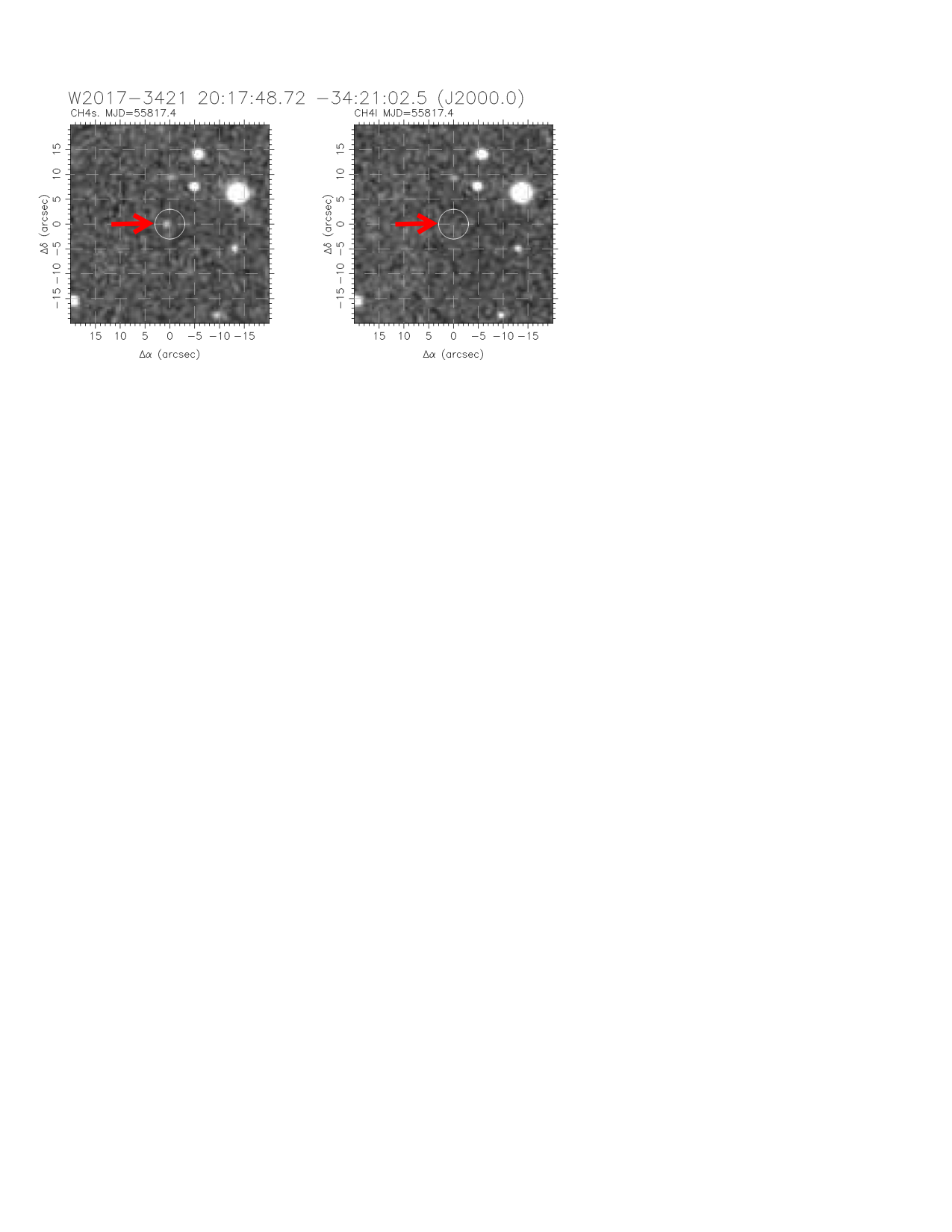}\qquad\includegraphics[width=7.0cm,trim=0.5cm 19.0cm 9.0cm 1.5cm,clip=true]{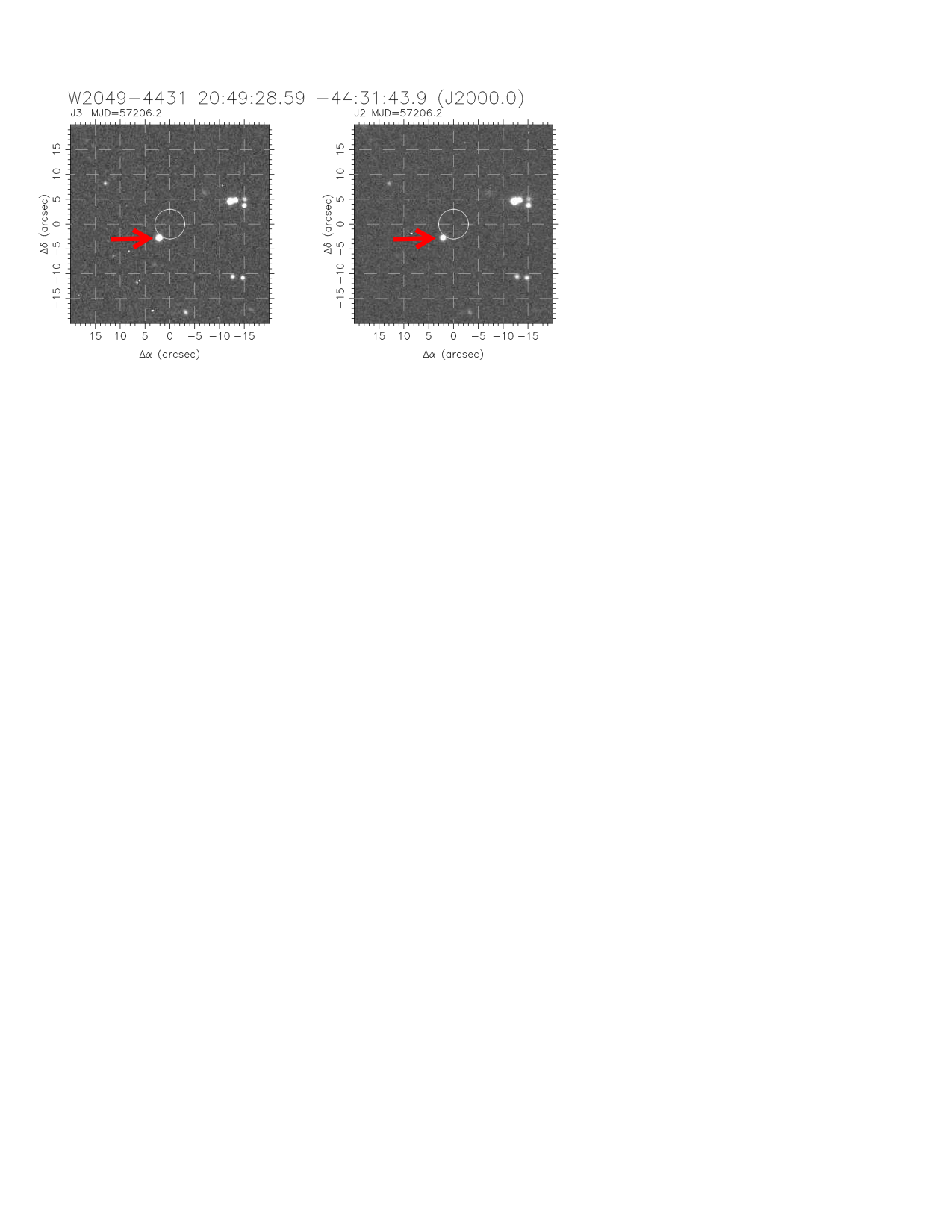}\\[-0.5cm]
   \caption{The single image of W1433-0837 is a J band one from the AAT, with the arrow indicating the object observed spectroscopically.)}
   \label{FC1a}
\end{figure*}

\begin{figure*}
\figurenum{1 (cont)}
\includegraphics[width=7.0cm,trim=0.5cm 19.0cm 9.0cm 1.5cm,clip=true]{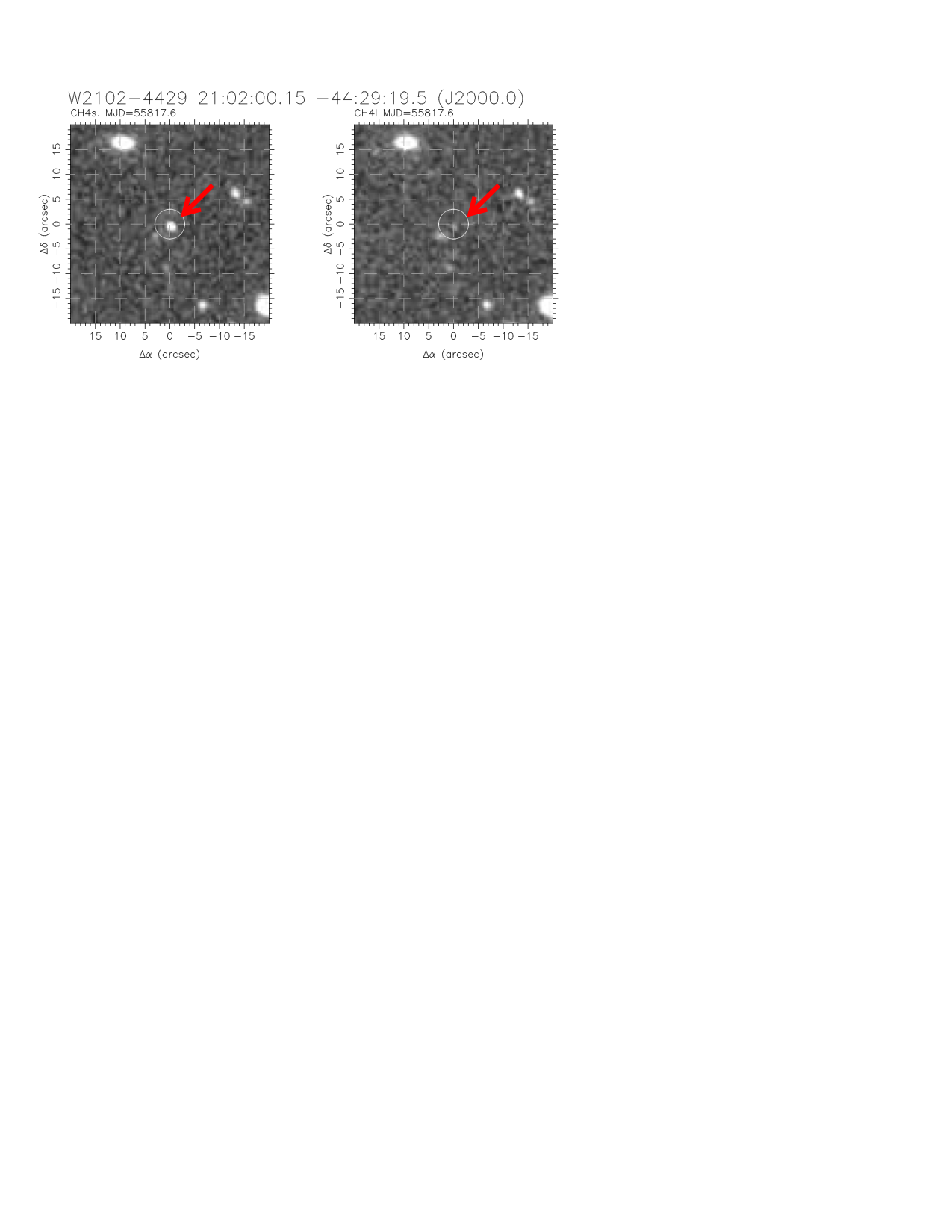}\qquad\includegraphics[width=7.0cm,trim=0.5cm 19.0cm 9.0cm 1.5cm,clip=true]{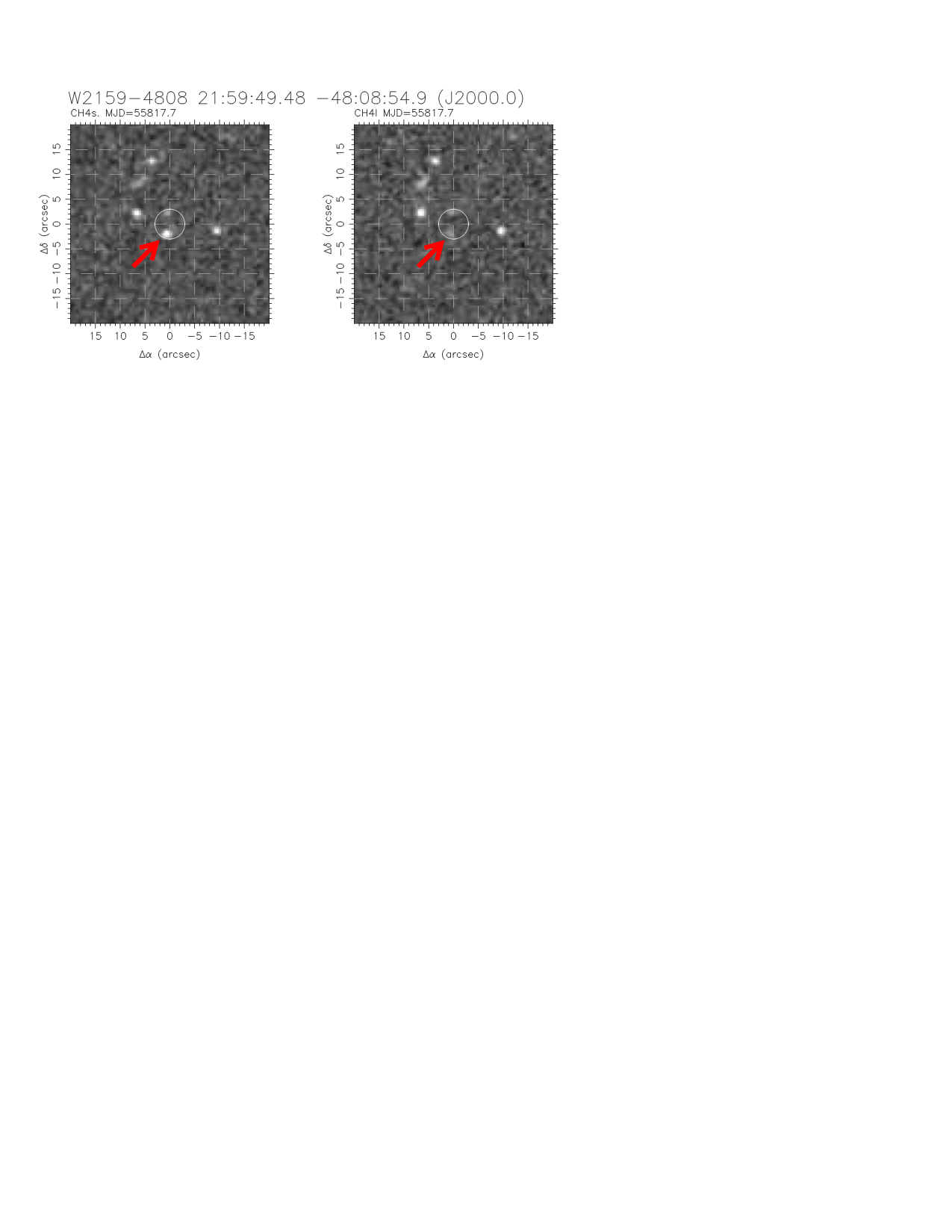}\\[-0.5cm]
\includegraphics[width=7.0cm,trim=0.5cm 19.0cm 9.0cm 1.5cm,clip=true]{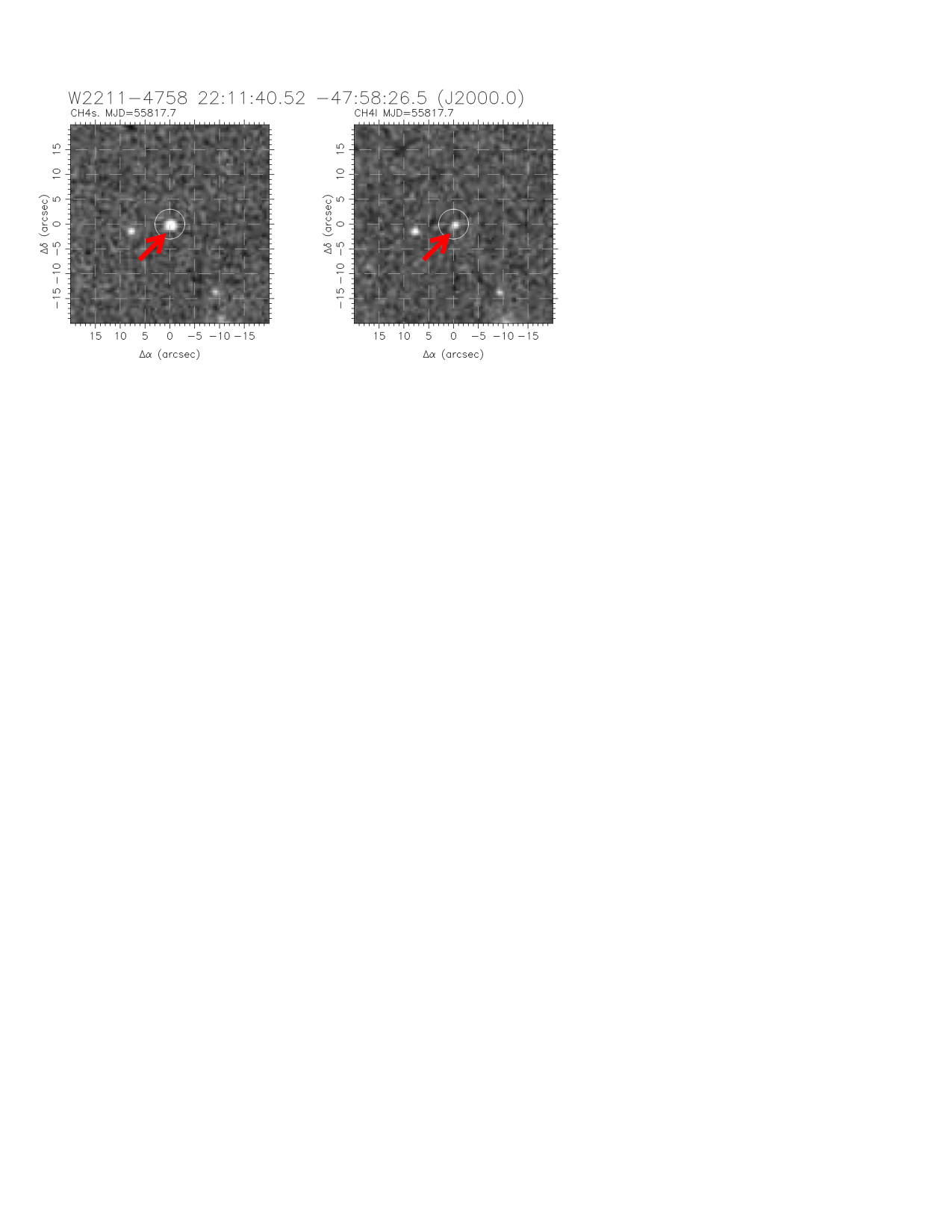}\qquad\includegraphics[width=7.0cm,trim=0.5cm 19.0cm 9.0cm 1.5cm,clip=true]{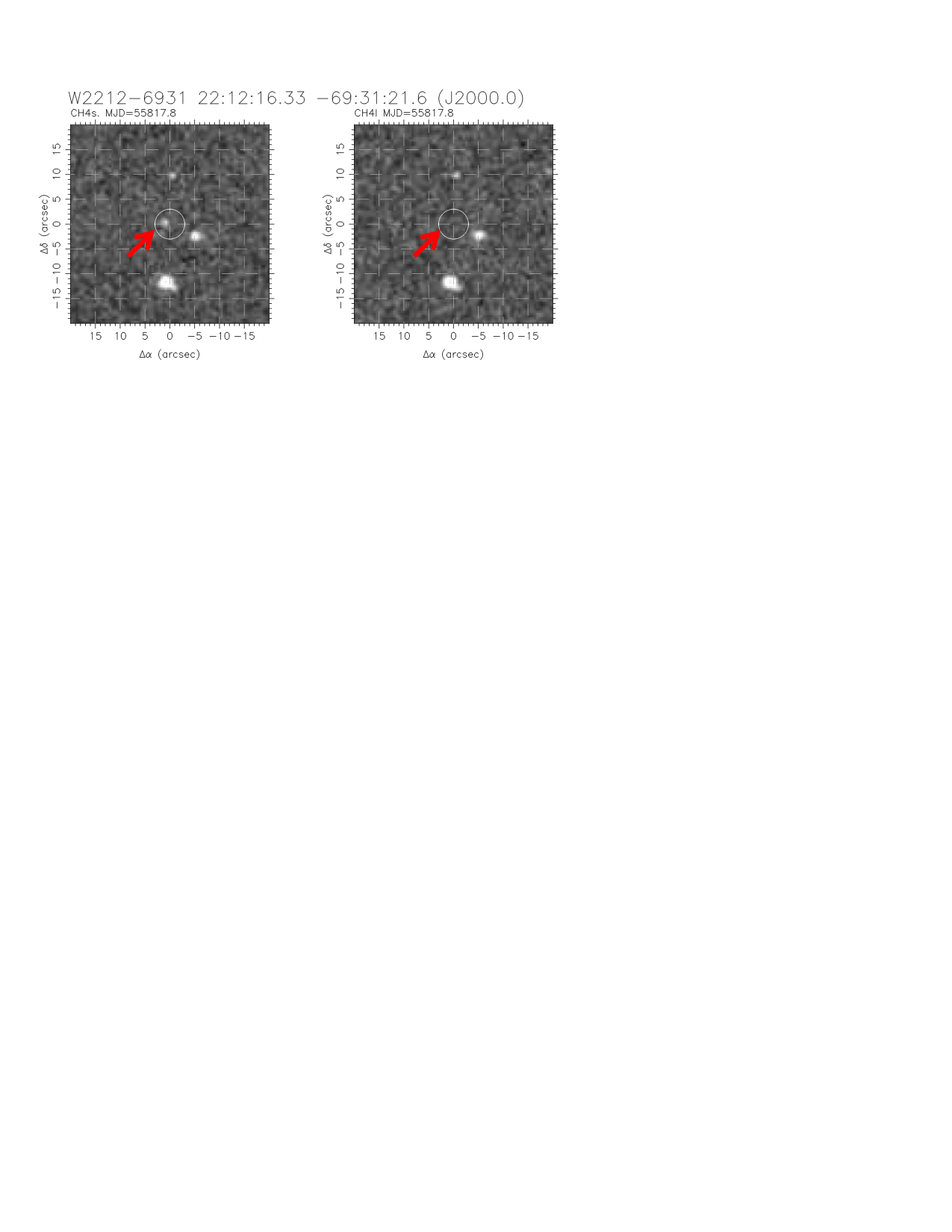}\\[-0.5cm]
\includegraphics[width=7.0cm,trim=0.5cm 19.0cm 9.0cm 1.5cm,clip=true]{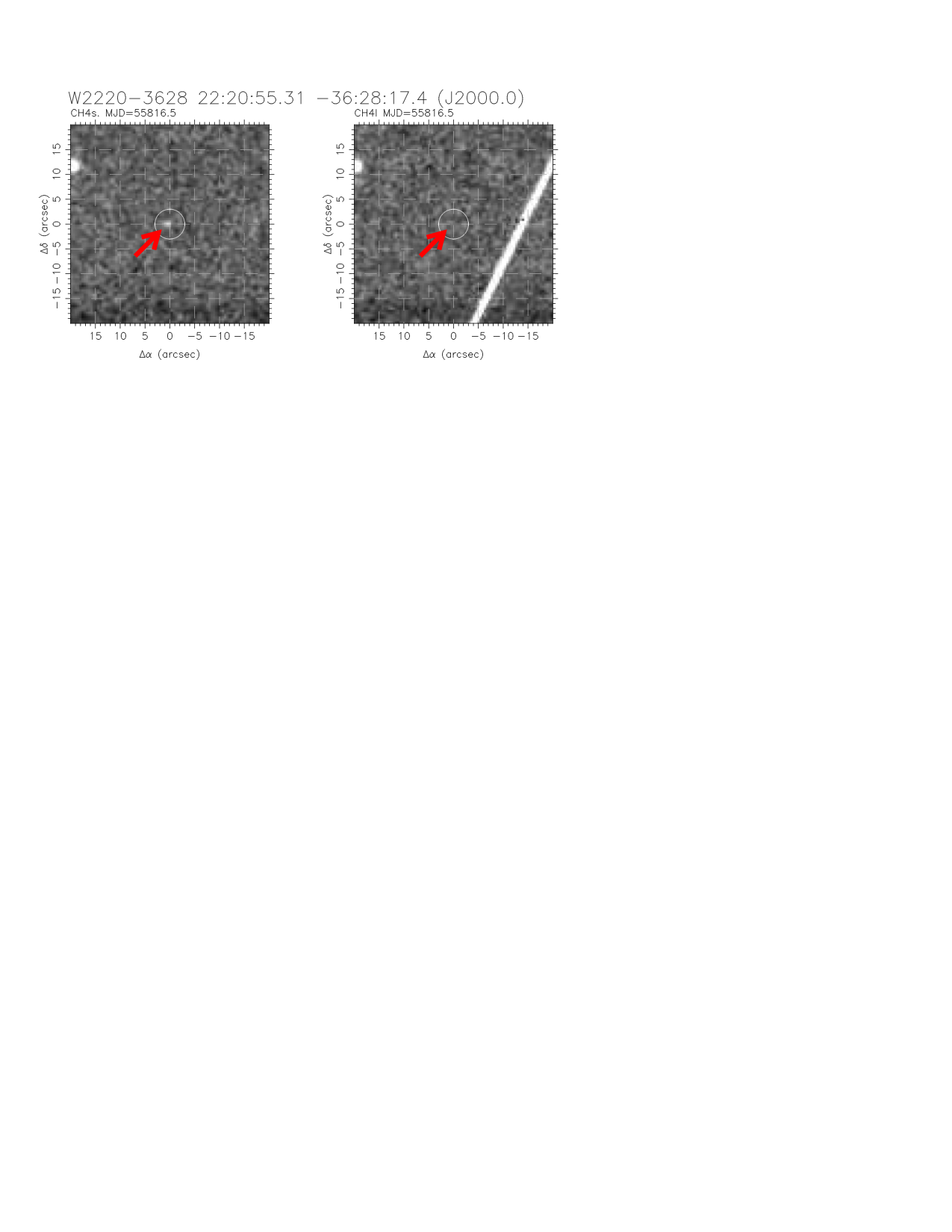}\qquad\includegraphics[width=7.0cm,trim=0.5cm 19.0cm 9.0cm 1.5cm,clip=true]{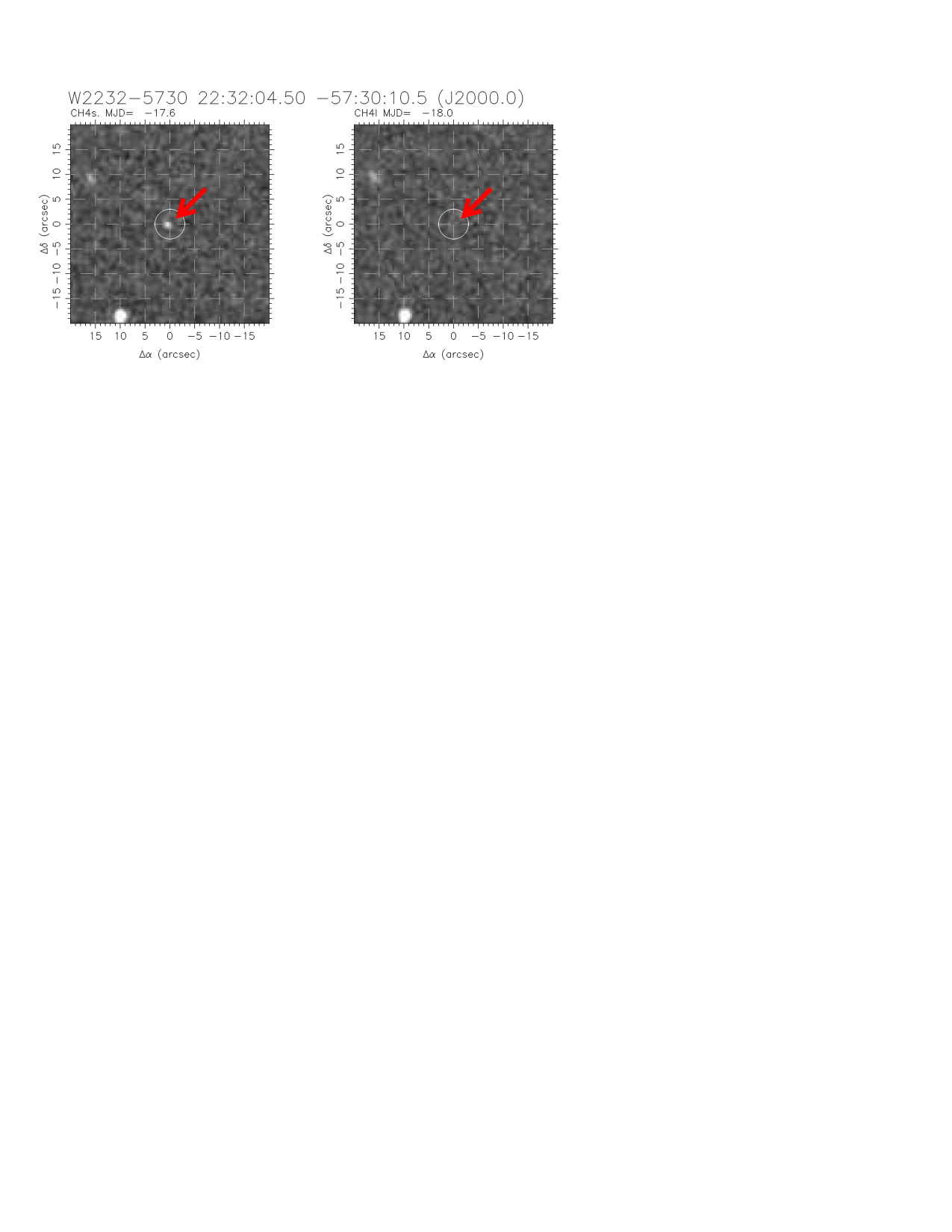}\\[-0.5cm]
\includegraphics[width=7.0cm,trim=0.5cm 19.0cm 9.0cm 1.5cm,clip=true]{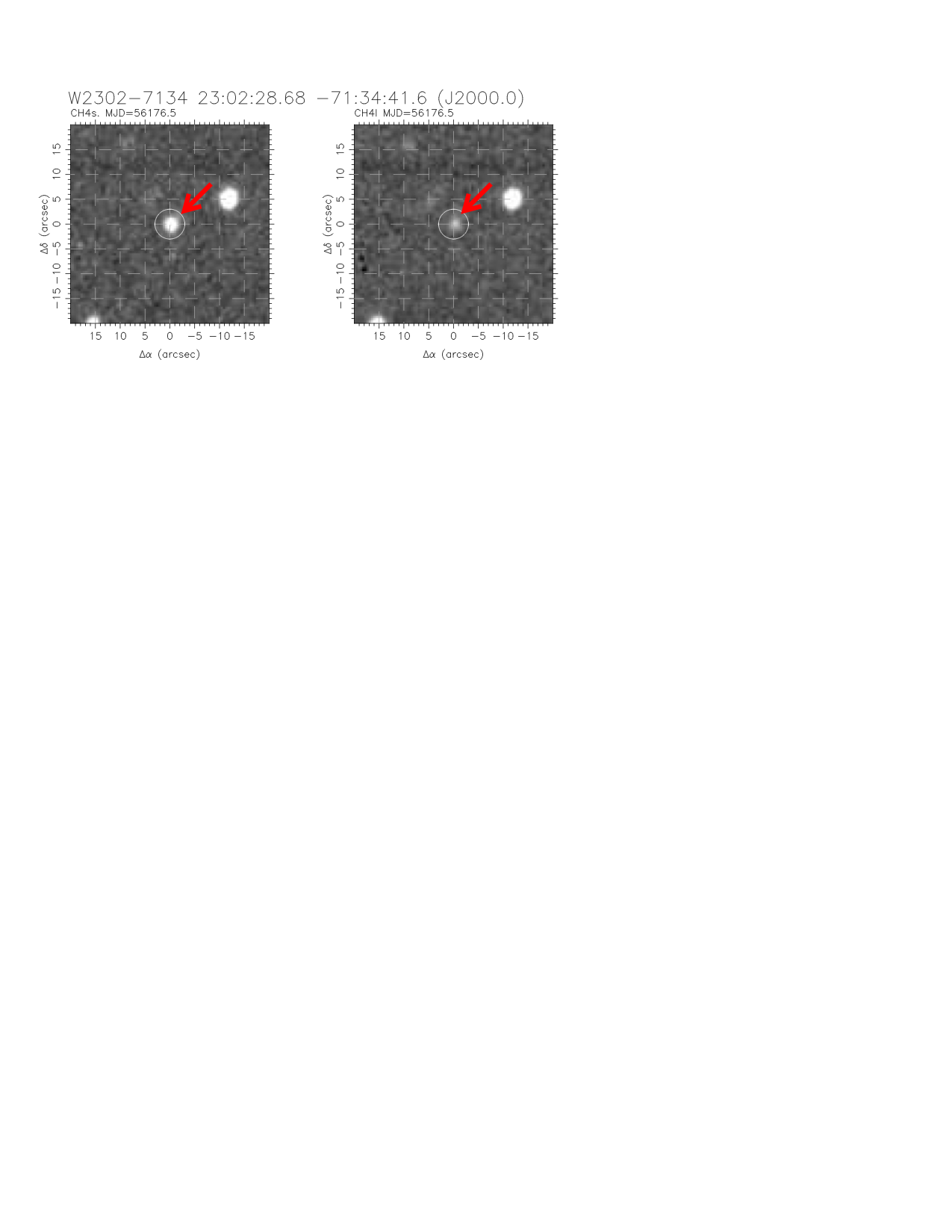}\qquad\includegraphics[width=7.0cm,trim=0.5cm 19.0cm 9.0cm 1.5cm,clip=true]{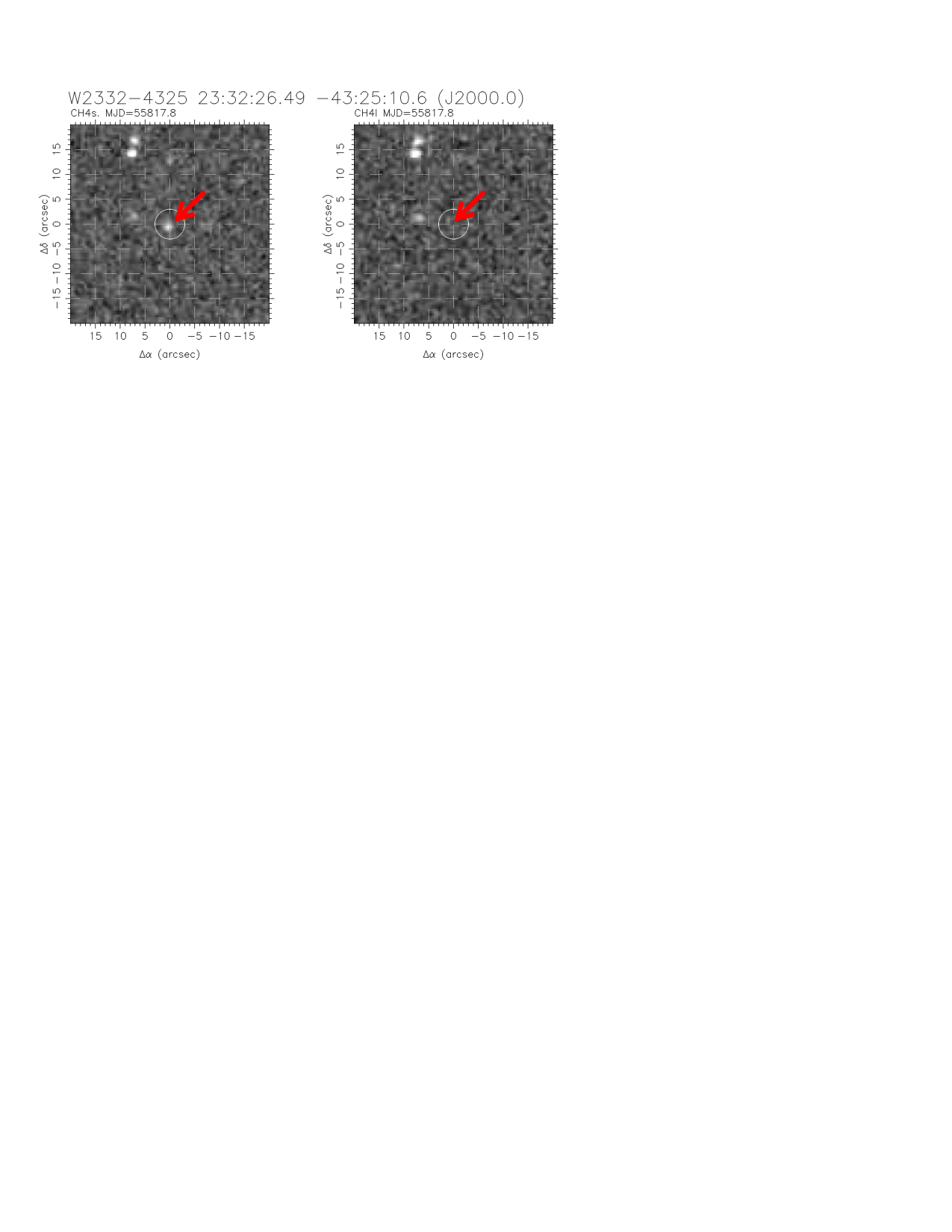}\\[-0.5cm]
\includegraphics[width=7.0cm,trim=0.5cm 19.0cm 9.0cm 1.5cm,clip=true]{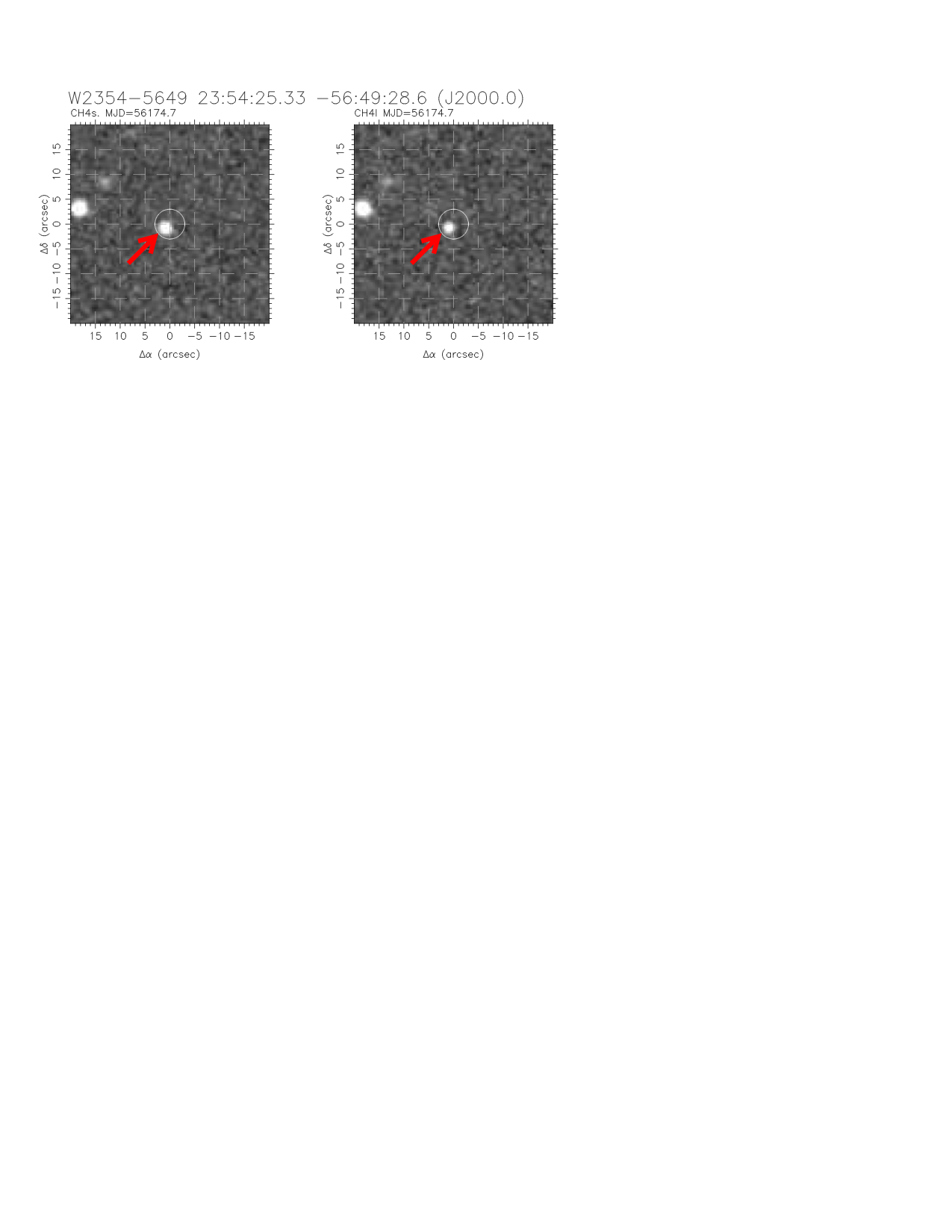}\qquad\includegraphics[width=7.0cm,trim=0.5cm 19.0cm 9.0cm 1.5cm,clip=true]{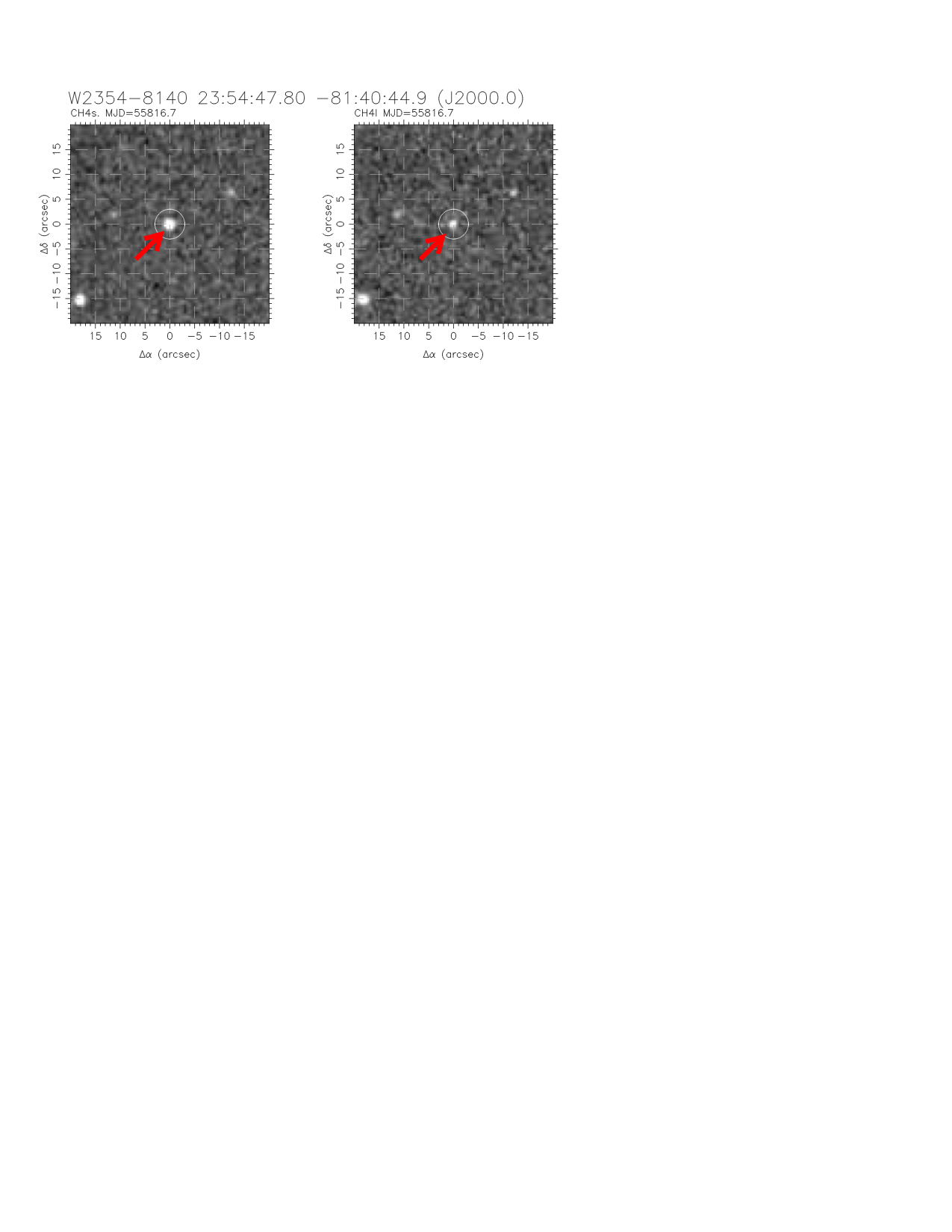}\\[-0.5cm]
   \caption{}
   \label{FC1b}
\end{figure*}

\setcounter{figure}{1}

\begin{figure*}
\includegraphics[width=7.0cm,trim=0.5cm 19.0cm 9.0cm 1.5cm,clip=true]{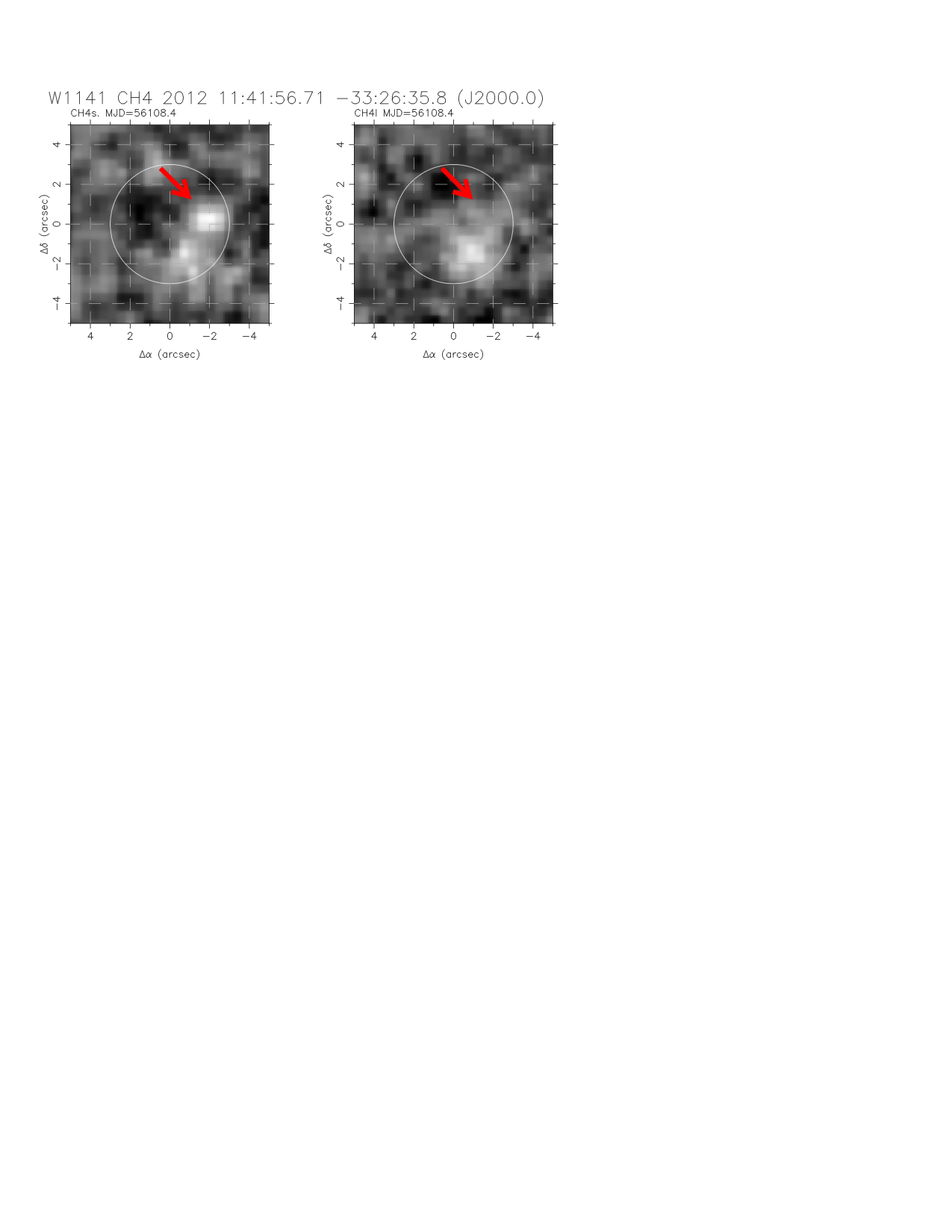}\qquad\qquad\includegraphics[width=7.0cm,trim=0.5cm 19.0cm 9.0cm 1.5cm,clip=true]{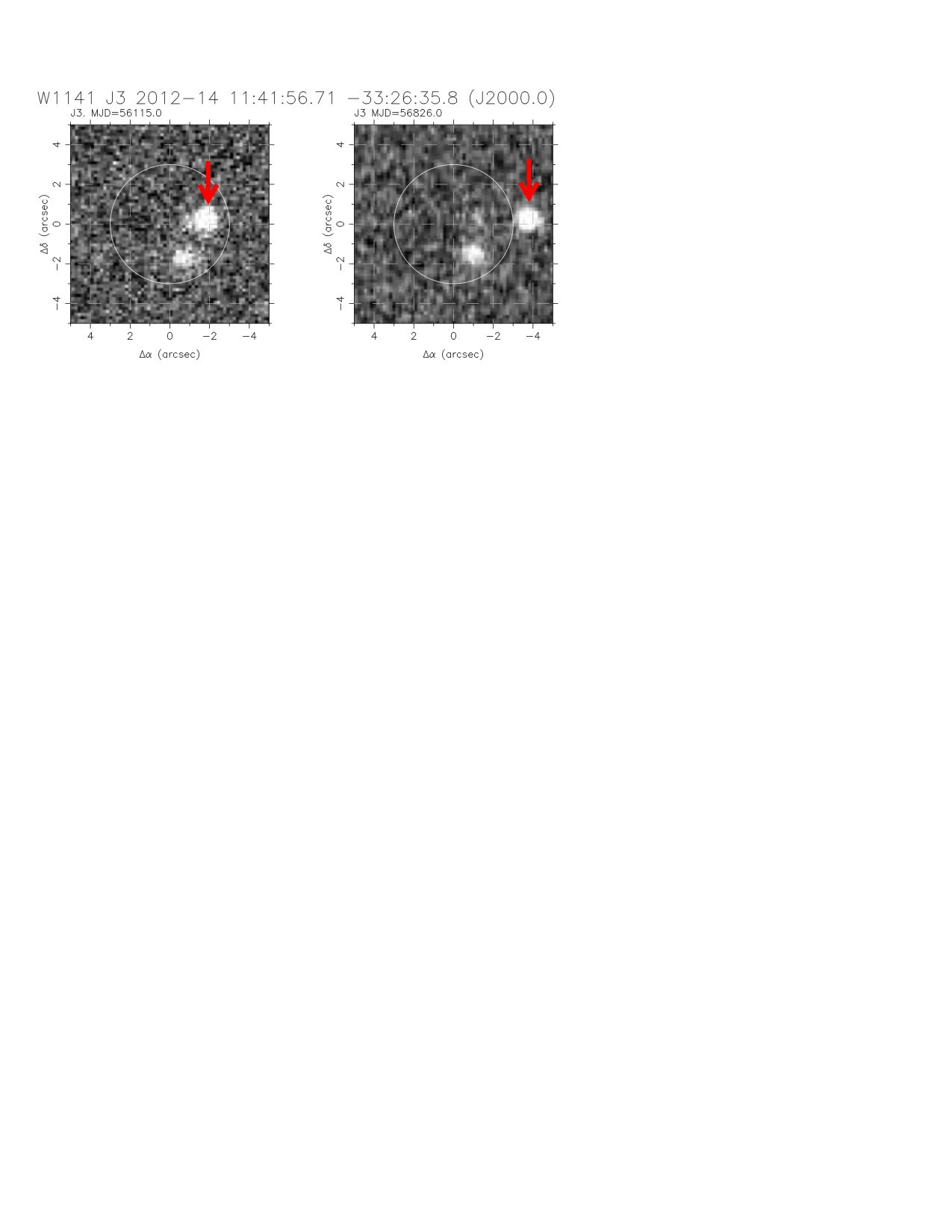}\\[-0.5cm]
   \caption{Zoomed images 10\arcsec\ on a side at the location of W1141-3326. \Ms\ and \Ml\ images from 2012 are shown on the left, and J3 images from 2012 and 2014 are shown on the right. W1141-3326's proper motion moved it 
   2\arcsec\ to the W between July 2012 and June 2014, and away from a group of background sources that clearly contaminated the \Ml\ photometry in 2012, producing a spuriously red \Msl\ colour.}
   \label{W1141}
\end{figure*}

\section{Observations -- Spectroscopy} \label{spectra}

Follow-up spectroscopy is the ``gold standard'' for confirming a T- or Y-dwarf identification. Spectroscopic observations have therefore been carried out by members of the \WISE Science Team brown dwarf collaboration on a variety of telescopes in parallel with our imaging program at the AAT and Magellan. We report here spectroscopy obtained on multiple nights between 2011 Sep 8 and 2017 Jan 05 (Table \ref{spec_log}).\\

\begin{deluxetable}{rllcc}
\tabletypesize{\scriptsize}
\tablecaption{WISE Cool Dwarf Spectroscopy\label{spec_log}}
\tablewidth{9.0cm}
\tablehead{
\colhead{WISE Designation} & \colhead{UT Date}    & \colhead{Telescaope}  & \colhead{Exp.} & \colhead{Sp.} \\
                           &                      &                       & \colhead{(s)}  & \colhead{Type}             
}
\startdata
~J001505.87$-$461517.6 &   2012 Jan 05         &  Magellan\tablenotemark{b}& 1014           & T8   \\
~J003231.09$-$494651.4 &   2012 Jan 05         &  Magellan\tablenotemark{b}& 1268           & T8.5  \\
~J024124.73$-$365328.0 &   2011 Sep 08         &  Keck\tablenotemark{d}    & 600            & T7   \\
AJ030237.53$-$581740.3 &   2017 Jan 05         &  Magellan\tablenotemark{c}& 2700           & Y0:   \\
~J064528.38$-$030248.2 &   2013 Mar 04         &  IRTF                     & 2400           & T6   \\
~J071301.84$-$585445.1 &   2012 May 06         &  Magellan\tablenotemark{b}& 1200           & T9   \\
~J091408.96$-$345941.5 &   2016 Mar 23		   &  Magellan\tablenotemark{c}& 900            & T8   \\
~J094020.10$-$220820.5 &   2013 Mar 04         &  IRTF                     & 1600           & T8   \\
AJ102313.22$-$315126.7 &   2016 Mar 23		   &  Magellan\tablenotemark{c}& 900            & T8   \\
~J105553.59$-$165216.3 &   2013 Mar 22         &  Magellan\tablenotemark{c}& 800            & T9:  \\
~J111239.24$-$385700.7 &   2012 May 06         &  Magellan\tablenotemark{b}& 1680           & T9   \\
~J114156.71$-$332635.8 &   2014 Mar 10         &  Magellan\tablenotemark{c}& 2000           & Y0   \\
~J143311.42$-$083736.4 &   2012 Jan 17         &  Magellan\tablenotemark{b}& 240            & T8   \\  
~J144806.48$-$253420.3 &   2014 Mar 10         &  Magellan\tablenotemark{c}& 800            & T8   \\
~J150115.92$-$400418.4 &   2014 Mar 04         &  IRTF                     & 2000           & T6   \\
AJ172907.10$-$753017.0 &   2016 Mar 23		   &  Magellan\tablenotemark{c}& 900            & T7   \\
~J173551.72$-$820900.1 &   2016 Mar 23		   &  Magellan\tablenotemark{c}& 600            & T6   \\
~J210200.15$-$442919.5 &   2011 Oct 09         &  Keck\tablenotemark{d}    & 4200           & T9  \\
~J215949.48$-$480854.9 &   2012 May 05         &  Magellan\tablenotemark{b}& 1400           & T9   \\
~J221216.33$-$693121.6 &   2012 May 06         &  Magellan\tablenotemark{b}& 1400           & T9.5 \\
~J223204.50$-$573010.5 &   2012 May 06         &  Magellan\tablenotemark{b}& 1400           & T9  \\
~J233226.49$-$432510.6 &   2011 Oct 09         &  Keck\tablenotemark{d}    & 4200           & T9:  \\
~J235425.33$-$564928.6 &   2016 Nov 18         &  Magellan\tablenotemark{c}& 1200           & T6 \\
\enddata
\tablenotetext{a}{\WISE designations follow Table \ref{ImageTable}}
\tablenotetext{b}{Data processing carried out with FIREHOSE package as described in the text.}
\tablenotetext{c}{Data processing carried out with Figaro package as described in the text.}
\tablenotetext{d}{These are the same spectra used to obtain spectral types by \cite{kirkpatrick2012}, albeit independently 
classified in this paper (Fig. \ref{SpFig}). In both cases we obtain the same types as Kirkpatrick et al.}
\end{deluxetable}

\subsection{Magellan/FIRE}
The Folded-port Infrared Echellette \citep[FIRE;][]{simcoe2008,simcoe2010} at
the 6.5m Walter Baade Telescope on Cerro Manqui at the Las Campanas Observatory, Chile,
uses a 2048$\times$2048 HAWAII-2RG array. In prism mode, it covers a wavelength range from
0.8 to 2.5\,$\mu$m at a resolution ranging from R=500 at J-band to R=300 at K-band for a slit
width of 0.6\arcsec. FIRE was used to obtain spectroscopy of T dwarf candidates on the nights listed in Table~\ref{spec_log}. 
FIRE reductions in this paper were carried out in two ways.
Roughly half the spectra were processed in the same manner as the spectroscopy presented in \cite{tinney2012a} using the Figaro
data reduction package, with telluric removal carried out using A0 spectra acquired immediately before
or after the science target at similar airmass. Remaining objects were processed using the FIREHOSE package for low-dispersion data 
following the procedure in the online ``cookbook"\footnote{See http://www.mit.edu/people/rsimcoe/FIRE/ob\_data.htm for details.}. 
For our faintest objects, pair subtraction prior to insertion into the pipeline 
greatly improved the accuracy of the sky-line fitting procedure. 
The combined spectrum was then corrected for telluric absorption and flux calibrated using  observations of an A0 V 
star and the technique described in \cite{vacca2003} and the XTELLCOR program from SpeXtool (see \citealt{cushing2004}).

\subsection{Keck/NIRSPEC}
The Near-Infrared Spectrometer \citep[NIRSPEC;][]{mclean1998,mclean2000} at the 10m
W. M. Keck Observatory on Mauna Kea, HawaiÕi, was used to obtain confirmation spectroscopy of several new T dwarfs.
For spectroscopy, NIRSPEC uses a 1024$\times$1024 InSb array.  The  NIRSPEC observations employed the 
42\arcsec$\times$0${\farcs}$57 slit, providing a resolution R$\sim$1500. Our brown
dwarf candidates were observed in the N3 configuration \citep[see][]{mclean2003} that covers
part of the J-band window from 1.15 to 1.35\,$\mu$m.
Data reduction made use of the publicly available REDSPEC package, with modifications to remove 
residuals from the sky-subtracted pairs prior to 1-D spectral extraction.  \\

\subsection{IRTF/SpeX}
\label{IRTF-section}

SpeX is a medium-resolution spectrograph and imager at NASA's 3m Infrared Telescope Facility (IRTF)
on Mauna Kea, Hawai'i. It uses a 1024$\times$1024 InSb array for its spectroscopic observations
(\citealt{rayner2003}). We used the prism mode with a
0$\farcs$5 wide slit to achieve a resolving power of $R\equiv
\lambda / \Delta \lambda \approx 150$ over the range 0.8-2.5\,$\mu$m.  A series of 200s 
exposures were typically obtained at two different positions along the 15$\arcsec$
long slit.  A0 dwarf stars were observed soon after or before the target and
at similar airmass, and used for telluric correction and flux calibration.  A set
of exposures of internal flat field and argon arc lamps were obtained for
flat fielding and wavelength calibration.\\

The data were reduced using Spextool (\citealt{cushing2004}) the
IDL-based data reduction package for SpeX.  The raw images were first
corrected for non-linearity, pair subtracted, and then flat fielded.
For some of the fainter sources, multiple pair-subtracted images were
averaged in order to facilitate tracing.  The spectra
were then optimally extracted (e.g., \citealt{horne1986}) and
wavelength calibrated using the argon lamp exposures.  Multiple spectra
were then averaged and the resulting spectrum was corrected for
telluric absorption and flux calibrated using observations of an A0
V star using the technique described in \cite{vacca2003}.\\

\subsection{Spectral Types}

\begin{figure*}
   \includegraphics[width=90mm,trim=1.0cm 0.5cm 7.0cm 3.5cm,clip=true]{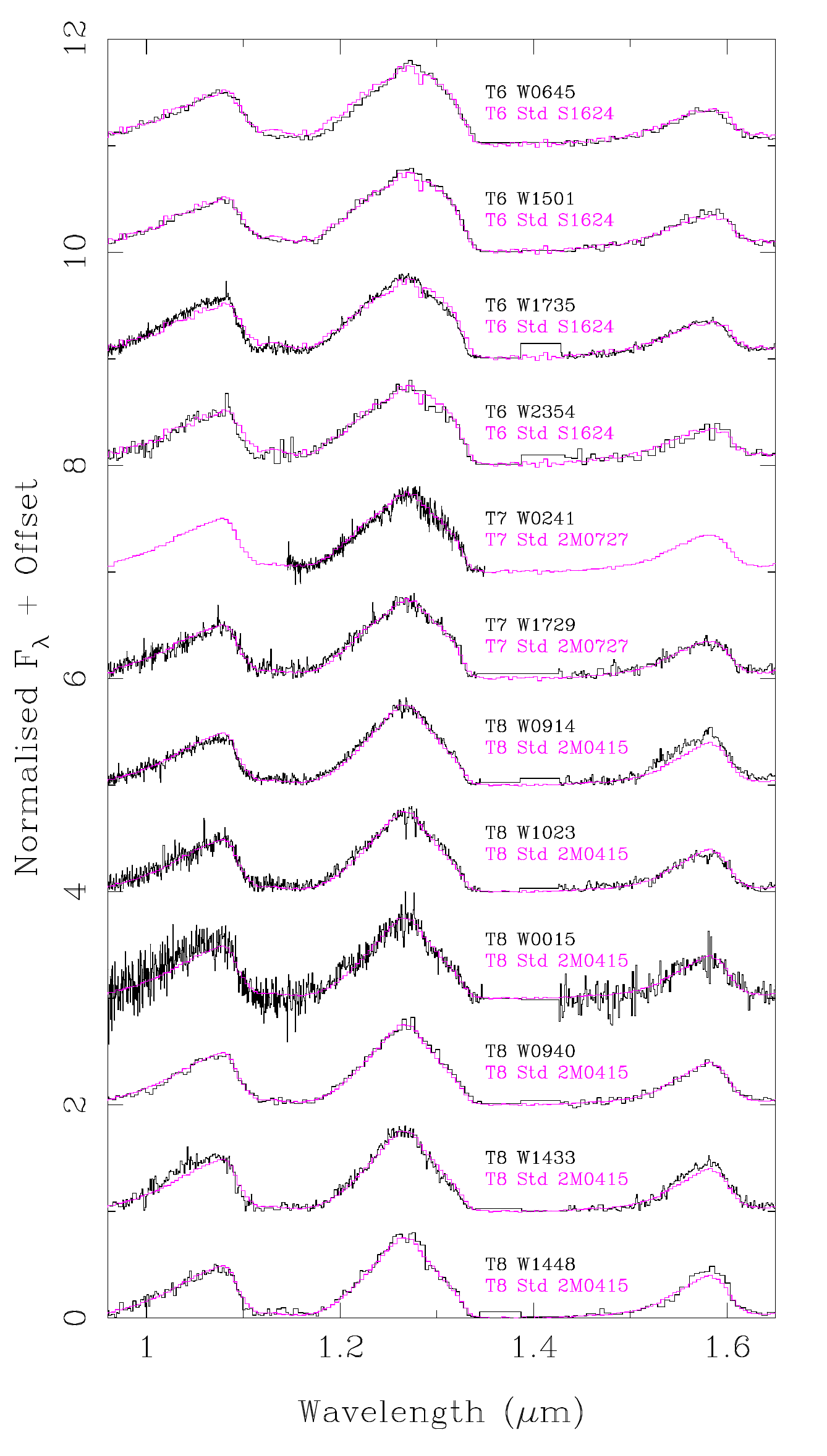}
   \includegraphics[width=90mm,trim=1.0cm 0.5cm 7.0cm 3.5cm,clip=true]{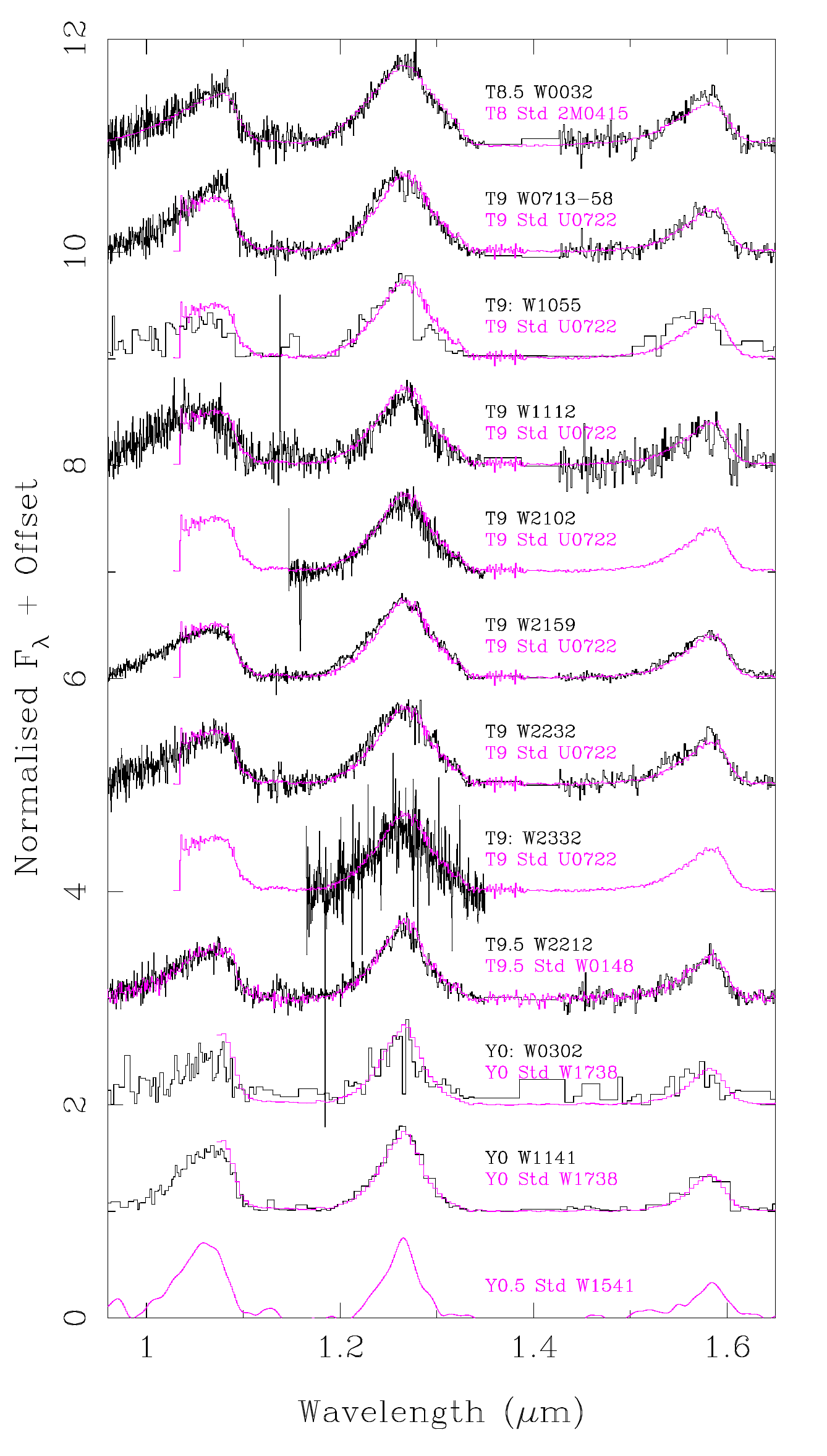}
   \caption{T and Y dwarf spectra listed in Table \ref{spec_log} plotted to demonstrate
   spectral classifications derived as described in the text. Each new spectrum {\em (black lines)} 
   is over plotted with the relevant spectral standard {\em (magenta lines)}. \label{SpFig}}
\end{figure*}

The objects listed in Table~\ref{spec_log} were spectrally classified using the 
near-infrared T0-to-T8 dwarf sequence of \cite{burgasser2006}, extended to later T and Y dwarfs by
\citet{cushing2011} and \citet{kirkpatrick2012}. (See Table \ref{SpecStd} for the specifically
adopted standard spectra). Assignment of  types
was performed by overplotting these standards onto the candidate spectra and determining by-eye which standard
provided the best match. In some cases two adjacent standards, such as T8 and T9, provided
an equally good match, so the candidate spectrum was assigned an intermediate type (in this example, T8.5). 
In Figure~\ref{SpFig} we show the  near-infrared spectra for all of our sources compared with the
relevant spectral standard. We consider these types to have an uncertainty of $\pm$0.5 subtypes.
Types which are more uncertain due to low signal-to-noise-ratio spectra (i.e. $\pm$1 sub-type) are
marked with a colon ``:'' next to the spectral type.\\

\begin{deluxetable}{lll}
\tabletypesize{\scriptsize}
\tablewidth{8.0cm}
\tablecaption{Adopted Spectral Standards.\label{SpecStd}}
\tablehead{
\colhead{Spectral}          & \colhead{Full}        & \colhead{Short}  \\     
\colhead{Type}              & \colhead{Designation} & \colhead{Designation}             
}
\startdata
   T5  &   2MASS J15031961$+$2525196		& 2M1503 \\
   T6  &   SDSS J162414.37$+$002915.6    & S1624 \\
   T7  &   2MASS J07271824$+$1710012     & 2M0727 \\
   T8  &   2MASS J04151954$-$0935066     & 2M0415 \\
   T9  &   UGPS J072227.51$-$054031.2	   & U0722 \\
   T9.5&   WISE J014807.25$-$720258.7	& W0148 \\
   Y0  &   WISEPA J173835.53$+$273258.9	& W1738 \\
   Y0.5&   WISEPA J154151.66$-$225025.2	& W1541 \\
   Y1  &   WISE J035000.32$-$565830.2	   & W0350 \\
\enddata
\end{deluxetable}

\section{Discussion} \label{discussion}

\subsection{An Updated \Msl\  Spectral Type Calibration}\label{newcal}

\begin{figure*}
   \includegraphics[width=160mm,trim=1.0cm 14.5cm 3.0cm 1.0cm,clip=true]{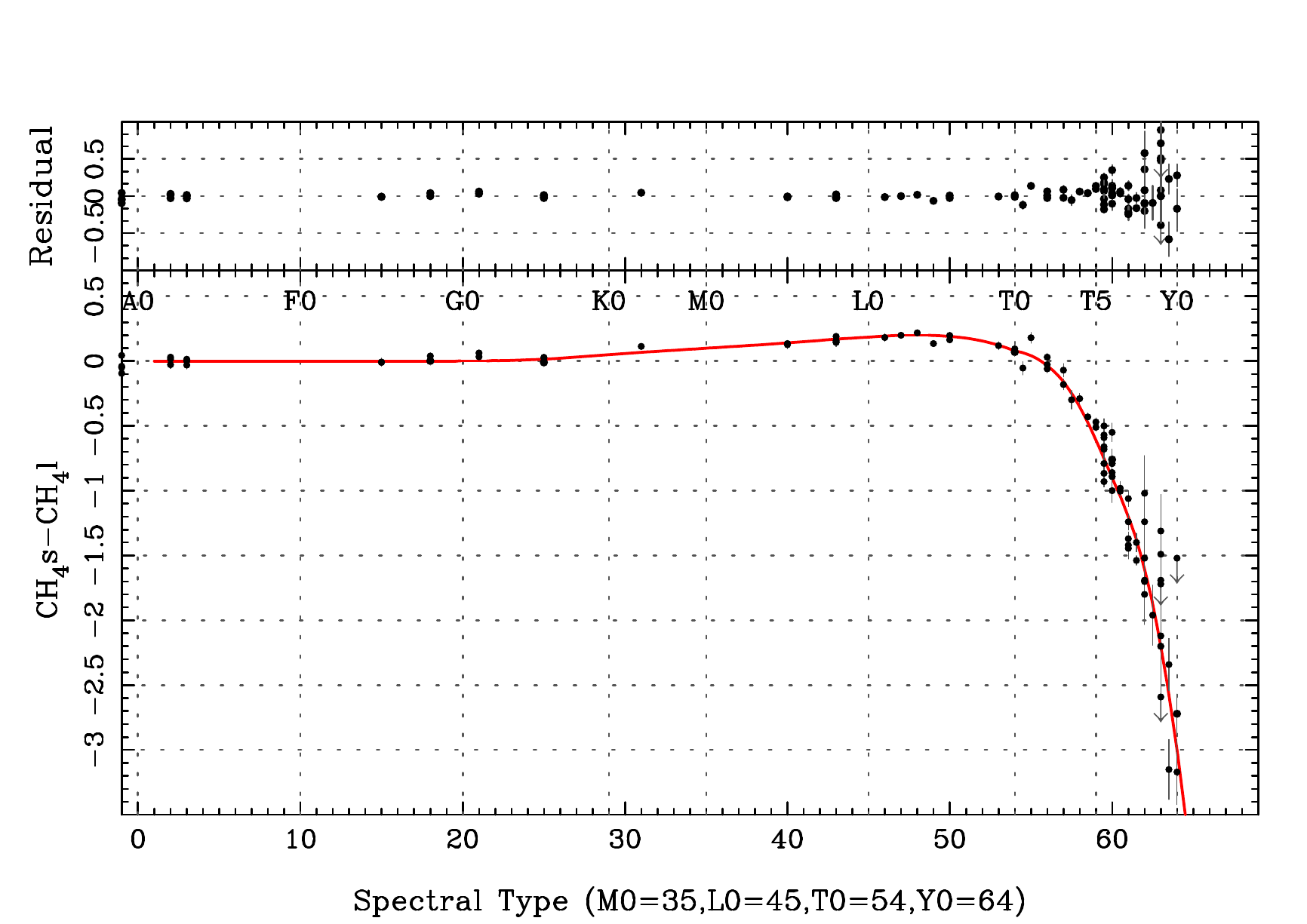}
   \includegraphics[width=160mm,trim=1.0cm 2.0cm  2.0cm 16.0cm,clip=true]{Tfit_CH4_Spt_2018.pdf}
   \caption{{\em Upper panel} -- Methane-sensitive \Ms$-$\Ml\ as a function of A-Y spectral type.
            The uncertainties plotted are the combination of photon-counting
            uncertainties, aperture correction uncertainties and photometric
            calibration uncertainties. Typical uncertainties on spectral types
            (not plotted) are $\pm$0.5. The plotted parametrisation is a spline
            constrained at the locations in Table \ref{MethaneSpectralTypeReln}.
            Root-mean-square (rms) scatter about the parametrisation is 0.18\,mag 
            for the whole range, 0.06\,mag for L0-T2 dwarfs and 0.12\,mag for
            T3-T8 dwarfs, and 0.45\,mag for T9-Y0 dwarfs. The corresponding
            spectral type scatters are 0.6\,sub-type for L0-T2 dwarfs and 0.43\,sub-type for
            T3-T8 dwarfs, and 0.6\,sub-type for T9-Y0 dwarfs. 
            {\em Lower right panel} -- Upper panel zoomed in on the T and Y
            dwarf region, with the previously published calibration relations due
            to Tinney et al. (2005) and Cardoso et al. (2016) also shown. 
            {\em Lower left panel} -- Comparison of the
            predicted spectral types from this \Ms$-$\Ml\ relation with observed
            spectral types (solid circles - \cite{tinney2005}; solid stars -- this paper).
            \label{MethaneSpectralType}}
\end{figure*}

\citet{tinney2005} have presented a calibration between the IRIS2 \Msl\ colour and spectral type. 
\cite{Cardoso2015} have used very similar filters on the NICS instrument on the Telescope Nazionale Galileo
and derived a calibration using the same functional form as \cite{tinney2005}, but with slightly different
values. With the additional 
methane imaging data for later-type objects available in Table \ref{ImageTable} we have updated this calibration 
to derive the relation shown in the upper panel of Fig. \ref{MethaneSpectralType}. A relation with a simple 
functional form (like that used by Tinney et al.) proved to be impossible to obtain, and so we have instead 
used a spline calibration, which we present as the sequence shown in Table \ref{MethaneSpectralTypeReln}. The 
scatter about this calibration over the whole spectral type range is 0.11\,mag. More importantly for the use of \Msl\ 
colours to estimate spectral types, the slope of the relation for types beyond T5 is very steep, so that typical 
measurement errors of $\pm$0.1\,mags in the colour map into uncertainties in  spectral type estimates of less 
than $\pm$0.3 sub-types for all objects later than T5. We  therefore adopt an uncertainty for our estimated T 
and Y types in Table \ref{ImageTable} of better than 0.5 sub-types.\\

The lower left panel of Fig. \ref{MethaneSpectralType} shows an expanded version of this plot, along with the
calibration relations due to \cite{tinney2005} and \cite{Cardoso2015}. The authors of the latter have
already noted that their calibration is systematically redder (in \Msl) for a given spectral type than the earlier calibration. 
Examination of
their Fig. 4 suggests that this may be due to either a fortuitous observation of a sample of objects lying
above the sequence of \cite{tinney2005}, or a systematic difference in the colour terms driven by the optics of
NICS on the Telescopio Nazionale Galileo, or both. (The filters and detectors in both IRIS2 and NICS are identical, so the
only the rest of the optical train can plausibly produce this difference.) The addition of further very late objects in our new
sample of T and Y dwarfs brings our new calibration and that of \cite{Cardoso2015} into much closer
alignment for all objects later than T7, while the differences at earlier types now seem consistent with cosmic scatter
about the calibration.\\

The lower right panel of Fig. \ref{MethaneSpectralType} plots the resulting estimated spectral types against actual spectral types,
for objects from both this paper and \cite{tinney2005}. In the vast majority of cases the estimated 
spectral types predicted based on \Ms$-$\Ml\ colour are in line with the spectroscopic types -- 70\% of the objects 
plotted have estimated types agreeing with their observed types to within $\pm$0.5 types, and 91\% agree to 
within $\pm$1 type. The most prominent exception is W1141-3326,  which (as already noted) was found to be {\em much} later when a 
spectrum was acquired (Y0), than predicted (T5), due to confusion with 
a background object. Of the 5 objects observed to deviate by more than 1 whole spectral 
type from their predicted type, four are in the sense that the observed type was {\em later} than predicted. 
Indeed only one of 53 objects was found to be earlier than predicted by more than a whole spectral type, which 
reinforces the power of these filters for robustly and rapidly identifying (and approximately typing) T- and 
Y-type dwarfs -- if imaging detects a methane absorption signature in an unresolved object (i.e. not
a galaxy) near the position of a \WISE or 2MASS 
candidate object, then it is almost certainly a legitimate match with the \WISE or 2MASS 
candidate, making it a T or Y dwarf. If the estimated type that results is in ``error'' it is more likely
to predict the object to be earlier, rather than later. \cite[A similar
result was found by][]{Cardoso2015}. These results 
give us  confidence in the T dwarf identifications even for the eight T dwarfs in Tables \ref{ImageTable} \&
\ref{MagTable} for which spectroscopy has not yet been obtained.\\

\begin{deluxetable}{cccccc}
\tabletypesize{\scriptsize}
\tablecaption{\Msl\ versus Spectral Type \label{MethaneSpectralTypeReln}}
\tablewidth{8cm}
\tablehead{
\colhead{SpT}  &\colhead{CH$_{\mathrm 4}$s$-$CH$_{\mathrm 4}$l} &\colhead{SpT}  &\colhead{CH$_{\mathrm 4}$s$-$CH$_{\mathrm 4}$l} &\colhead{SpT}  &\colhead{CH$_{\mathrm 4}$s$-$CH$_{\mathrm 4}$l}  }
\startdata
  A0   &  0.000  & M4   &  0.132   & T2   & -0.035 \\
  A5   &  0.000  & M6   &  0.150   & T3   & -0.157 \\
  F0   &  0.000  & M8   &  0.169   & T4   & -0.350 \\
  F5   &  0.000  & L0   &  0.185   & T5   & -0.595 \\
  G0   &  0.001  & L2   &  0.199   & T6   & -0.900 \\
  G5   &  0.015  & L4   &  0.198   & T7   & -1.250 \\
  K0   &  0.050  & L6   &  0.178   & T8   & -1.600 \\
  K3   &  0.076  & L8   &  0.123   & T9   & -2.200 \\
  M0   &  0.099  & T0   &  0.080   & Y0   & -3.000 \\
  M2   &  0.115  & T1   &  0.043   & Y1   & -4.000 \\
\enddata                                                                                                                                             
\end{deluxetable}

\subsection{J3--J2 versus Spectral Type}
\label{J3J2_SpT}

To observe fainter cool dwarfs (i.e. J$\gtrsim$21 candidates where methane observations in the H-band with \Ms,\Ml\ filters
 on the 4m AAT became problematic) we made use of methane sensitive filters in the J-band 
 (i.e. the J3 and J2 filters installed in FourStar).
 These J-band filters see a substantially lower sky background than the equivalent H-band filters, while 
 the blue colours of late T and Y dwarfs mean they see essentially the same flux as in the H-band. 
 This combination makes it feasible to target fainter objects down to J$\sim$22-24 -- and in one extreme case
 to J3$\sim$25 \citep{faherty2014}.\\
 
We take this opportunity to note that the faintest ground-based near-infrared detection of a Y-dwarf 
(WISE J085510.83-071442.5 at J3=24.8$^{+0.5}_{-0.35}$)
 reported by \citet{faherty2014} has been the subject of some debate \citep{Schneider,Luhman2016}.
The claimed discrepancy here is almost certainly not real, but rather a result of  issues
associated with conversion to a standard MKO J passband from the measured bandpasses (in order of decreasing width): 
HST filter F125W 1.10-1.40\,$\mu$m for Schneider et al.; J3=1.21-1.37\,$\mu$m for Faherty et al.; and  HST filter
F127M 1.24-1.31\,$\mu$m for Luhman \& Esplin.
 Faherty et al. measured J3=24.8$^{+0.5}_{-0.35}$, following which Schneider et al. reported F125W=26.41$\pm$0.27 and
Luhman \& Esplin  a mean value of F127M=24.45$\pm$0.1 from 3 observations (all Vega-magnitudes).
 When one considers the different bandpasses, these magnitude differences make sense. F125W includes substantial
 water vapour and methane absorption in its wider band-pass than J3. So the F125W Vega-magnitude should be fainter
 than J3.  In turn J3 includes slightly more
 molecular absorption than F127M, which selects out a Y-dwarf flux peak. The significant ``discrepancy'' that
 Schenider et al. note with the result of Faherty et al. is not in the observed detections in these
 three bands, but rather in J magnitudes derived after conversion from the measured bandpass. For the Faherty et al. J3
 detection, this was done using an empirical colour correction based on (hotter) T and Y dwarfs, which
 is clearly not appropriate for this very cool object. \\
 
\begin{figure}
   \includegraphics[width=90mm,trim=1.0cm 1.0cm 1.0cm 1cm,clip=true]{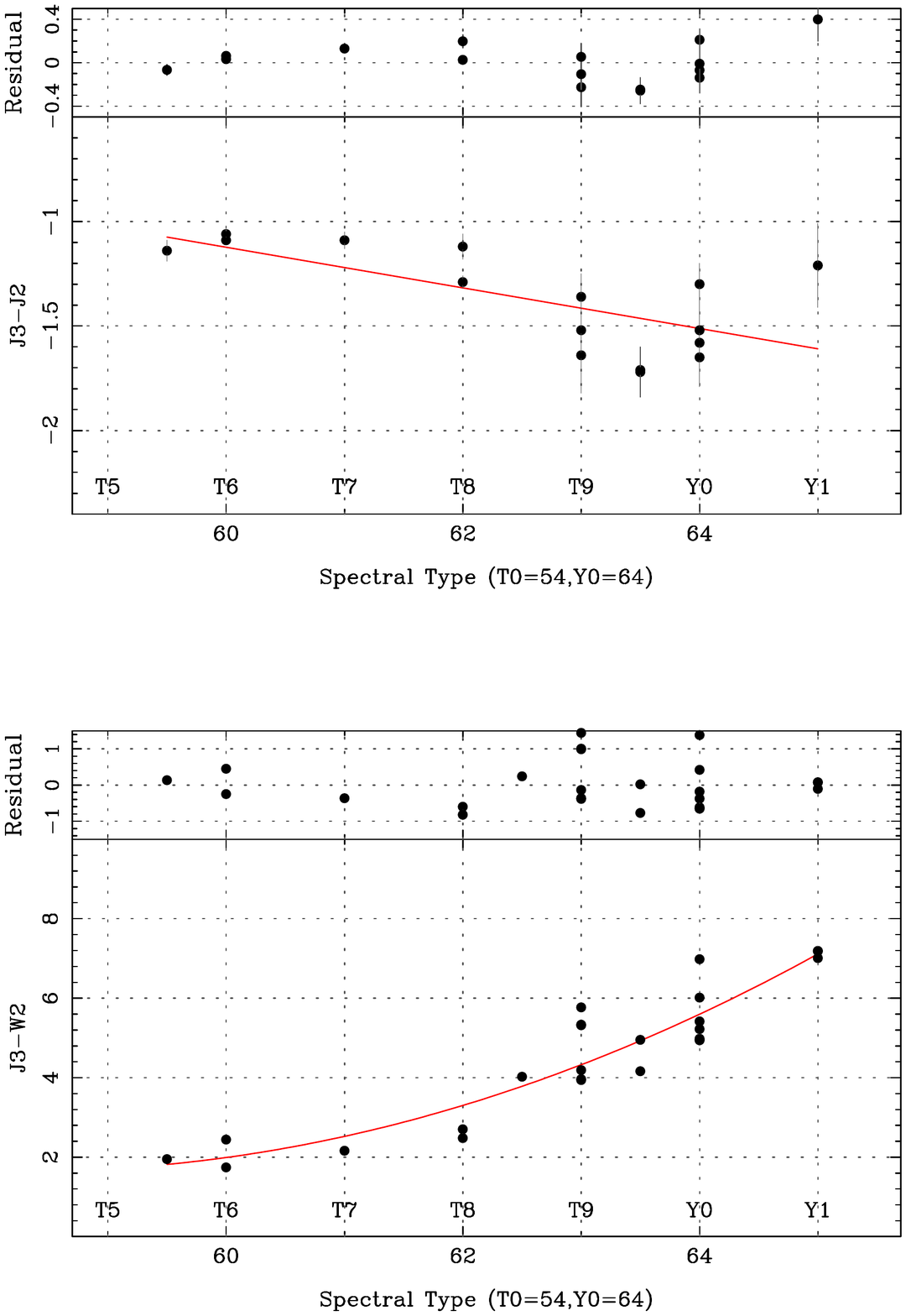}
   \caption{{\em Upper panel} -- Methane-sensitive  J3$-$J2 as a function of T5.5-Y1 spectral type.
            The uncertainties plotted are the combination of photon-counting
            uncertainties, aperture correction uncertainties and photometric
            calibration uncertainties. Typical uncertainties on spectral types
            (not plotted) are $\pm$0.5. The plotted linear fit
            is described in \S\ref{J3J2_SpT}, and has rms 0.19\,mag.\\[3pt]
            {\em Lower panel} -- Temperature-sensitive  J3$-$W2 as a function of T5.5-Y1 spectral type.
            Along with the objects in Table \ref{MagTable}, we also show nine objects with J3 and W2 
            (but without J2) from \cite{tinney2014a}. The uncertainties are as for the upper panel. 
            The plotted parametrisation is a quadratic fit described in \S\ref{J3W2_SpT}, with 
            rms 0.65\,mag.
            \label{J3J2SpectralType}
            }
\end{figure}

We have J2 and J3 photometry for a smaller range of spectral types than we have \Ms\ and \Ml\ -- i.e. only spanning 
T5.5 to Y1. These data are plotted in the upper panel of Fig. \ref{J3J2SpectralType}. 
Recalling that our J3$-$J2 system is defined so that
objects in the A-G spectral type range will have J3$-$J2=0, the observation of a significantly negative J3$-$J2
colour does clearly distinguish T and Y dwarfs from much hotter field stars. The simple linear fit to these data 
in the figure is parametrised as a function of a modified numeric spectral type $n^\prime$, such that $n^\prime = n-60$
(making $n^\prime = 4$ equivalent to a Y0), as
$$ {\mathrm J3}-{\mathrm J2} =  -1.12 - 0.097\,n^\prime, $$
with rms scatter of 0.19\,mag.\\

 Unfortunately, it would appear that the ability to observe fainter targets in J3$/$J2 comes at the penalty of 
 obtaining less information on the spectral types of those objects. The J3$-$J2 colour 
 appears to ``saturate'' at  J3$-$J2 $\approx -1.5$ for late-T and Y dwarfs -- a range
 of spectral types over which \Msl\ continues to become more-and-more negative for later objects. Moreover
 Y dwarfs are only $\sim$0.5 magnitudes more negative in J3$-$J2 than mid-T dwarfs, and the scatter 
 about any trend is substantial at $\pm$0.18\,mag.
 The equivalent numbers for  \Ms$-$\Ml\ are $\sim3$\, mag and $\pm$0.45\,mag.
 This means that while J3$-$J2 can unequivocally identify a very cool brown dwarf's near-infrared counterpart (given the
prior information that a cool dwarf is expected at that position from a large survey like \WISE), 
it cannot provide a very good estimate of {\em how} cool  that brown dwarf is.\\

\begin{figure}
   \includegraphics[width=90mm,trim=1.0cm 14.0cm 1.0cm 1cm,clip=true]{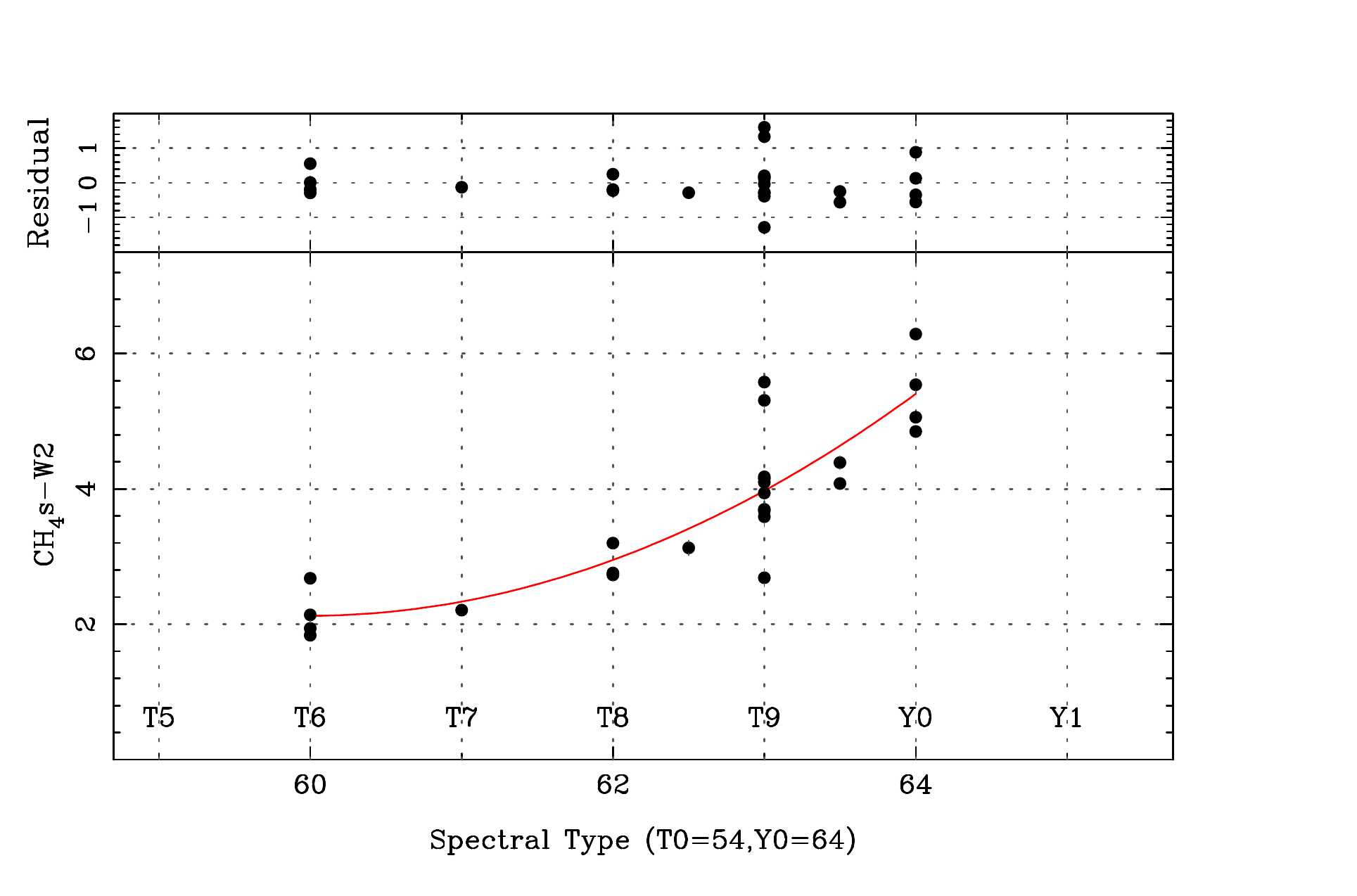}
   \caption{Temperature-sensitive  \Ms$-$W2 as a function of T6-Y0 spectral type.
            Uncertainties are as for Figure \ref{J3J2SpectralType}. 
            The plotted parametrisation is a quadratic fit described in \S\ref{J3W2_SpT}, with 
            rms 0.60\,mag.
            \label{CH4sJ2SpectralType}
            }
\end{figure}

\subsection{J3$-$W2 versus Spectral Type}
\label{J3W2_SpT}

However, all is not lost, because the ability of J3$-$J2 to unequivocally associate a near-infrared
source with a {\em WISE} thermal-infrared one, means that its J3$-$W2 colour is therefore determined. 
J3$-$W2 is primarily sensitive to effective temperature by sampling stellar flux in gaps between
the strongest molecular absorptions over a long wavelength baseline, 
as has already been shown for J$-W2$ by \cite{cushing2011} and \cite{kirkpatrick2012} -- as distinct to the
methane colours \Msl\ and J3$-$J2 which are sensitive to effective temperature by
measuring the strength a specific molecular absorption.\\

As the lower panel of Fig. \ref{J3J2SpectralType} demonstrates, J3$-$W2 shows a pronounced trend with spectral
type, and while the scatter about a quadratic fit to these data is substantial (we again parametrise 
this fit as a function of a modified numeric spectral type $n^\prime$ such that $n^\prime = n-60$ meaning a Y0 dwarf has $n^\prime = 4$)
$$ {\mathrm J3}-{\mathrm W2} =  1.99 + 0.4090\,n^\prime + 0.12313\,n^{\prime 2}, $$
with rms=0.65\,mag, it also spans a large range in J3$-$W2 of over 5.5\,mag. This means that
its discriminating power for assigning an estimated spectral type is similar to that of the 
\Ms$-$\Ml\ colour over this spectral type range. For convenience we show the value of this parametrisation
as a function of spectral type in Table \ref{J3W2SpectralTypeReln}.\\

In Figure \ref{CH4sJ2SpectralType} we show the equivalent plot to the lower panel of Figure \ref{J3J2SpectralType},
but based on \Ms--W2 colour instead of J3--W2. \Ms--W2 colour has a similar ``lever arm'' on spectral
type as J3--J2, spanning a  smaller range in colours (3.5\,mag.), with a slightly smaller rms of 0.60\,mag.
The equivalent polynomial fit is
$$ {\mathrm CH}_4{\mathrm s}-{\mathrm W2} =  2.13 + 0.00519\,n^\prime + 0.20357\,n^{\prime 2}. $$

It is important to note, however, that observation of a candidate \WISE counterpart in J3 or \Ms\ 
(or indeed J or H) alone {\em is not sufficient to make a cool brown dwarf identification}. J2 or \Ml\
is essential to obtain
a clear identification of methane absorption. Without that identification, the association of a chance 
(usually faint) background source will invariably result in a large,  {\em but completely spurious}, J3$-$W2
or \Ms--W2 colour. Both J3 and J2 methane sensitive bands are to identify a near-infrared, methane-absorbing counterpart,
following which J3$-$W2 or \Ms--W2 can provide an estimate of how cool that object is.\\

\begin{deluxetable*}{ccccccccc}
\tabletypesize{\scriptsize}
\tablecaption{J3$-$W2, \Ms$-$W2 versus Spectral Type\label{J3W2SpectralTypeReln}}
\tablewidth{8cm}
\tablehead{
\colhead{SpT}  &\colhead{J3$-$W2} &\colhead{J3$-$W2} &\colhead{SpT}  &\colhead{J3$-$W2} &\colhead{J3$-$W2}&\colhead{SpT}  &\colhead{J3$-$W2} &\colhead{J3$-$W2} }
\startdata
  T5   &  1.70  & \nodata & T8   &  3.30   & 2.95 & Y0   &  5.60 & 5.41   \\
  T6   &  1.98  & 2.13    & T9   &  4.32   & 3.97 & Y0.5 &  6.33 & \nodata \\
  T7   &  2.52  & 2.34    & T9.5 &  4.93   & 4.64 & Y1   &  7.11 & \nodata \\
\enddata                                                                                                                                             
\end{deluxetable*}

\subsection{New Y Dwarfs}
\label{interest}

W0302-5817 is one of two new Y dwarfs presented for the first time in
this paper. Its spectrum (obtained in 2017 Jan) is of low signal-to-noise but
does clearly indicate a Y0 spectral type. This allows us to make a distance
estimate using both its J3 and W2 photometry. The mean correction between $J_{MKO}$ and J3 for cool dwarfs
($J_{MKO}$$-$J3 = 0.20$\pm$0.03) and the median $J_{MKO}$ absolute magnitude for Y0
dwarfs \citep[][M$_{J}$=20.32$\pm$1.25, giving a median M$_{J3}$=20.12$\pm$1.25]{tinney2014a}, 
gives a distance for W0302-5817 lying in the range 24-7.5pc. A better
estimate can be obtained using W2 (the median absolute magnitude for a Y0 has much lower scatter -- 
M$_{W2}$ = 14.65$\pm$0.35), which predicts d=17.5$\pm$3.5\,pc. This places W0302 on the outer edge of the
20\,pc sample of nearby brown dwarfs.\\

W1141-3326 is the second new Y dwarf presented here. As noted earlier, despite being
observed with \Ms,\Ml\ filters in mid-2012, it took some time to obtain spectra for this object, because 
confusion with background sources meant its \Msl\ colour was that of a mid-T dwarf, and so it
was not made a high priority (see Fig. \ref{W1141}).
A high quality spectrum was eventually obtained in 2014 March, and it is an excellent match to
the template Y0 spectrum. W1141-3326's W2 photometry indicates a likely 
distance in the range 11.4-8.0\,pc. This is a distance fully consistent with
the preliminary trigonometric distance presented in \cite{tinney2014a} of 9.5$\pm$0.4\,pc.\\

Both W1141-3326 and W0302-5817 are targets of our on-going parallax program with 
FourStar on the Magellan Baade telescope \citep{tinney2014a}.\\

\subsection{Comparison with Extant Spectroscopy}

Table \ref{Others} compares previously published
spectral types for objects where we present new spectra, as well as
objects where we identify the object as a T dwarf but do not have spectroscopy. For W1433$-$0837
and W1448$-$2534 we obtain the same spectral types from completely independent spectra and typing
processes, while for W2212$-$6931 we obtain a type different by only 0.5 sub-types,
which we consider to be consistent -- especially as examination of Fig. 8 in \cite{Schneider2015}
suggests that W2212$-$6931 has the narrowest 1.3\,$\mu$m peak of of all the T9 objects shown,
and so is possibly the closest to T9.5 in that group on their classification system.\\

We have not been able to obtain a spectrum for W0404$-$6420, and the T9 classification of
\cite{Schneider2015} is 2 sub-types later from that estimated by our \Msl\ photometry.
This is 3-4 times larger larger than the 0.43-0.6\,sub-type scatters observed in our
calibration (Fig. \ref{MethaneSpectralType}), which reinforces our view that while a methane
absorption detection is robust for identifying a cool brown dwarf, and the colour provides
an estimate of the spectral type, spectroscopy remains the ``gold standard' for a firm
classification. \\

\begin{deluxetable*}{cccccc}
\tabletypesize{\scriptsize}
\tablecaption{Comparison with Extant Spectroscopy\label{Others}}
\tablewidth{0pt}
\tablehead{
\colhead{Object}  &\colhead{CH$_{\mathrm 4}$s$-$CH$_{\mathrm 4}$l ``Type''} &\colhead{SpT (this paper)}  &\colhead{SpT (other)} &\colhead{Reference}  }
\startdata
W0404$-$6420 & T7.1$\pm$0.5 & \nodata      & T9 & Schneider et al. 2015 \\
W1055$-$1652 & $>$T8.2      & T9$\pm$0.5   & T9.5 & Martin et al. 2018\\
W1433$-$0837 & \nodata      & T8$\pm$0.5   & T8 & Lodieu et al. 2012\\
W1448$-$2534 & T6.4$\pm$0.5 & T8$\pm$0.5   & T8 & Thompson et al. 2013\\
W2212$-$6931 & Y0.2$\pm$0.5 & T9.5$\pm$0.5 & T9 & Schneider et al. 2015\\
\enddata                                                                                                                                             
\end{deluxetable*}


\section{Conclusion}

Our results show that -- despite using a modest 4m-class telescope like the AAT on targets at J$\sim$20 -- 
methane imaging is an effective technique for refining cool dwarf candidate lists arising
from an external survey. These are magnitudes at which near-infrared
spectroscopy on a 4m telescope would be almost impossible, or at least prohibitively expensive.
Methane imaging observations make both identifications and first-estimates of the spectral type
in a single observation -- without having to obtain spectra
for the multiple candidate targets that can usually be found in the substantial positional error boxes that arise
from a survey like \WISE. \\

Cool brown dwarfs from a large area, but shallow, survey like \WISE will reside quite close to the Sun, and almost invariably
have significant proper motions. As such the position error box to search in a follow-up program grows with time.
Indeed it has been our experience, that substantial proper motions (i.e. $>$ 0.2\arcsec/year) are so ubiquitous for \WISE brown dwarfs, that once follow-up 
extended more than a few years beyond the baseline of
the \WISE mission, any objects which {\em do} positionally match with the \WISE source to better than about
half an arcsecond, are invariably found to {\em not} be cool brown dwarfs, but rather background sources. \\

We have shown that methane imaging observations -- either in the H-band using \Ms\ \& \Ml\ filters or in the
J-band using J2 \&\ J3 filters -- can rapidly and efficiently identify  {\em and preliminarily classify} cool
brown dwarf candidates that arise from large, all-sky surveys like {\em WISE}.\\

We have presented data identifying 21 new T dwarfs and 2 new Y dwarfs from the {\em WISE} All-Sky Survey
using methane imaging \citep[in addition to the Y dwarf W1639 previously published by][]{tinney2012a}.
In many cases these identifications were made for objects using a 4m-class telescope (the AAT) 
for objects at J$\gtrsim$20 -- magnitudes at which near-infrared spectroscopy on such a 
telescope would be either impossible or prohibitively expensive.\\

We present a further 5 late T dwarfs (W0309-5016, W0628-8057, W2017-3421,
W2211-4758) and 3 early T dwarfs (W0042-5840, W2302-7134, W2354-8140)
with methane identifications, for which typing spectroscopy is required.\\

\acknowledgements

CGT gratefully acknowledges the support of ARC Australian Professorial Fellowship grant DP0774000 and ARC Discovery
Outstanding Researcher Award DP130102695.  We are grateful for the extraordinary support we  have
received from the AAT's technical staff  -- K. Fiegert, Y. Kondrat, S. Lee, R. Paterson,  and 
D. Stafford. Australian access to the Magellan Telescopes was supported through the National Collaborative 
Research Infrastructure Strategy of the Australian Federal Government. 
Travel support for Magellan and AAT observing was provided by the Australian Astronomical Observatory.\\

This publication makes use of data products from the Wide-field Infrared Survey Explorer, which is a joint project of the University of California, Los Angeles, and the Jet Propulsion Laboratory/California Institute of Technology, funded by the National Aeronautics and Space Administration. This publication also makes use of data products from 2MASS, which is a joint project of the University of Massachusetts and the Infrared Processing and Analysis Center/California Institute of Technology, funded by the National Aeronautics and Space Administration and the National Science Foundation. This research has made extensive use of the NASA/IPAC Infrared Science Archive (IRSA), which is operated by the Jet Propulsion Laboratory, California Institute of Technology, under contract with the National Aeronautics and Space Administration. This research has also benefitted from the M, L, and T dwarf compendium housed at DwarfArchives.org, whose server was funded by a NASA Small Research Grant, administered by the American Astronomical Society. \\

\facilities{AAT (IRIS2), Magellan:Baade (FourStar, FIRE), Keck::II (NIRSPEC), IRTF (SpeX), WISE, CTIO:2MASS}
\software{Figaro, FIREHOSE (http://www.mit.edu/people/rsimcoe/FIRE/ob\_data.htm), Spextool (Cushing et al. 2004), ORACDR (http://www.jach.hawaii.edu/JACpublic/UKIRT/software/oracdr)}

\end{document}